\documentclass[a4paper,useAMS,usenatbib]{mn2e}

\newcommand{\tff}{$\theta_v=\,\,$45$^{\circ}$}

\newcommand{\dd}{$^{\circ}$}

\newcommand{\mf}{magnetic fields}

\newcommand{\hy}{Hydra~A}
\newcommand{\meanrm}{$\left< RM \right>$}

\newcommand{\cy}{Cygnus~A}

\newcommand{\fig}[1]{Figure~\ref{#1}}
\newcommand{\figs}[1]{Figures~\ref{#1}}

\usepackage{graphicx}
 \usepackage{subfigure}

\title[Interaction of FR~II radio jets with magnetised intra-cluster medium]
{Interaction of Fanaroff-Riley class~II radio jets with a randomly magnetised
  intra-cluster medium}
\author[M. Huarte-Espinosa, M. Krause and P. Alexander]{M. 
Huarte-Espinosa,$^{1,2,5}$\thanks{E-mail: martinhe@pas.rochester.edu} 
M. Krause,$^{3,4}$ and P. Alexander,$^{2,5}$ \\
$^1$Department of Physics and Astronomy, University of Rochester, 600 Wilson Boulevard, Rochester, NY, 14627-0171 \\
$^2$Astrophysics Group, Cavendish Laboratory, 19~J.~J.~Thomson~Ave., Cambridge CB3 0HE, UK \\
$^3$Universit\"atssternwarte M\"unchen, Scheinerstr.~1, 81679 M\"unchen, Germany \\
$^4$Max-Planck-Institut f\"ur Extraterrestrische Physik, 
Giessenbachstrasse, 85748 Garching, Germany \\
$^5$Kavli Institute for Cosmology Cambridge, Madingley Road, Cambridge CB3 0HA, UK}
\begin{document}

\date{Received \today}
\pagerange{\pageref{firstpage}--\pageref{lastpage}} \pubyear{2009}
\maketitle
\label{firstpage}

\begin{abstract}
A combination of three-dimensional (3D) magnetohydrodynamics (MHD)
and synthetic numerical simulations are presented to follow the evolution of 
a randomly magnetised plasma that models the intra-cluster medium (ICM), under 
the isolated effects of powerful, light,
hypersonic and bipolar Fanaroff-Riley class~II (FR~II) jets. We
prescribe the cluster magnetic field (CMF) as a Gaussian random field
with a 
%
	Kolmogorov-like
   energy spectrum. 
%
 Both the power of the 
jets and the viewing angle that is 
used for the synthetic Rotation Measure
(RM) observations
are investigated. 
We find the model radio sources
introduce and amplify fluctuations on the RM statistical properties
which we analyse as a function of time as well as the viewing angle.
The average RM and the RM standard deviation are increased by the
action of the jets.
Energetics, RM statistics and magnetic
power spectral analysis consistently show that the effects
also correlate with the jets' power, and that the lightest, fastest jets 
produce the strongest changes in their environment. 
We see jets distort and amplify the CMFs
especially near the edges of the lobes and the jets' heads.
This process
leads to a flattening of the RM structure functions 
at scales comparable to the source size. The edge features we find are
similar to ones observed in \hy.
The results show that jet-produced RM
enhancements are more apparent in quasars than in radio galaxies. Globally,
jets tend to enhance the RM standard deviation which may lead to overestimations
of the CMFs' strength by about 70\%. This study means to serve as a 
pathfinder for the SKA, EVLA and LOFAR to follow the evolution of cosmic \mf.
\end{abstract}


\section{Introduction}
\label{intro}

Combinations of X-ray and radio observations have 
revealed the 
large-scale interactions between radio jets from active galactic
nuclei (AGN) and the intra-cluster medium (ICM; 
see e.g. \citealp{carilli94,blanton03,mcnamara00,mcnamara05,nulsen05}).
X-ray surface brightness depressions, or X-ray cavities, are seen
in the ICM, the position of which correlates with that of the synchrotron
emitting radio galaxy lobes at GHz-frequencies, e.g. NGC~1275 in Perseus
\citep{boehringer93,schmidt02}, Cygnus~A \citep{carilli94,smith02}, \hy
~(\citealp{nulsen02}, and references therein) and several other clusters
\citep{birzan08}. AGN inject large amounts of energy into
the ICM in the form of relativistic, magnetised and collimated plasma jets.
These seem to be active for periods of some~tens~of~Myr \citep{alexander87,kaiser99,vlj2}. 
The jets drive strong shocks in the ICM which is consequently
heated (\citealt{binney95,kaiser99}; Reynolds, Heinz \& Begelman 2001;
\citealt{alexander02,vlj1}). 
Jet plasma and \mf\ inflate 
a cavity, the ``cocoon'', which expands and displaces the ambient 
medium plasma \citep{scheuer74,kaiser97,hrb98}. The material inside cocoons
is responsible the observed radio synchrotron emission.

The interaction between AGN jets and the ICM has been modelled as a
hydrodynamical process
\citep{clarke97,kaiser99,vicent01,churazov01,alexander02,bruggen02,basson,vlj2}.
It is well established however that \mf\ thread the ICM.  Radio
polarimetry is the major source of information about cluster magnetic
fields (CMFs). The AGN radio emission is typically
\hbox{$\sim$\,10--50\,\%} polarized \citep{bridle84} and the
intervening magnetised ICM gives rise to Faraday rotation
(\citealp{kim91}, and references therein; for a review see
\citealp{feretti08}), which is characterized by the Rotation
Measure (RM). Maps of this quantity have been produced using radio
sources located within the ICM or behind it, as well as using
cluster radio halos (for a review see \citealp{carilli02}). The
distribution in RM images, along with the assumption that CMFs vary
over a single scale, has lead to estimates of the strength of
CMFs within \hbox{5--30$\,\mu$G} in the cool-core clusters \cy,
\hy, A1795~and 3C~295~(see \citealp{dreher87,taylor93,ge93,allen01},
respectively), within \hbox{2--8$\,\mu$G} in the non-cool core clusters
Coma, A119, 3C129~(see \citealp{feretti95,feretti99,taylor00},
respectively) and also in A2634~and A400 \citep{eilek02}. The RM also suggests
that CMFs located in the cores of clusters are stronger than 
those in their outer regions \citep{kim91,clarke01}.
A correlation between the RM values and the mass deposition rates of cluster
cooling flows has been detected as well \citep{soker90}.

  The structure of CMFs seems to be mainly related to turbulence. This has been inferred
  using the spatial variation of observed Faraday rotation images
  and their statistical properties \citep{tribble91b,ensslin03},
  in particular, the shape of the RM structure functions
  \citep{laing08,guidetti10}.  \citet{murgia04} used the RM images
  of A119 and modelled its \mf\ with a three-dimensional multi-scale
  tangled geometry.  They used spatial variations of the field characterized
  by a magnetic power spectrum following a power law in Fourier
  space. Murgia et~al. found that both the mean and the standard
  deviation of the RM, which are used to estimate the strength of
  CMFs, depend on the magnetic spectral index, $n$. They conclude
  that $n\simeq\,$2 produces the best fit to the observations in~A119.
  With a similar approach \citet{govoni06} reported spectral power
  law slope fits \hbox{of $n\,\simeq\,$2} at the core of~A2255, and
  of $n\,\simeq\,$4 at the outer parts. Also, indexes 
     close to 
  %
  %
  11$/$3 
  have been estimated in~A2382
  \citep{guidetti08}, A400,~A2634 and Hydra~A \citep{vogt03,vogt05b}.

  Models of the evolution of CMFs suggest that the fields' structure is
  shaped by the dynamics of the ICM via magnetic flux freezing and
  ideal MHD which is a good approximation in clusters (see
  \citealp{jones08}, and references therein). In this context, CMFs
  should follow the ICM advection closely, particularly the one driven
  by AGN jets. Turbulence in the ICM is expected to be produced during cluster
  formation and mergers
  \citep{ruzmaikin89,roettiger99,dolag02,schindler02,dubois08,ryu08}, as
  well as by AGN outflows (\citealp{ensslin06,ruszkowski07}, and
  references therein). Turbulent ICM flows are thus believed to
  amplify CMFs from weak seed fields to the observed microGauss regime
  by the formation of local small-scale dynamos (see
  \citealp{schekochihin04}, for a numerical review).

Characteristic depolarization gradients are often observed
between the main hotspots of several
powerful radio galaxies (see
e.g. \citealp{laing88}). \citet{laing88} and \citet{garrington88}
have explained such features using path length differences
between each of the radio lobes and the edge of the observed X-ray emitting 
ICM that contains them. The RM map
of the Fanaroff-Riley Class~I (FR~I, \citealp{FR}) 
radio source \hy\ shows opposite signs
in the approaching and the receding radio lobes \citep{taylor93}.
\citet{laing08} have described these RM fluctuations 
with synthetic three-dimensional Faraday screens characterized by
(i) a random distribution of \mf, (ii) a spherically symmetric gas distribution, (iii)
axisymmetric prolate-spheroid cavities resembling the geometry
of the Faraday screen projected by \hy. These 3D differential Faraday
screen models are consistent with the radio source expansion framework
of \citet{scheuer74} and also with observations of X-ray cavities. The
RM gradients in question, therefore, imply that AGN jets make their
way by pushing the ICM gas and its magnetic fields at once. 
%
	This process was recently studied by \citet{guidetti11}
	with high resolution multi-frequency
   images of Faraday rotation and depolarization of the radio
   galaxies 0206+35, M~84, 3C~270 and 3C~353. The latter is an
   extended Fanaroff~Riley Class~II (FR~II) radio source such as
	the ones that we model in this paper. Guidetti et~al. reported,
	for the first time, well defined RM bands, with
   very little small-scale structure, perpendicular to the major
   axis of the radio lobes of these galaxies. Also, they use a simple 
	time-independent model of both the radio source compression and 
	the ambient medium magnetic fields to generate the RM bands 
	in synthetic observations.

Observations of CMFs are limited by the resolution
and sensitivity of telescopes. Many questions about these
fields are still open (see \citealp{beck06}, for a review).
At least for powerful FR~II sources there seems to be little mixing
of ICM plasma with the synchrotron-emitting one;
the observed RM distribution is caused by plasma external to the
radio source \citep{dreher87,carilli02}. The interaction of expanding
radio sources with the magnetised ICM should happen locally to the
sources. An important question here is how significant this 
process is when interpreting RM observations as indicators of the properties
of both the CMFs and the ICM gas themselves. 
The SKA \citep{huarte09,krause09}, EVLA, e-MERLIN, LOFAR and XEUS (the new
generation of telescopes) are planned to have ultra-high sensitivity
and sub-arc-second resolution which will allow a better understanding
about the cosmological evolution and structure of CMFs. e.g. we will
understand important details about the ICM turbulence (see e.g.
\citealp{laing08,guidetti10}) and its heat conduction
\citep[see e.g.][]{schekochihin05,bogdanovic09}.

This is the second of two papers in which we use numerical simulations
to study the evolution of magnetic fields
%
   in Fanaroff-Riley class~II jets and their immediate cluster surroundings.
%
In this work we introduce a combination of 3DMHD and synthetic RM
numerical simulations. We follow the large-scale 
interaction of powerful radio jets from an FR~II
radio source with the \mf\ in the core of a non-cool core cluster.
%
The structure, the evolution and the polarised synchrotron emission of the radio sources themselves are
discussed in our earlier paper (Huarte-Espinosa, Krause \& Alexander, 2011; Paper~1).
Here we concentrate on the 
structure of the ICM and the feedback from radio sources.
%
%
	We only follow the evolution of the sources while jets are active.
%
The behaviour of both energetics and magnetic
power spectra (in Fourier space) of CMFs is explored as a function of
position in the two-dimensional parameter space defined by the velocity
and density of the 
jets. Synthetic RM maps are produced at different 
viewing angles to investigate the observational signatures of 
expanding radio sources.

This paper is organised as follows: in Section~\ref{simul} we briefly
describe the formalism of ideal MHD, the numerical methods we use
and write about our implementation of the ICM, the CMFs
and AGN jets. The results of our simulations are presented in
Section~\ref{results} along with their interpretation and analysis
from the point of view of energetics, RM statistics and magnetic
power spectra in Fourier space. Both the implications and
applications of our work are discussed in Section~\ref{discussion}.
There, we also examine the well-studied case of \hy\ and 
finish by commenting about our model assumptions. Finally, in
Section~\ref{conclu} we summarise the main results of our work and
present our conclusions.

\section[]{Simulations}
\label{simul}

\subsection[]{Governing equations}
\label{mhd}

To describe the dynamics of the plasma in the ICM and AGN radio
jets, we use the system of nonlinear time-dependent hyperbolic equations
of ideal compressible MHD. In three~dimensions and
non-dimensional conservative form, these are given by:

\begin{eqnarray} 
   \label{eq:mass}
   \frac{\partial\rho}{\partial t} + \nabla\cdot(\rho\mbox{\bf V}) &=& 
   \dot{\rho}_\mathrm{j}
   \\
   \label{eq:momentum}
   \frac{\partial (\rho {\bf V})}{\partial t} + 
   \nabla\cdot \left( \rho {\bf V V} + p + B^2/2 - 
   {\bf B B}\right)  &=& \rho {\bf g} + \dot{\bf P}_\mathrm{j}~~~~
   \\
   \label{eq:energy}
   \frac{\partial E}{\partial t} + \nabla\cdot\left[\left(E
   +p + B^2/2 \right)
   {\bf V}-{\bf B}({\bf V}\cdot{\bf B})\right] &=& 
   \dot{E}_j 
   \\
   \label{eq:induction}
   \frac{\partial {\bf B}}{\partial t} - \nabla\times( {\bf V}
   \times {\bf B}) &=& 0, 
\end{eqnarray}
\noindent where $\rho$, $p$, {\bf V} and {\bf B} are the plasma density,
thermal pressure, flow velocity and magnetic fields, respectively.
In (\ref{eq:energy}), \hbox{$E=p/(\gamma-1)+\rho V^2/2+B^2/2$} and
is the total energy density, while $\gamma$ is the 
ratio of specific heats.  In the right hand side of (\ref{eq:mass}),
(\ref{eq:momentum})~and~(\ref{eq:energy}) we have source terms to
implement jets via the injection of mass, $\dot{\rho}_j$, momentum,
$\dot{P}_j$, and kinetic energy, $\dot{E}_j$ (see Section~\ref{jets}),
as well as a Newtonian gravitational acceleration, ${\bf g}$, in order to
keep the initial plasma in a magneto-hydrostatic equilibrium.

\subsection[]{Code and implementation}
\label{code}

We solve the equations of ideal MHD in three~dimensions
using the numerical code Flash~3.1 \citep{fryxell00}.
 Flash's new
multidimensional unsplit constrained transport (CT) solver is employed to 
maintain the divergence of 
magnetic fields down 
\hbox{to $\la$\,10$^{-12}$} \citep{lee08}.
A diffusive HLLC solver \citep{hllc} prevents spurious low pressure
and density values from appearing in the grid.
We use a Courant-Friedrichs-Lewy parameter of~0.25 and
periodic boundary conditions in all the domain's faces.
These boundary conditions prevent numerical noise from polluting
the magnetic spectrum in the grid (Section~\ref{cmfs}).
Our computational domain is a cube with edges 
$|{\bf x}| \le 1/2$~computational units 
(which represent a volume of 
200\,kpc$^3$), Cartesian coordinates and a uniform grid with 
200$^3$~cells.

We present six simulations to study the effects of active AGN jets
on the \mf\ in the core of a non-cool core cluster. Five simulations, which we refer
to as \emph{jets-simulations}, are designed to experiment with the
power of the jets. 
We do this by varying their velocities and densities. These jet-simulations 
are the same 
%
	that we present 
%
in Paper~1. The other simulation,
the \emph{no-jets} model,
is carried out in order to follow the evolution of
the ICM plasma without any perturbations throughout the same
simulation timesteps as the jet-simulations. All of these models
are labelled descriptively and summarised in Table~1. See Paper~1
for a discussion about the low densities and hypersonic velocities
of the jets.

\setcounter{table}{0}
\begin{table}
\centering
    \begin{minipage}{80mm}
   \caption{Simulations and parameters.}
   \begin{tabular}{@{}lccccc@{}}
   \hline
   Simulation
      &$v_j$\footnote{Time average average jets velocity in the nozzle. 
			It is equal to the external Mach~number.}
         &$\eta$\footnote{Time average average jets vs ambient density 
                          contrast in the nozzle.}
	    &$L_j$\footnote{Jets power from equation~(11) in Paper~1.}
               &$t_e$\footnote{Simulations' ending times.} \\    
   name  
      &[Mach] 	
          &
	    &[$\times$10$^{38}$\,W]  
               &[Myr]  \\    
   \hline           
   lighter-slow		 &~40    &0.004   &~~\,4.6    &14.1   \\
   light-slow		    &~40    &0.020   &~\,17.2  &~\,8.3   \\
   lighter-fast		 &~80    &0.004   &~28.1    &~\,7.1   \\
   light-fast		    &~80    &0.020   &128.8    &~\,4.4   \\
	lighter-
	%
	   faster
	%
	&130    &0.004   &112.8    &~\,4.7   \\
   no-jets 	 	 &\ldots   &\ldots    &\ldots     &all of \\
			 & & & &the above \\ 
   \hline
   \end{tabular}
   \end{minipage}
\label{table1}
\end{table}
 
\subsection[]{Initial conditions} 
\label{init}


We implement the cluster gas using an ideal gas equation of state,
a King density profile \citep{king72} and a state of 
magneto-hydrostatic equilibrium with a background Newtonian central 
gravity field (see Paper~1 for details).
   %


\subsubsection[]{Cluster \mf} 
\label{cmfs}

The magnetic ﬁeld within the cluster is set up as an isotropic
random ﬁeld with a power law energy spectrum
Following \citet{tribble91b} and \citet{murgia04}, 
we generate a cubic grid in Fourier space.
For each cell, we define three components of a vector potential which takes the
form ${\bf \tilde{A}}({\bf k})\,=\,{\bf A}({\bf k})e^{i {\bf \theta}({\bf
k})}$, where ${\bf k}$ is the frequency vector
($k^2=k_x^2+k_y^2+k_z^2$) and $i$ is the unitary complex
number, while ${\bf A}$ and ${\bf \theta}$ are the vector's amplitudes and
phases, respectively.  We draw ${\bf \theta}({\bf k})$ from a uniform
random distribution within 0 and 2$\pi$, and ${\bf A}({\bf k})$ is also
randomly distributed but has a Rayleigh probability distribution
\begin{equation}
   P(A,\theta)dAd\theta = \frac{A}  {2 \pi |A_k|^2}
   exp \left( - \frac{A^2}  {2 |A_k|^2} \right) dAd\theta,
   \label{eq:Rayleigh}
\end{equation}
\noindent where we choose the power law Ansatz
\begin{equation}
   |A_k|^2 \propto k^{- \zeta},
   \label{eq:zeta}
\end{equation}
for a given slope $\zeta$.

We transform to real space by taking the inverse fast Fourier Transform
\citep{recipes} of ${\bf \tilde{A}}({\bf k})$. The resulting
magnetic vector potential ${\bf A}({\bf x})$ is multiplied by the plasma density
radial profile (see Paper~1, Section~2.3.1). This multiplication implements the magnetic
flux freezing by generating \mf, the strength of which follows the plasma density
and pressure profiles.

We then take the curl of this vector to get the magnetic field which
is normalised so that the ICM thermal pressure is approximately
ten~times larger than its magnetic pressure. i.e. 
\begin{equation}
   \beta_m = \frac{p}{{\bf B}^2/2} \sim 10,
   \label{betaEQ}
\end{equation}
\noindent which is a reasonable value in this context \citep{carilli02}.

This procedure yields solenoidal \mf\ which are tangled at scales
of order our computational resolution, and characterized by spatial 
variations that follow a magnetic power spectrum with a power law of the form
\begin{equation}
   |B|^2 \propto k^{- \zeta + 2} = k^{-n}.
   \label{eq:POW}
\end{equation}
\noindent We choose a Kolmogorov-like three-dimensional turbulent
slope \hbox{$n= -$11$/$3}, based on the work of 
\citet{vogt03,vogt05b} and \citet{guidetti08}. 
%
   %
	   Note that other values for $n$ would also have been possible (compare 
	   Section~1, above). 
	%
	This choice 
	%
	   may also be
	%
	motivated by theoretical studies of the
   cosmological evolution of cluster magnetic fields
   (e.g. \citealp{bruggen05,ryu08};
	but compare also \citealp{schekochihin04,xu10}).
%
The topology of these
fields is (of course) constrained by the approximation of ideal MHD and
thus we do not model physical magnetic reconnection, but the simulations
do present numerical resistivity.
%
   We note that the Fourier method implicitly imposes maximum and
   minimum scales.

\subsection{Jets}
\label{jets}

By implementing source terms to the equations~(\ref{eq:mass}),
(\ref{eq:momentum})~and~(\ref{eq:energy}) we inject mass, momentum
and kinetic energy to the central grid cells located within a
control cylinder of radius, $r_j$, and height, $h_j$, 
%
	which are 
%
resolved by 3~and
\hbox{8\,cells}, respectively. Inside this ``nozzle'' we update the plasma
density and $x$-velocity using the
constant source terms $\dot{\rho}_\mathrm{j}$ and $\dot{v}_\mathrm{j}$.

Bipolar back-to-back jets are continuously injected
until they reach the computational boundaries. The simulations 
are stopped at this point. The nozzle plasma pressure, $p_j$,
takes the constant value of the central ambient pressure (i.e.
\hbox{$\rho_c/\gamma$}), while ${\bf B}, v_y, v_z$
and $E$ evolve according to the MHD equations 
(\ref{eq:mass})\,--\,(\ref{eq:induction}).  
This implementation is similar to that of \citet{omma04},
but generalised here to the MHD case. 

The power of the jets is given in Table~1 and computed using
the equation~(11) in Paper~1.

\subsubsection{Cocoon contact surface}
\label{cocoon}

We implement a passive tracer, $\tau({\bf x},t)$, which is
injected along with the jets' plasma to distinguish it from the one of
the ambient medium. Defined as a real number within (1$\times$10$^{-10}$,.99),
$\tau$ initially takes these extreme values in the ICM and in the jets'
nozzle, respectively.  The tracer is advected with the jets,
and a comparison of the distribution of both $\tau$ and $\rho$ allows us
to identify the contact surface of the cocoon with an accuracy of about~4
computational cells. 

\section{Results and analysis}
\label{results}

To analyse our simulations in the context of galaxy cluster cores,
we scale the plasma variables in the following way. The size of the
computational domain is given by \hbox{$L=$ 200\,kpc}, the ICM speed of
sound as \hbox{$c_s=\,$1.66$\times$10$^{3}$\,km\,s$^{-1}$}
%
%
\hbox{--\,which} corresponds to a temperature of 10$^8$\,K\,-- and
the cluster central density as \hbox{$\rho_c=
$1$\times$10$^{-23}$\,kg\,m$^{-3}$}.  Each computational cell is
one kiloparsec long, the central cluster radius is \hbox{$r_c=$
160\,kpc} and the nozzle's radius and height take the values
\hbox{$r_j=$ 3\,kpc} and \hbox{$h_j=$ 8\,kpc}, respectively.
%
  For the given ICM properties, the initial RMS magnetic field
  \hbox{is $\sim\,$20\,microGauss.}
%

\subsection{Evolution of the plasma}
\label{visit}

\begin{figure*}
       \includegraphics[width=\textwidth]{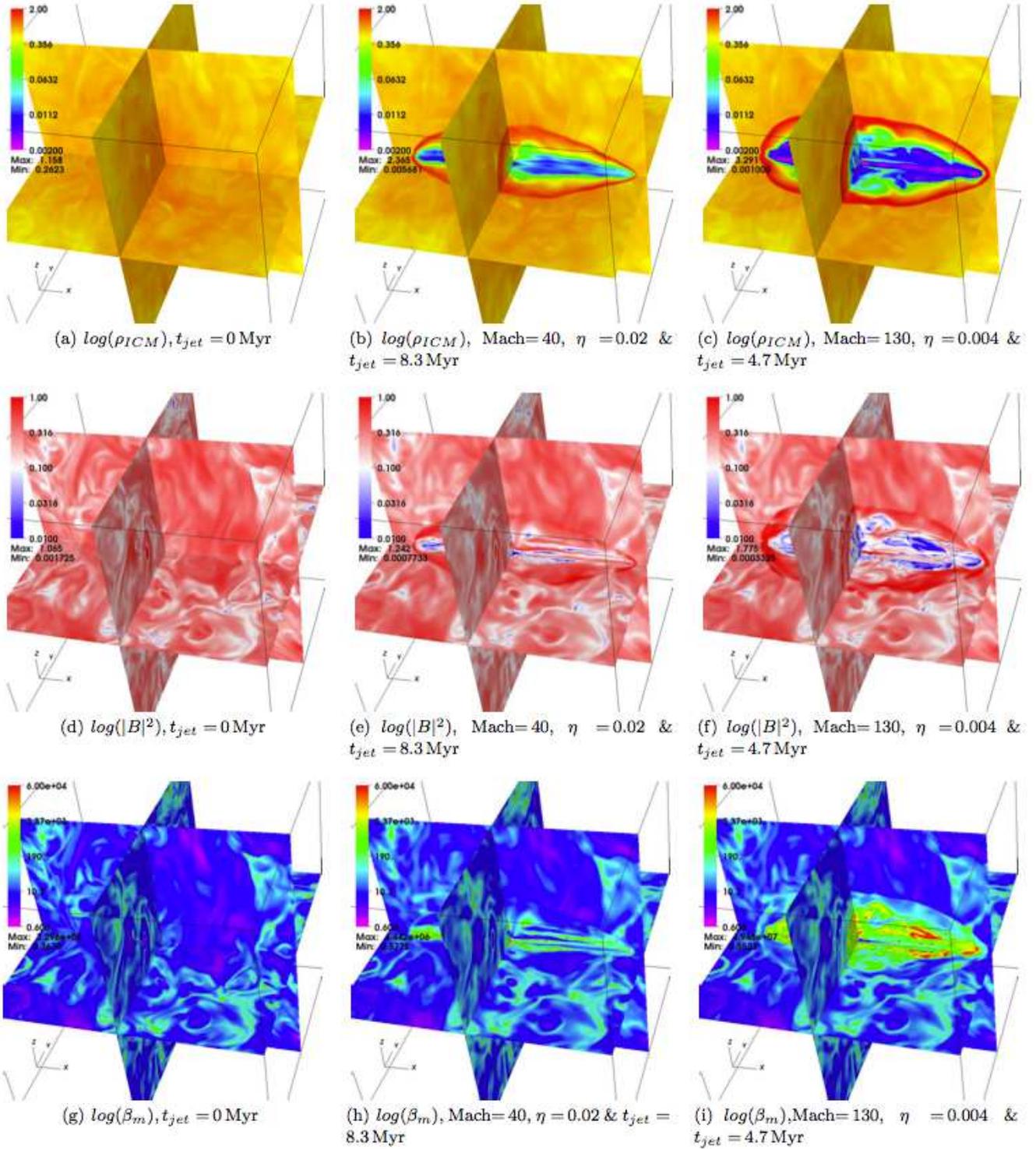}
  \caption{Logarithmic false colour
maps of the plasma density (top row), the magnetic fields strength
(middle row) and the magnetic beta parameter (bottom row). 
The top and middle rows are shown in computational units. The left
column shows the ambient plasma after one relaxation
time (equivalent to 118.5\,Myr), but
just before the injection of jets. The middle column shows the
plasma after the effects of the light-slow source at the
end of the simulation. The right column shows the ambient medium
after the injection of the lighter-faster
source at the end
of the simulation.
}
     \vspace*{0pt}
\label{colormaps}
\end{figure*}

We let the magnetised ICM plasma described in Section~2.3 to relax
for $\sim118\,Myr$ before we inject the jets.  This timescale
represents one thermal crossing time (i.e. the ratio of the
computational domain's length over the gas sound speed). During
this phase the magnetic energy of the cluster dominates the kinetic
energy which was initially zero. CMFs produce local, subsonic and
homogeneous 
%
	random
%
motion on the plasma.  
%
%
We see the three-dimensional magnetic power spectrum
of the plasma preserves the initial Kolmovogov turbulent profile.
Mild losses of about half an~order of~magnitude develop in the spectrum
at \hbox{scales $\la$\,10\,kpc} due to numerical diffusion.
%
   We only follow this phase in order to reduce the initial differences
   between the magnetic and the thermal pressure profiles of our initial
	setup (Section~2.3),  
	but we do not intend to simulate the evolution of the ICM and its 
	magnetic fields at any time before the injection of our jets.

The left column in Figure~1 shows the ICM plasma at the end of the
relaxation phase. These are logarithmic false colour maps of the plasma
density (top row), the magnetic field strength (middle row) and
the magnetic beta parameter (bottom row).  The injection of the
jets starts at this point; at $t_{jet}=\,$0\,Myr.  Then we see that our
simulations produce the basic fluid dynamical structures found in
previous jet simulations (e.g. see \citealt{basson}; \citealt{omma04};
\citealp{reynolds01,vlj2}).  A strong bow~shock is formed in the
ambient medium, behind which a contact discontinuity bounds the
cocoon inflated by the jets' former plasma and \mf.  The middle
column in Figure~1 shows the plasma after the effects of the
light-slow source at the end of that simulation. In the right column,
we present the plasma after the injection of the lighter-faster 
source at the end of that simulation (see Table~1). Note the dependence
of the cocoon geometry, which is marked by yellow surfaces in Figures~1(b)
and~1(c), on the density of the jets.  A full discussion of the 
source properties and evolution is covered in Paper~1.

\begin{figure}
  	\includegraphics[width=84.00mm]{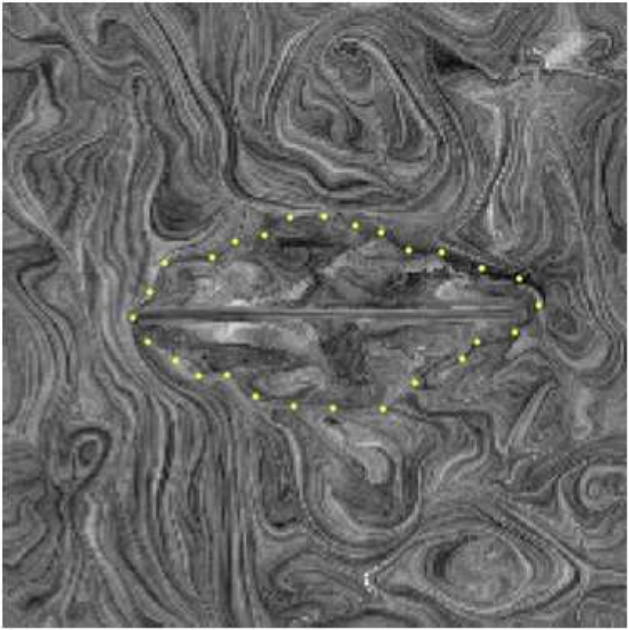}
  \caption{Central slice thought our cubic computational domain showing
the geometry of cluster magnetic field lines under the effects of the
``lighter-fast'' source at $t_{\rmn{jet}}=\,$5.3\,Myr. 
The cocoon is marked
with yellow dots and jets are in the plane of the image, horizontally.
We note the grayscale has no relation to the strength of the fields
at all.
}
     \vspace{0pt}
\label{figTopo}
\end{figure}

As the sources develop, the ambient plasma is compressed and the shocked
region between the cocoons and the bow~shocks expands. During this process
the CMF component that is normal to the direction
of the cocoon expansion is compressed and stretched (\fig{figTopo}).
The strength
of the CMFs is thus amplified, particularly ahead of the jets' working
surfaces, and shows local enhancements above the mean field intensity
--Figures~1(e) and~1(f).
Such enhancements reach factors up to about 7.4, 7.7,
10.9, 7.5 and 12.2, respectively, for the jet-simulations in the
order in which they appear in Table~1. 

We see (Figure~1, bottom row) that jets produce considerable gradients
in $\beta_m$ inside the sources; the cocoons and the ambient medium
show values of order~10000 and~10, respectively. Yet the global
change of $\beta_m$ in the ambient medium is fairly low, $\beta_m
\sim\,$10 throughout the simulations.

\subsection[]{Energetics}
\label{energetics}

\begin{figure}
  \centering
        \includegraphics[width=.52\textwidth]{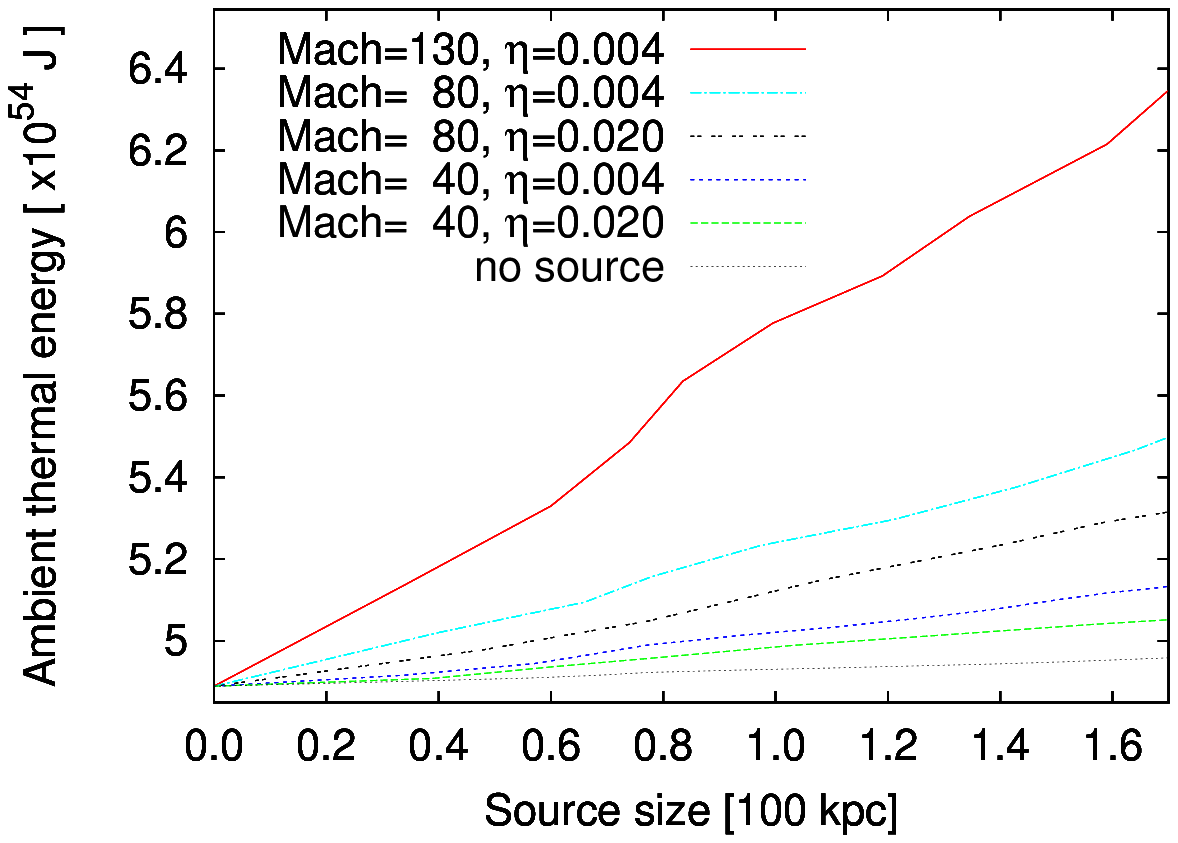} \\ 
        \includegraphics[width=.52\textwidth]{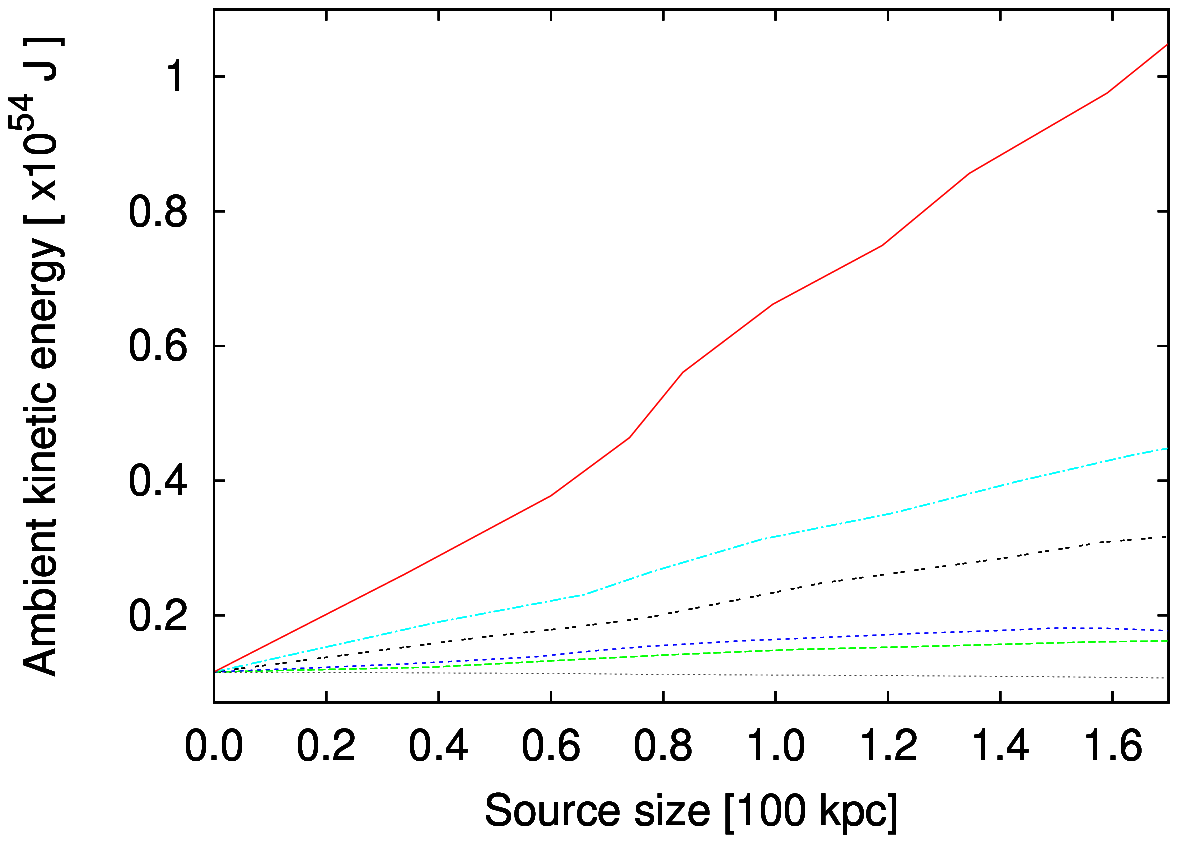} \\
        \includegraphics[width=.52\textwidth]{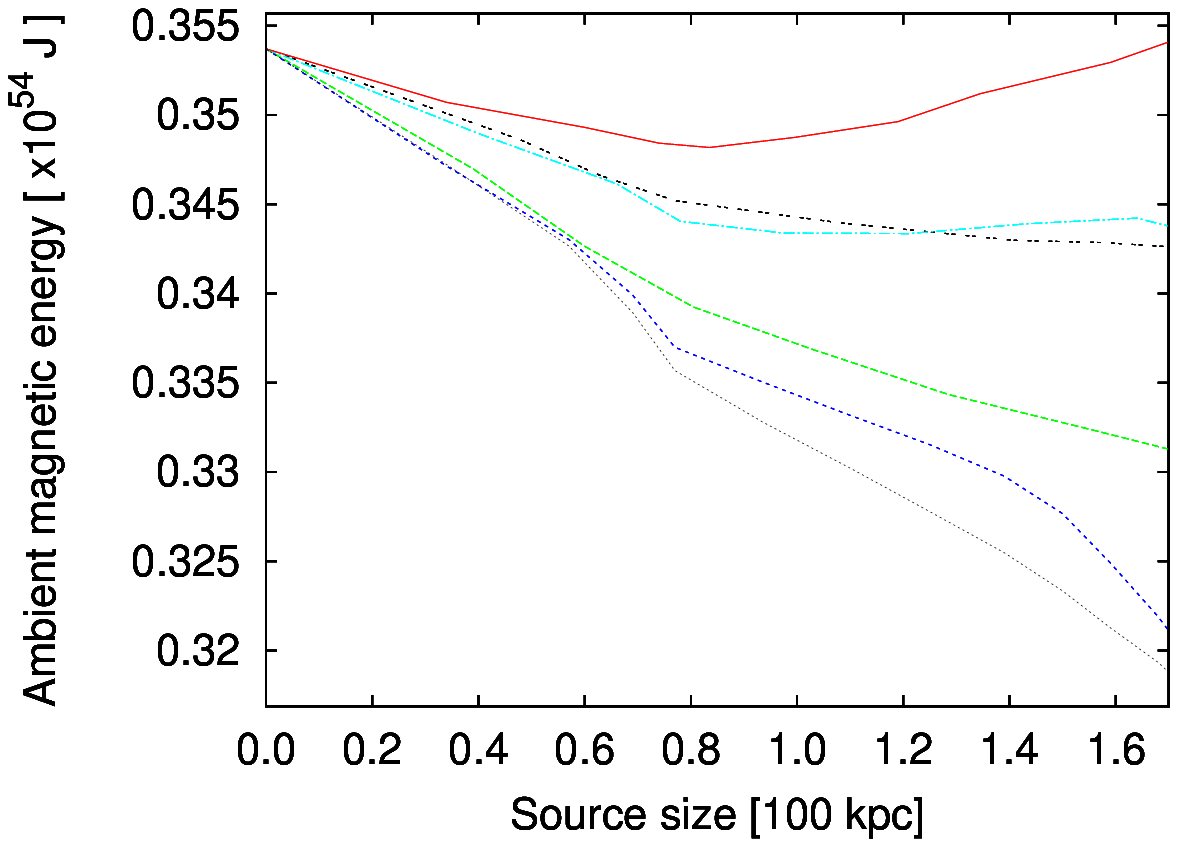} 
     \vspace*{0pt}
 \caption{ICM energetics as a function of source size measured 
 from hotspot to hotspot.}
\label{fig:energy}
\end{figure}

We have calculated the energetics of the ICM as a function of the
source size which is measured from hotspot to hotspot.  Figure~\ref{fig:energy}
has three panels which show the thermal, the kinetic and the
magnetic energies of the ICM, from top to bottom. Each panel has
six profiles corresponding to the simulations in Table~1.
We see that in the absence of jets (black dotted curve) the ICM 
thermal energy increases slightly in
time due to dissipation of magnetic energy which falls \hbox{by
$\sim\,$10\%} 
%
	due to numerical diffusion.
%
When jets are present,
on the other hand, the total energy of the ICM increases consistently.
The energy transfer from jets to the ambient medium is predominantly
thermal and proportional to their velocity. Jets
with a density contrast of 4$\times$10$^{-3}$ are more efficient in 
this process than those with a density contrast of 2$\times$10$^{-2}$
(Figure~\ref{fig:energy}).
%
%
This behaviour is characteristic of very light jets which inflate fat
cocoons that provide large areas over which to affect the surrounding
ICM. In contrast, the forward momentum of relatively heavier jets leads to
smaller cocoons.  Also, both the pressure and the speed of sound inside 
the cocoons with lighter jets are higher than those inside the cocoons 
with relatively heavier jets \citep{vlj1}.

The 
decay of the ICM magnetic energy is slowed down, but
not suppressed, by the Mach-40 jets (see the green and dark-blue curves
in Figure~\ref{fig:energy}). In contrast, the faster jets are able to
   significantly impede 
the decay by the time that their cocoons extend for $\sim\,$70\,kpc
(from hotspot to hotspot).  As the lightest, faster sources continue
expanding (see the red and light-blue curves) they increase 
 the energy
of the CMFs in proportion to the jet velocity.  This increase however
tends to flatten modestly after the cocoons of these sources 
are~$\sim\,$130\,kpc long.

\subsection[]{Rotation measure}
\label{RMsection}

We use the information 
from our simulations to produced synthetic RM maps
and compare the predictions of our models 
with the observations. Our maps are produced by computing
\begin{equation}
   \rmn{RM} = \rmn{812} \int\limits_{D_0/\rmn{kpc}}^{D/\rmn{kpc}} 
   \left( \frac{ n_e }{ \rmn{cm}^{-3}} \right) 
   \left( \frac{ B_{\parallel} }{\mu\,\rmn{G}} \right) dl 
   \, \rmn{rad/m}^2, 
   \label{RM}
\end{equation}
\noindent along the line of sight, 
where $D_0, D, n_e$ and $B_{\parallel}$ are the location 
of the computational domain's boundary, the location of the cocoon's contact surface
(the spatial distribution of which is given by the tracer of the jets;
Section~\ref{cocoon}), the ICM electron density and the CMF component that is
parallel to the line of sight, respectively. We note that the magnetic
fields inside the sources are not considered for this analysis, and hence, 
there is no internal rotation.

The viewing 
or inclination 
angle, $\theta_\mathrm{v}$,
is measured from the jets' axis to the line of sight. 
Jets point towards the observer when $\theta_\mathrm{v}=\,$0\dd, and higher
viewing angles are obtained by rotating 
the simulation data cubes 
perpendicularly to the jets, anticlockwise. 
Thus jets are on the plane of the sky, horizontally, 
at $\theta_\mathrm{v}=\,$90\dd.  For a consistent comparison at
all viewing angles, the computational cells 
which are located 
at radii larger than 100\,kpc
are discarded from our cubic computational domain of 200\,kpc$^3$
(Figure~\ref{RMmapsEX}). The geometry of the Faraday screens, or the RM
integration volumes, depends on both the shape of the cocoons' contact surface
--\,which is different in each simulation\,-- and the
viewing angle. Observationally, the measured screens 
are modified by noise and the finite telescope resolution as well \citep{ensslin03}.

\subsubsection{No-source RM maps}
\label{RMmapsB}

\begin{figure*}
\begin{center}
\includegraphics[width=\textwidth]{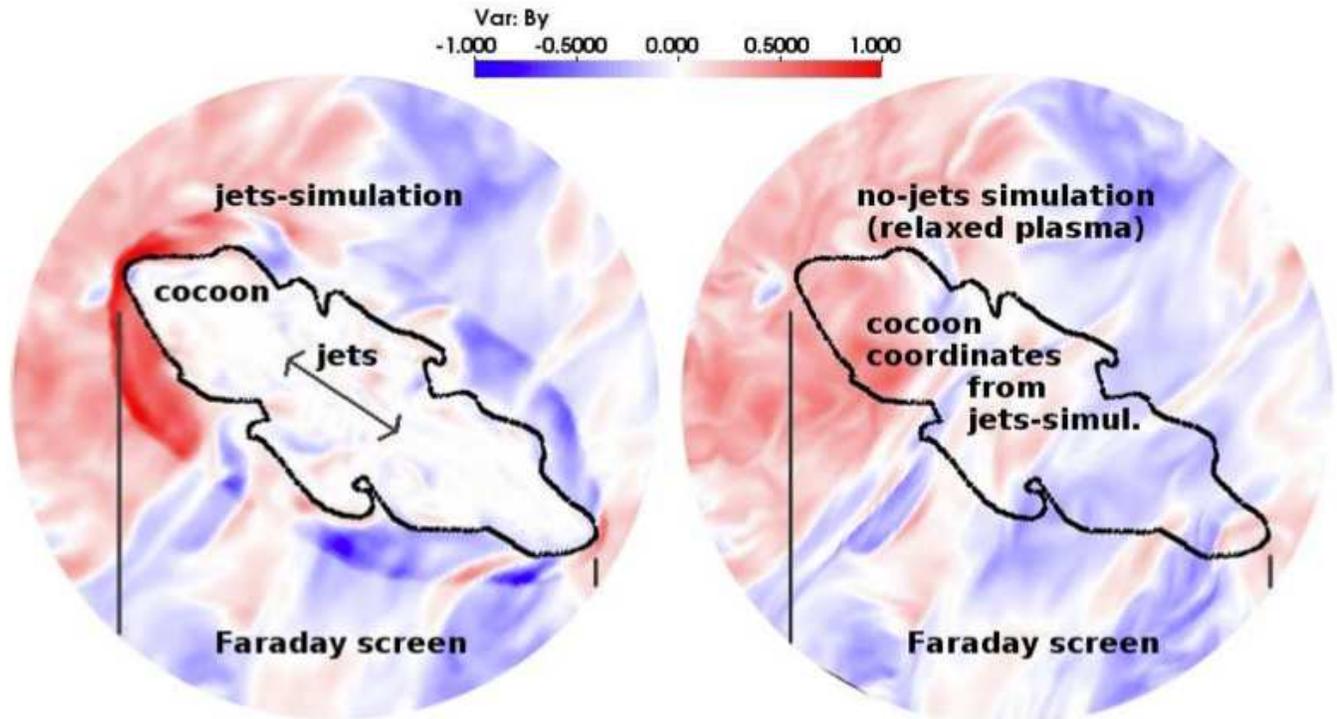} \\
  \caption{
Sketch of the geometry employed for the no-source RM maps. The linear colour scale shows the $y$~component
of the CMFs normalized to their maximum value in the $X$--$Z$ plane through the centre of the cubic
computational domain. No-source RM images
are produced using (\ref{RM}); gas density and magnetic field factors
are taken from the no-jets simulation, whereas the integration
limits are taken from a jets-simulation. See Section~3.3.1 for details.}
     \vspace*{00pt}
\label{RMmapsEX}
\end{center}
\end{figure*}

  We are 
  only
  interested in the statistical changes of 
  %
	  the RM which are 
  %
  caused by the
  expansion of FR~II sources. In our simulations however we also
  see modest, yet detectable, RM variations that are caused by subsonic
  %
  random 
  %
  flow in the ICM, independently of the source
  expansion (see top of Section~\ref{visit}). In order to interpret
  the mechanical work specifically done by the sources on the ICM's gas and magnetic
  fields, we produce \emph{``no-source''}
  RM maps. To this end we use equation~(\ref{RM}), but combine
  information from different simulations in the following way:
  \begin{itemize}
    \item $n_\mathrm{e}$ and $B_{\parallel}$ are taken from the no-jets 
    simulation (Section~\ref{code}),
    \item $D_0$ and $D$, which define the geometry of the Faraday Screen,
    are taken from the jets-simulations (see Table~1).
  \end{itemize}
  This idea is illustrated in Figure~\ref{RMmapsEX}.  Hence
  ``no-source'' RM maps give the RM structure which would have been
  observed against an evolving radio source had it {\emph{\bf not}}
  affected the ICM dynamically.

  We have produced no-source RM maps at $t_{\rmn{jet}}=\,$0, 4.4,
  4.7, 7.1,~8.3 and~14.1\,Myr (i.e. 
  just before the injection of jets and at the five~timesteps 
  that correspond to the end of the jets-simulation,
  $t_\mathrm{e}$; see Table~1).

\subsubsection{Synthetic RM maps}
\label{RMmaps}

\begin{figure*}
     \subfigure[No-source RM map at $t=\,$0\,Myr   (left), 
	             No-source RM map at $t=\,$7.1\,Myr (middle),
	                       RM map at $t=\,$7.1\,Myr (right).]
        {\label{RM-lighter-fast-90}
   \includegraphics[width=\textwidth]{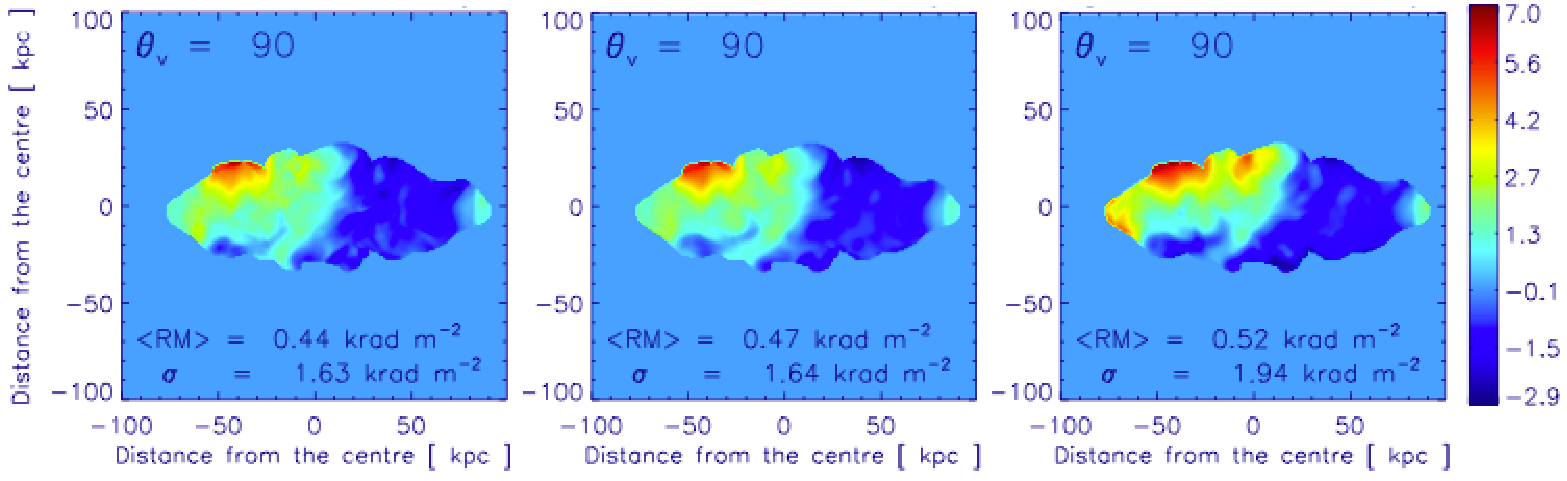}}
     \subfigure[RM at $y=-$25~(left), $y=$0~(middle) and $y=$25\,kpc~(right).
	  The black, red and green curves correspond to the left, the middle and
	  the right panels in (a), respectively.]
        {\label{RM-lighter-fast-90-cut}
  \includegraphics[width=.34\textwidth,bb =.5in 0in 6.8in 6.52in,clip=]
{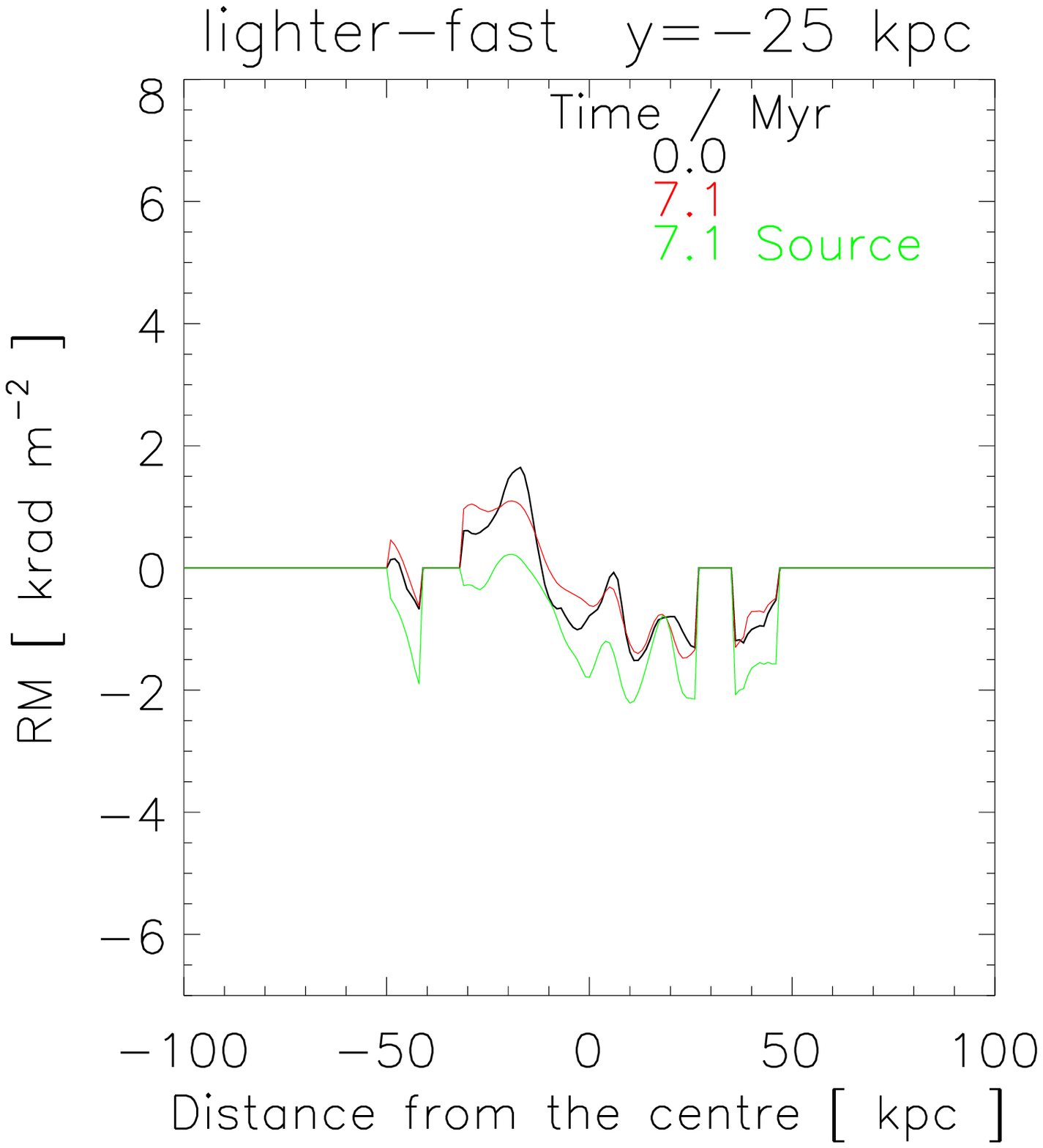}
  \includegraphics[width=.3\textwidth,bb =1.2in 0in 6.8in 6.51in,clip=]
{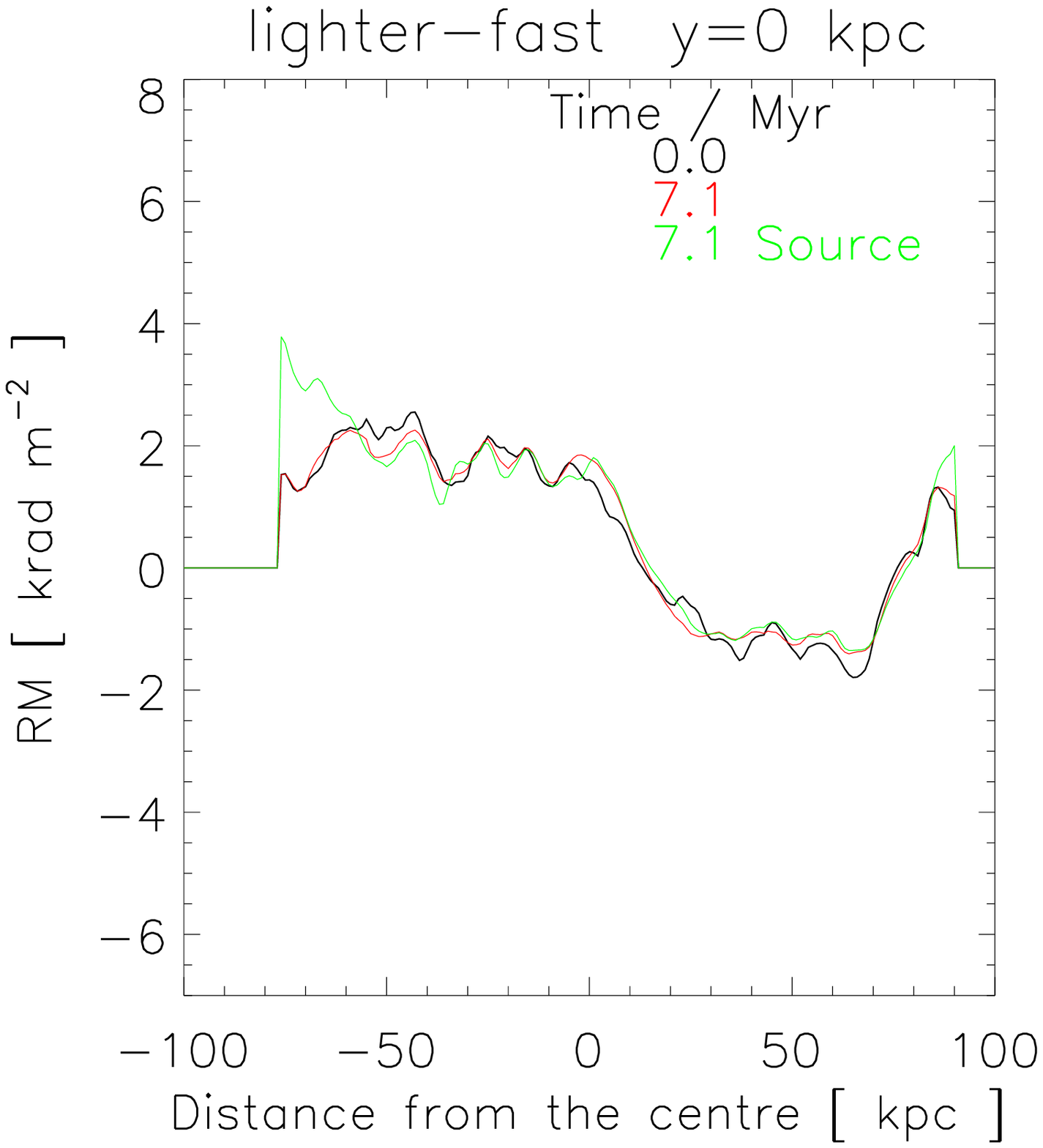}
  \includegraphics[width=.3\textwidth,bb =1.2in 0in 6.8in 6.51in,clip=]
{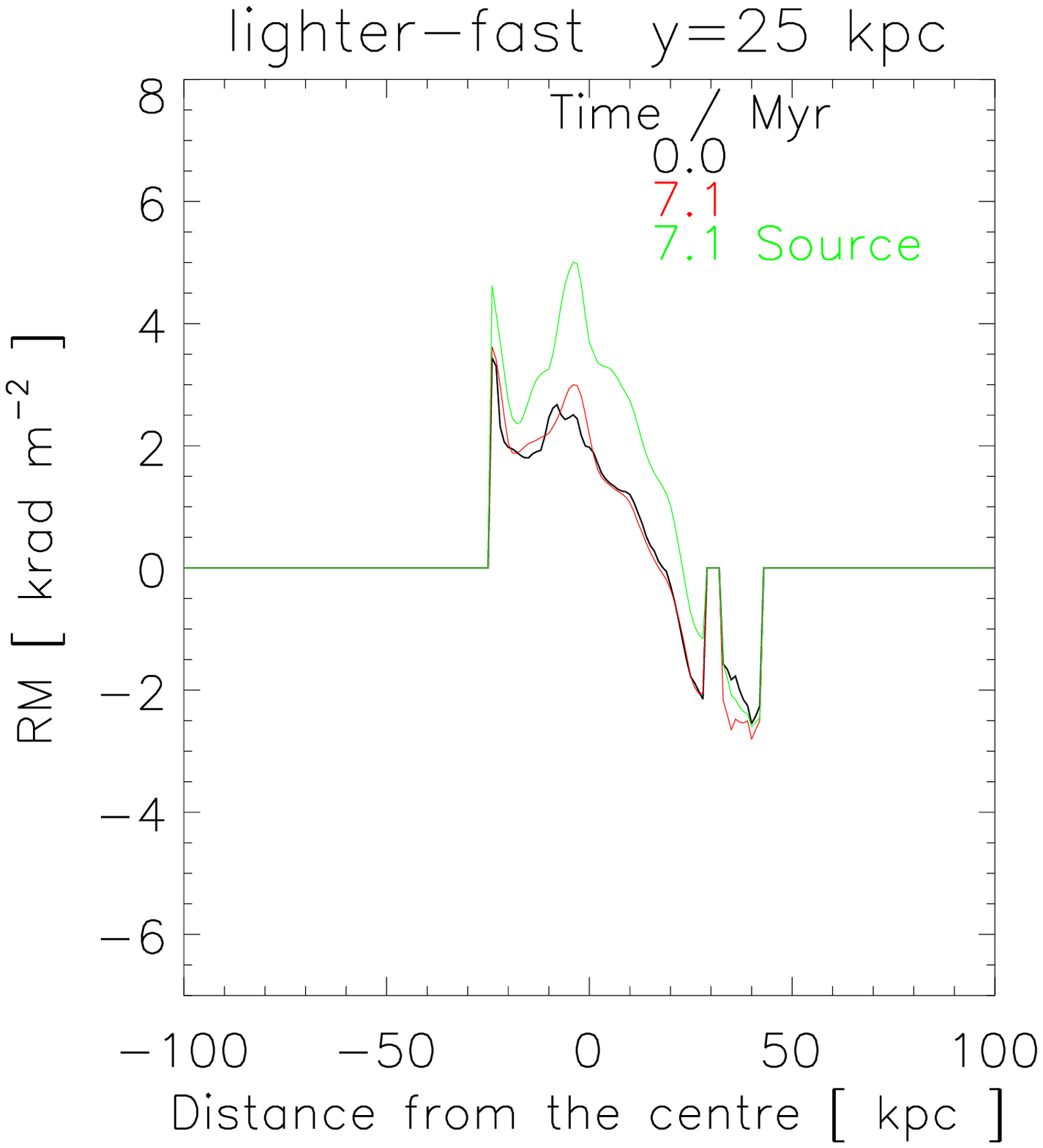}}
     \subfigure[RM histograms of the three maps in~(a). Colors are as in panel~(b).]
        {\label{RM-lighter-fast-90-hist}
  \includegraphics[width=.4\textwidth,bb =0.6in 0.25in 5.6in 5.505in,clip=]
{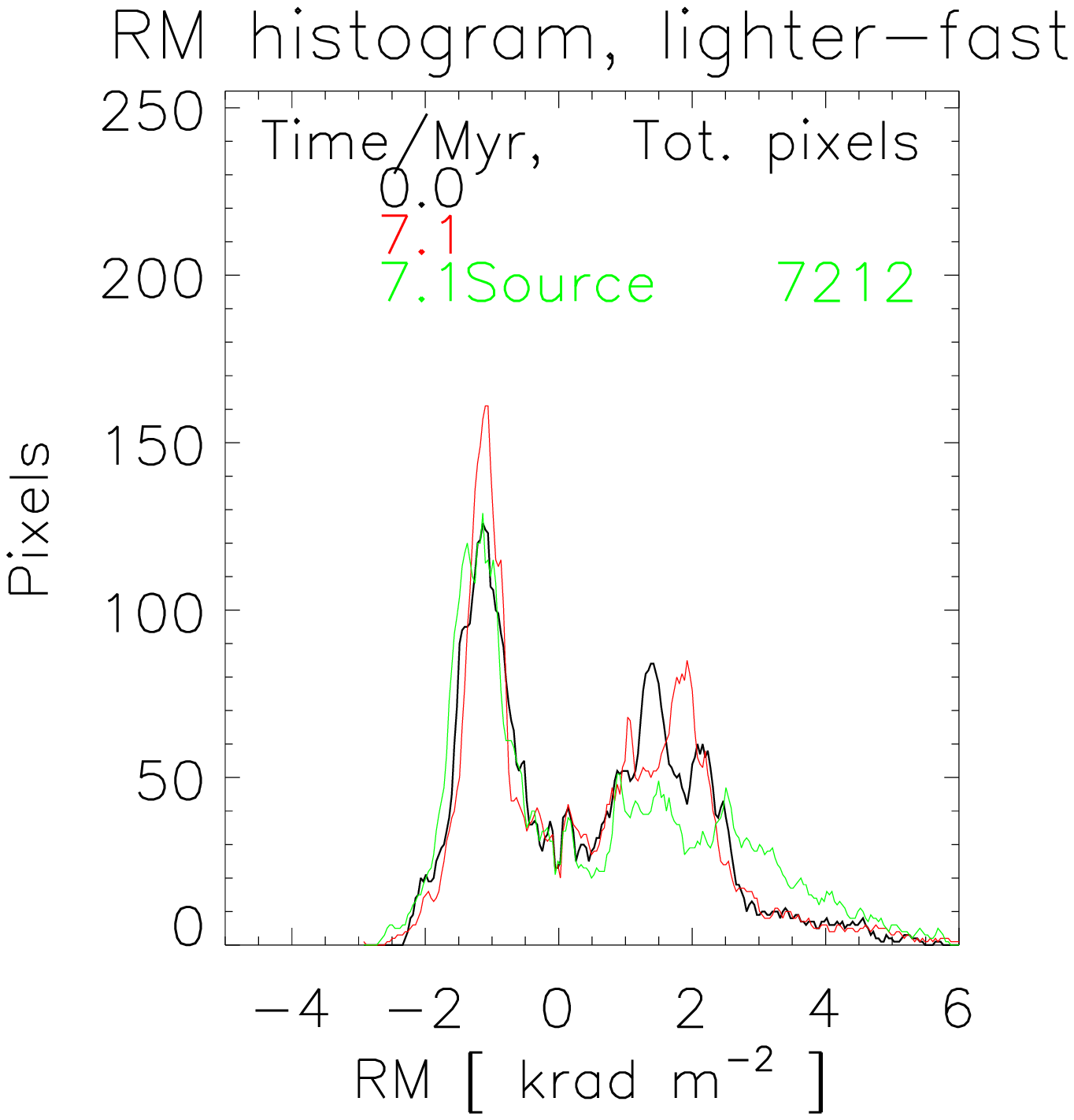}}
     \vspace*{0pt}
  \caption{RM maps and histograms of the lighter-fast simulation at
\hbox{$\theta_v=\,\,$90$^{\circ}$}.  {\bf (a)}~From left to right:
the no-source RM map at $t_{\rmn{jet}}=\,\,$0\,Myr (using the cocoon
coordinates of the lighter-fast source at $t_{\rmn{jet}}=\,\,$7.1\,Myr);
the no-source RM map at $t_{\rmn{jet}}=\,\,$7.1\,Myr; 
the RM map at $t_{\rmn{jet}}=\,\,$7.1\,Myr. 
{\bf (b)}~Each panel shows RM profiles corresponding
to horizontal lines through the three maps in~(a). 
}
\label{RM-lighter-fast90}
\end{figure*}

\begin{figure*}
  \centering
     \subfigure[No-source RM map at $t=\,$0\,Myr   (left), 
	             No-source RM map at $t=\,$7.1\,Myr (middle),
	                       RM map at $t=\,$7.1\,Myr (right).]
        {\label{RM-lighter-fast-45}
   \includegraphics[width=\textwidth]{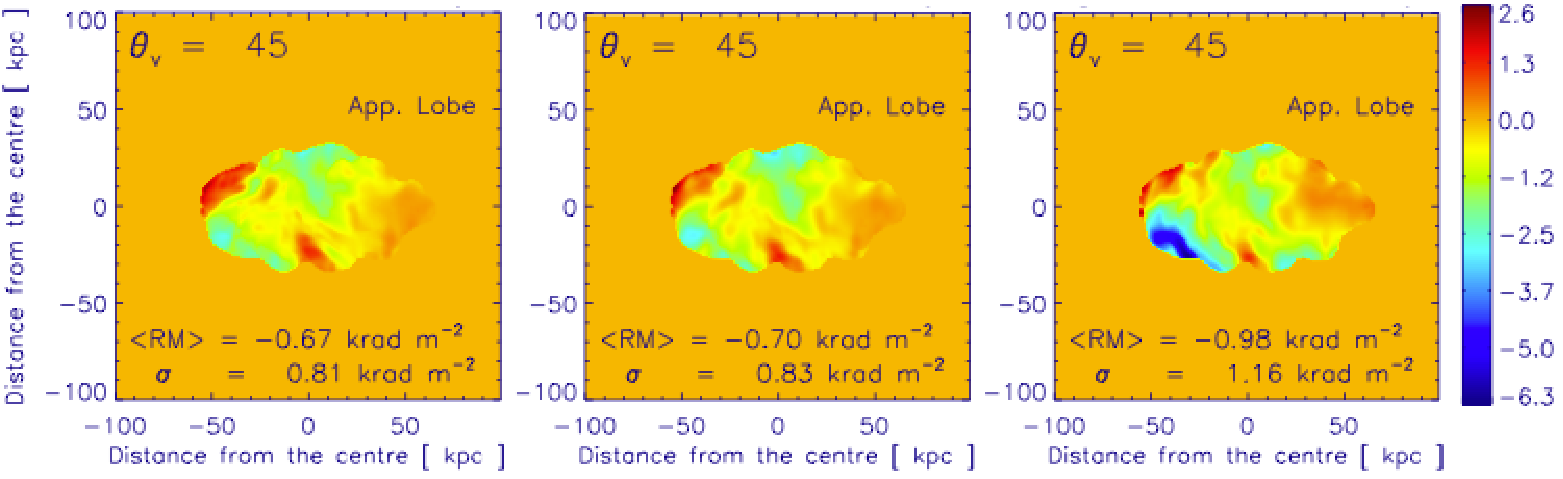}}
     \subfigure[RM at $y=-$25~(left), $y=$0~(middle) and $y=$25\,kpc~(right).
	  The black, red and green curves correspond to the left, the middle and
	  the right panels in (a), respectively.]
        {\label{RM-lighter-fast-45-cut}
  \includegraphics[width=.34\textwidth,bb =.5in 0in 6.8in 6.52in,clip=]
{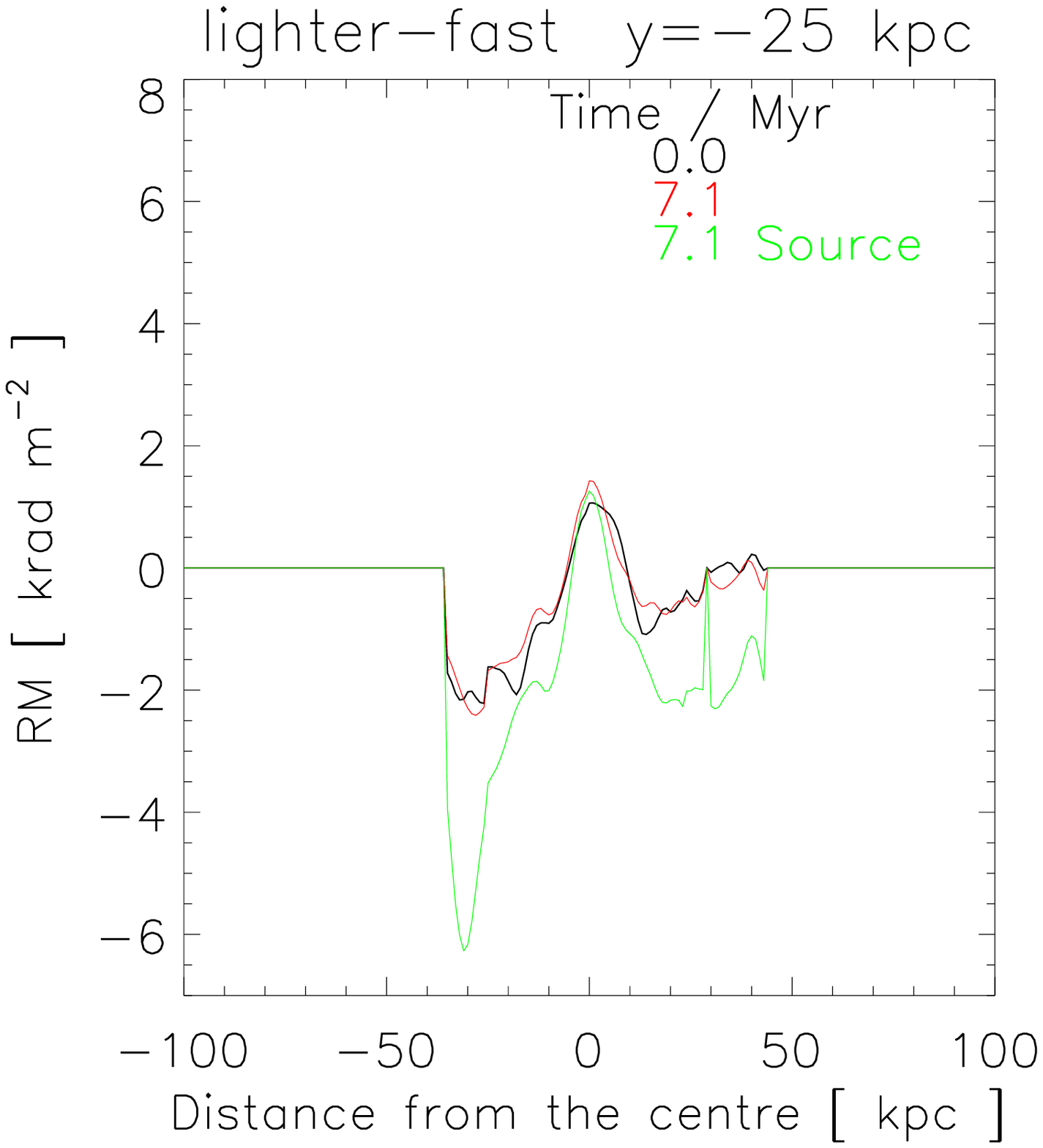}
  \includegraphics[width=.3\textwidth,bb =1.2in 0in 6.8in 6.51in,clip=]
{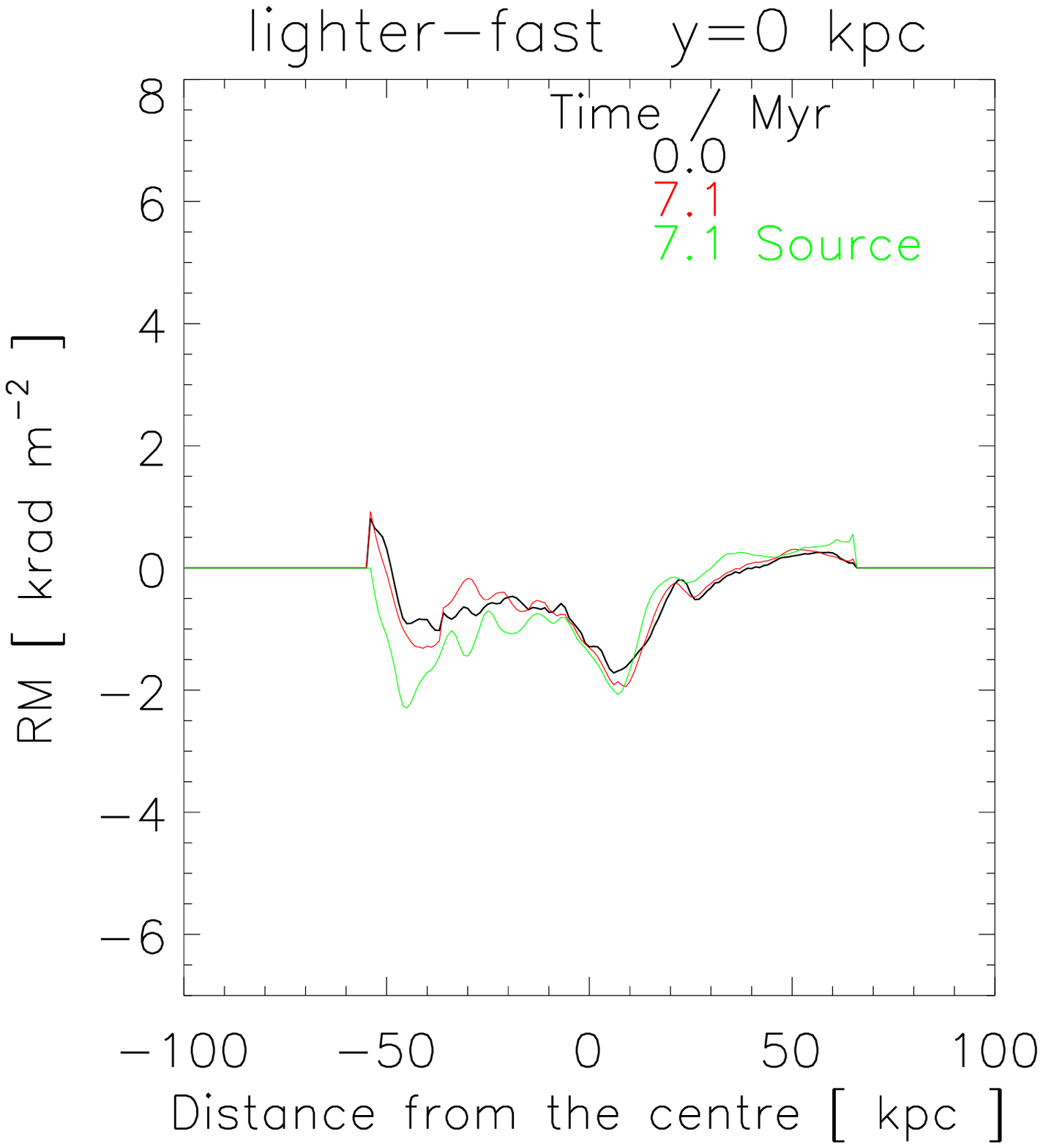}
  \includegraphics[width=.3\textwidth,bb =1.2in 0in 6.8in 6.51in,clip=]
{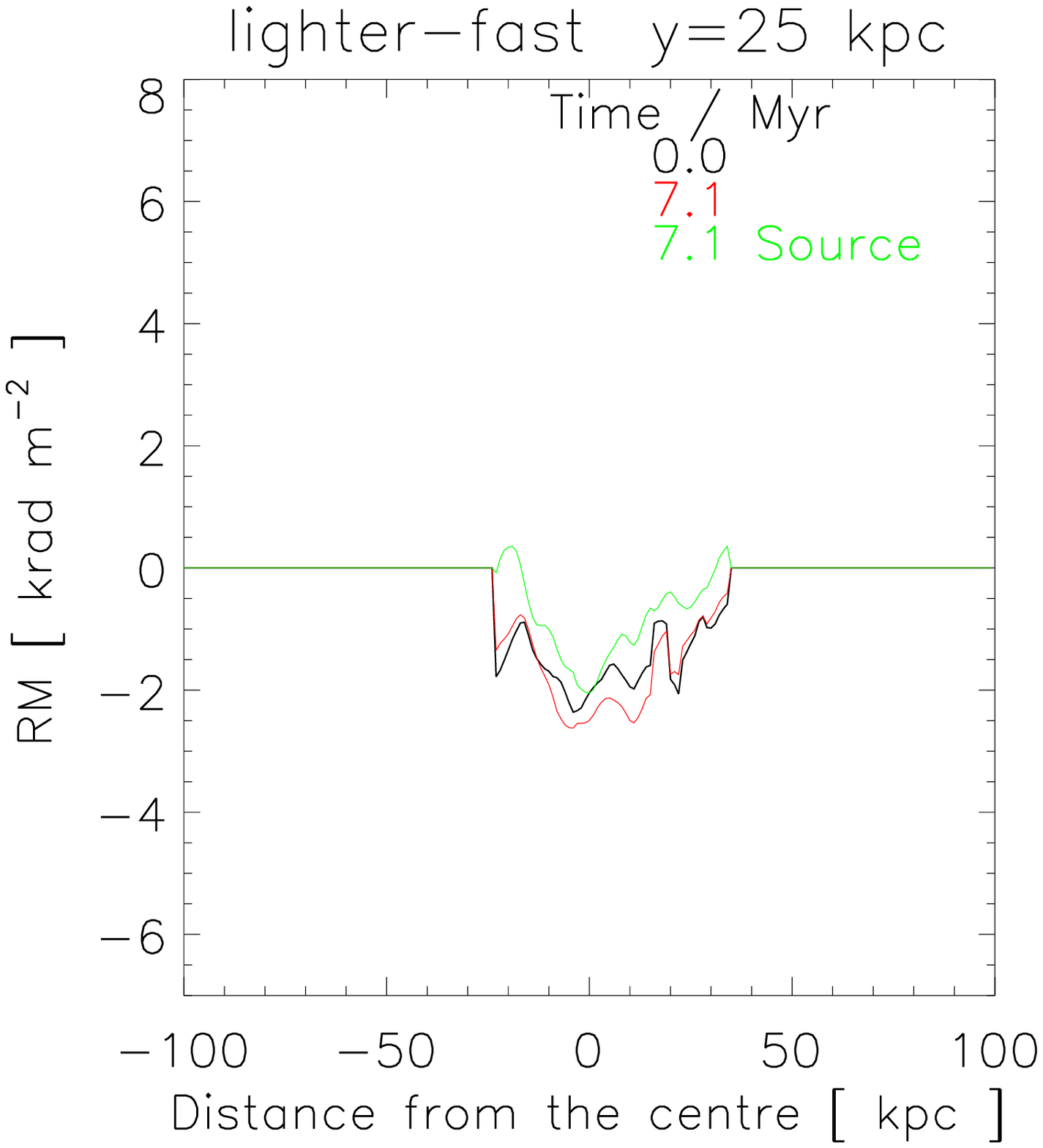}}
     \subfigure[RM histograms of the three maps in~(a). Colors are as in panel~(b).]
        {\label{RM-lighter-fast-45-hist}
  \includegraphics[width=.4\textwidth,bb =0.6in 0.25in 5.6in 5.505in,clip=]
{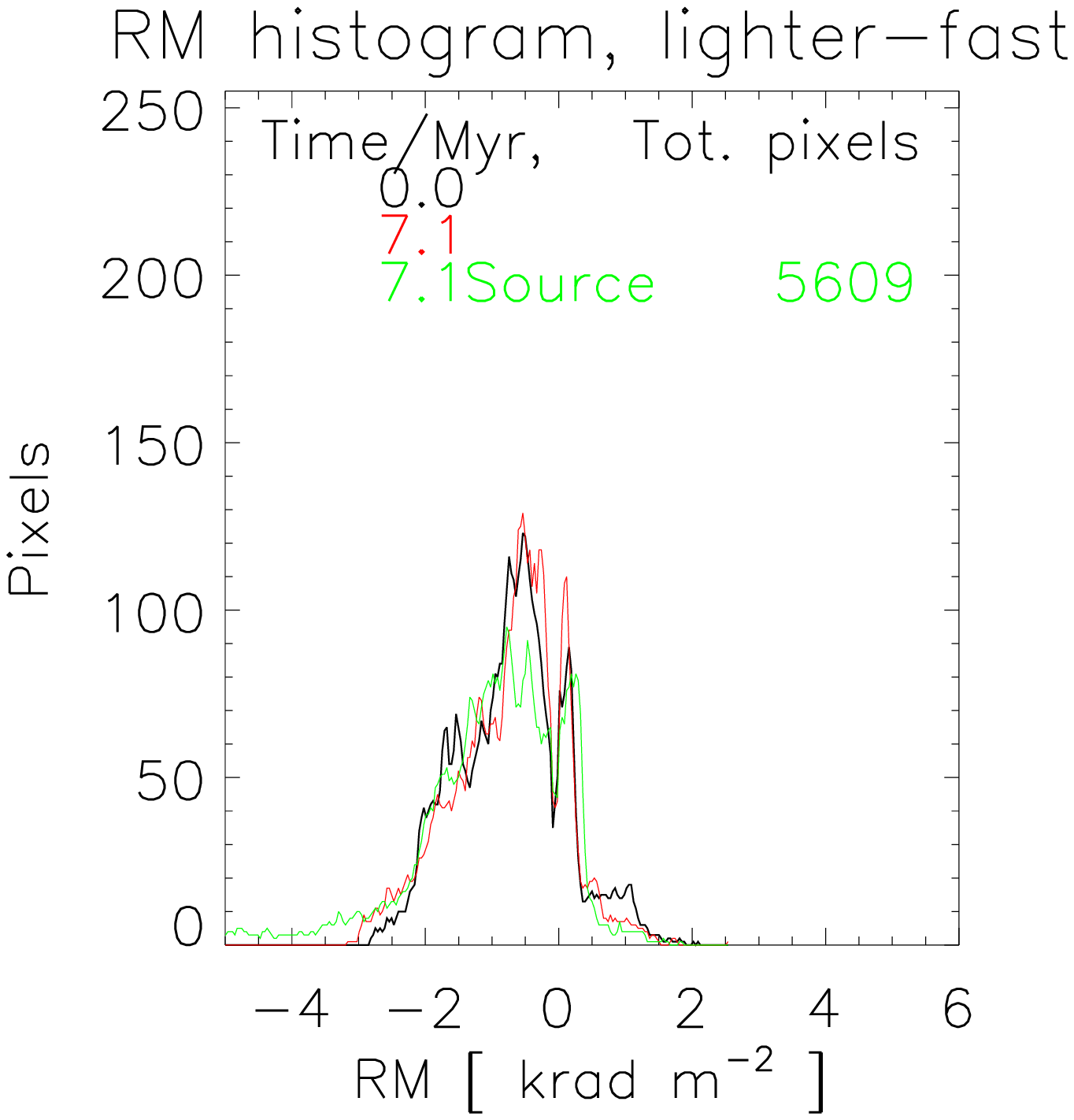}}
     \vspace*{0pt}
  \caption{Same as \fig{RM-lighter-fast90} but for
\hbox{$\theta_v=\,\,$45$^{\circ}$}. 
The approaching lobe is on the positive distance range.
}
\label{RM-lighter-fast45}
\end{figure*}

\begin{figure*}
  \centering
     \subfigure[No-source RM map at $t=\,$0\,Myr   (left), 
	             No-source RM map at $t=\,$4.7\,Myr (middle),
	                       RM map at $t=\,$4.7\,Myr (right).]
        {\label{RM-lighter-relativistic-90}
   \includegraphics[width=\textwidth]{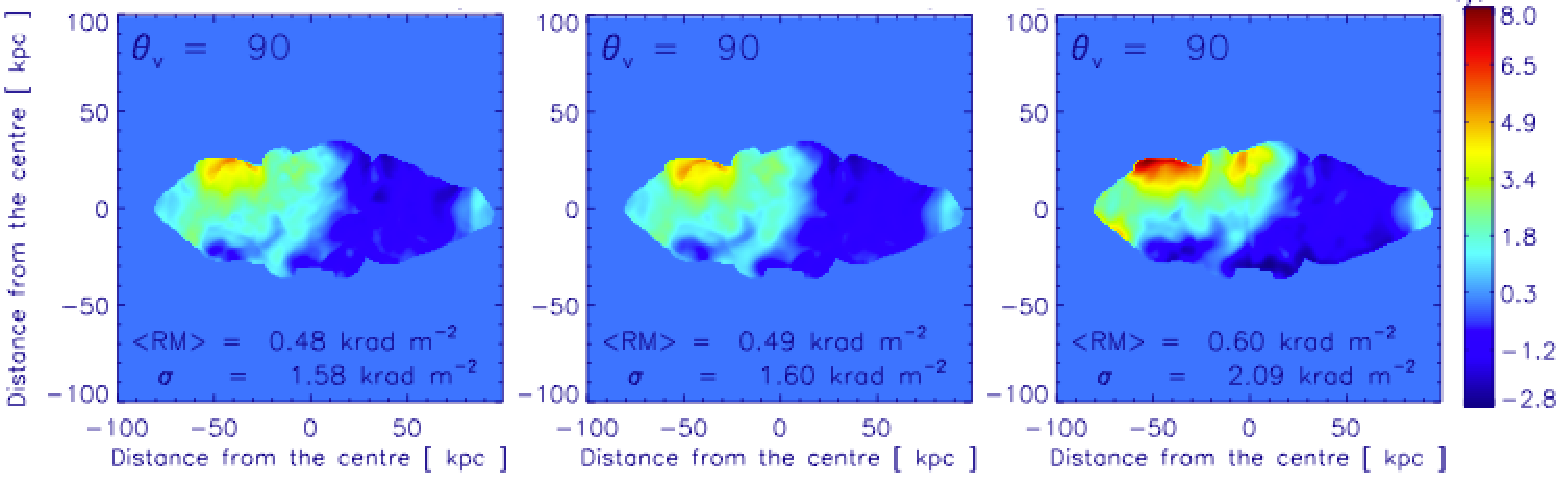}}
     \subfigure[RM at $y=-$25~(left), $y=$0~(middle) and $y=$25\,kpc~(right).
	  The black, red and green curves correspond to the left, the middle and
	  the right panels in (a), respectively.]
        {\label{RM-lighter-relativistic-90-cut}
  \includegraphics[width=.34\textwidth,bb =.5in 0in 6.8in 6.52in,clip=]
{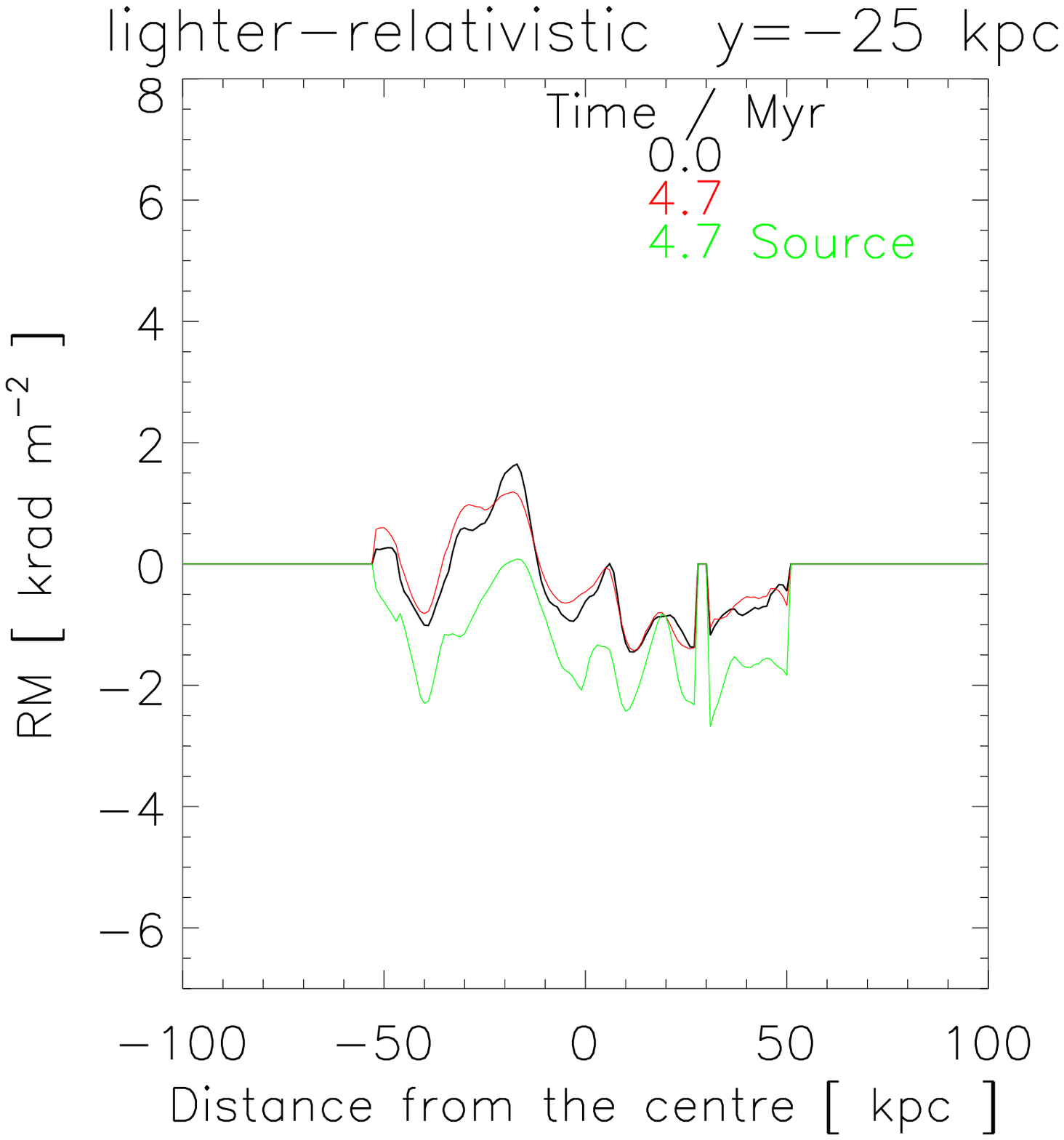}
  \includegraphics[width=.3\textwidth,bb =1.2in 0in 6.8in 6.51in,clip=]
{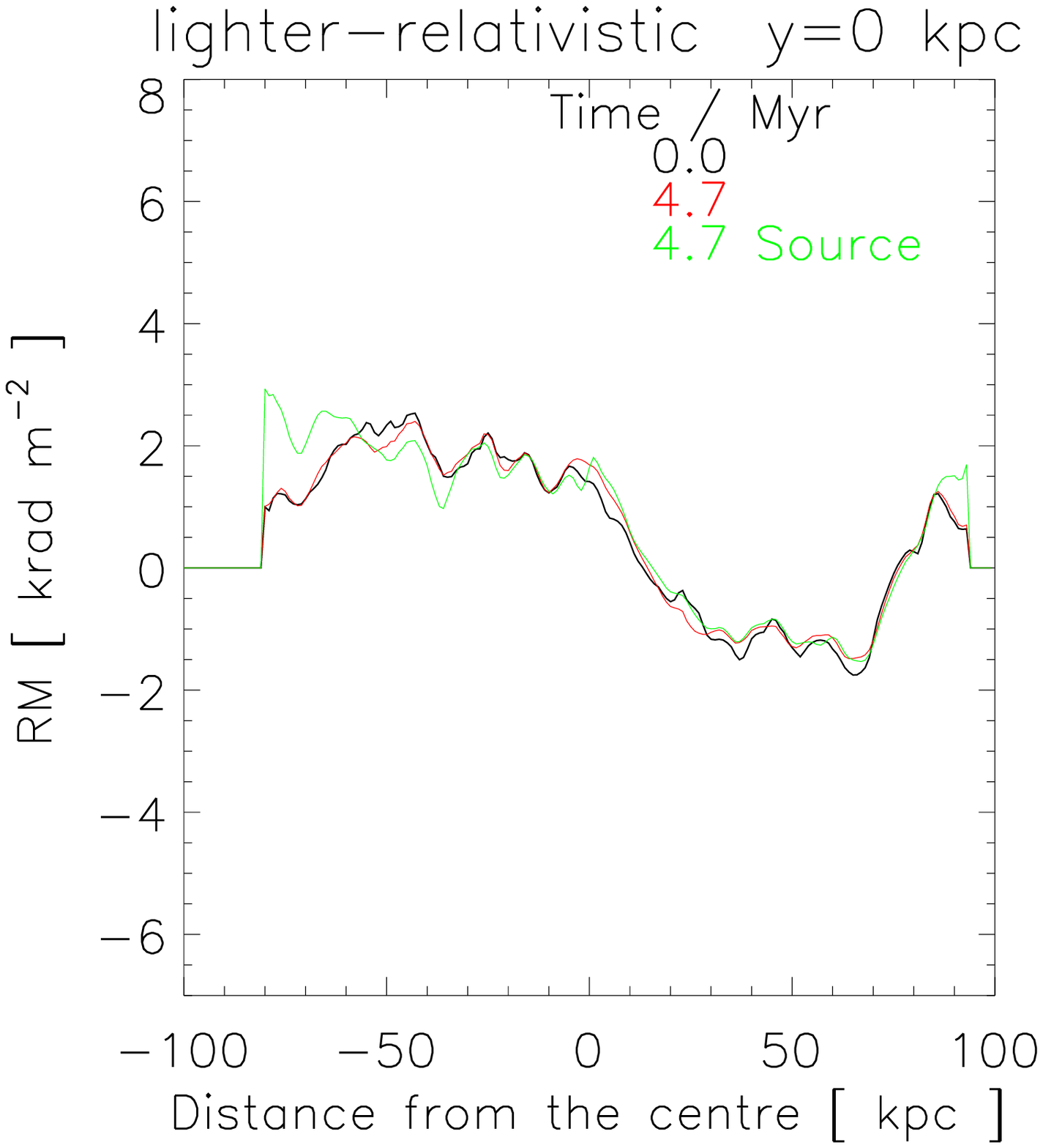}
  \includegraphics[width=.3\textwidth,bb =1.2in 0in 6.8in 6.51in,clip=]
{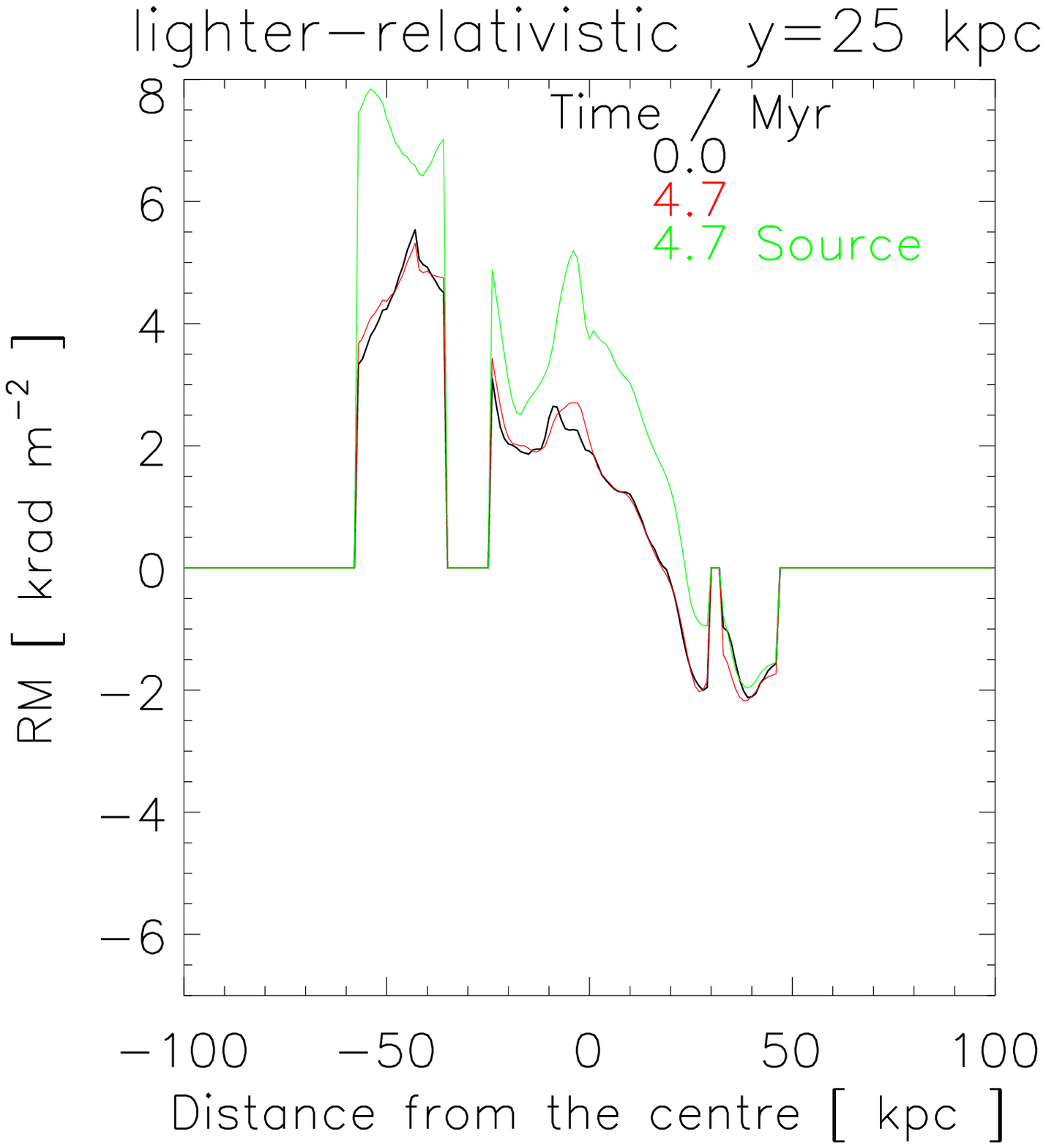}}
     \subfigure[RM histograms of the three maps in~(a). Colors are as in panel~(b).]
        {\label{RM-lighter-relativistic-90-hist}
  \includegraphics[width=.4\textwidth,bb =0.6in 0.25in 5.6in 5.505in,clip=]
{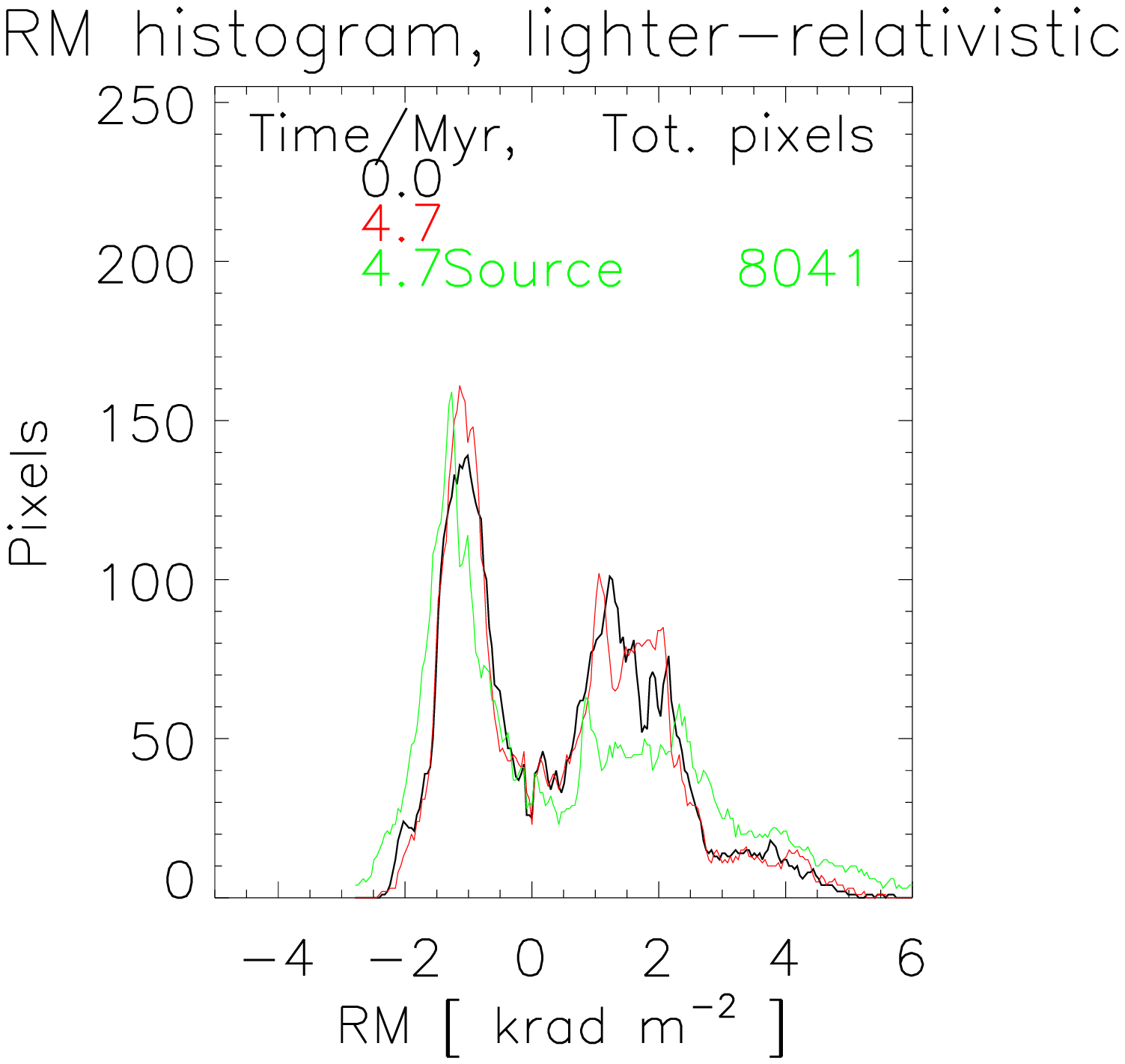}}
     \vspace*{0pt}
  \caption{Same as \fig{RM-lighter-fast90} but for the lighter-faster 
jets.
}
\label{RM-lighter-relativistic90}
\end{figure*}

\begin{figure*}
  \centering
     \subfigure[No-source RM map at $t=\,$0\,Myr   (left), 
	             No-source RM map at $t=\,$4.7\,Myr (middle),
	                       RM map at $t=\,$4.7\,Myr (right).]
        {\label{RM-lighter-relativistic-45}
   \includegraphics[width=\textwidth]{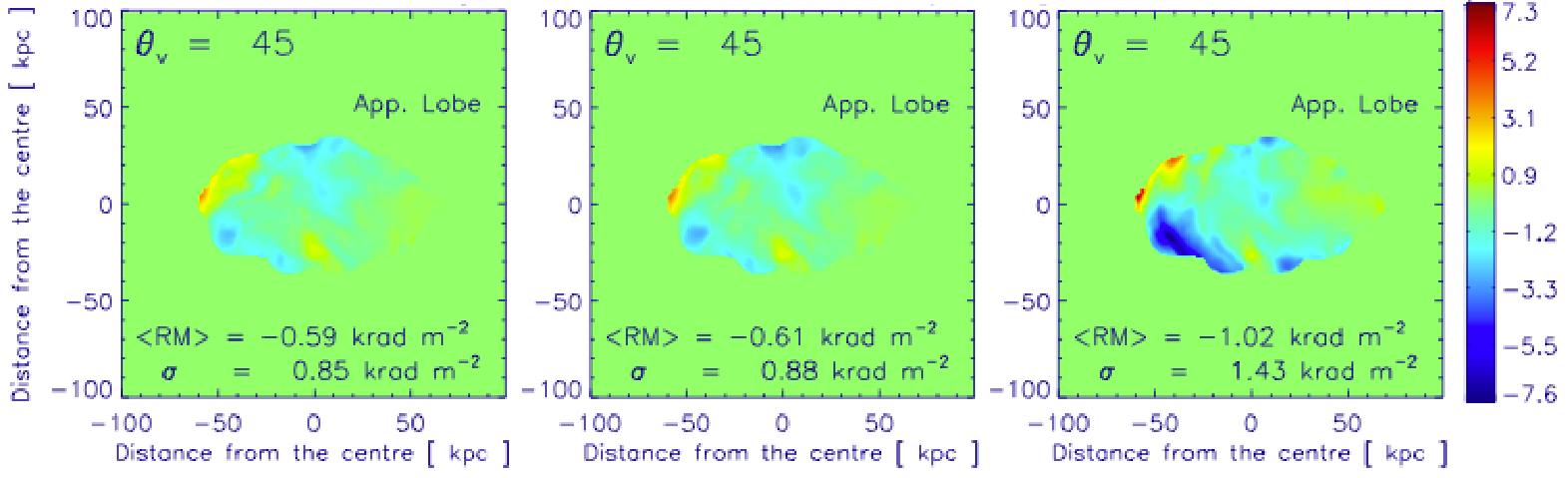}}
     \subfigure[RM at $y=-$25~(left), $y=$0~(middle) and $y=$25\,kpc~(right).
	  The black, red and green curves correspond to the left, the middle and
	  the right panels in (a), respectively.]
        {\label{RM-lighter-relativistic-45-cut}
  \includegraphics[width=.34\textwidth,bb =.5in 0in 6.8in 6.52in,clip=]
{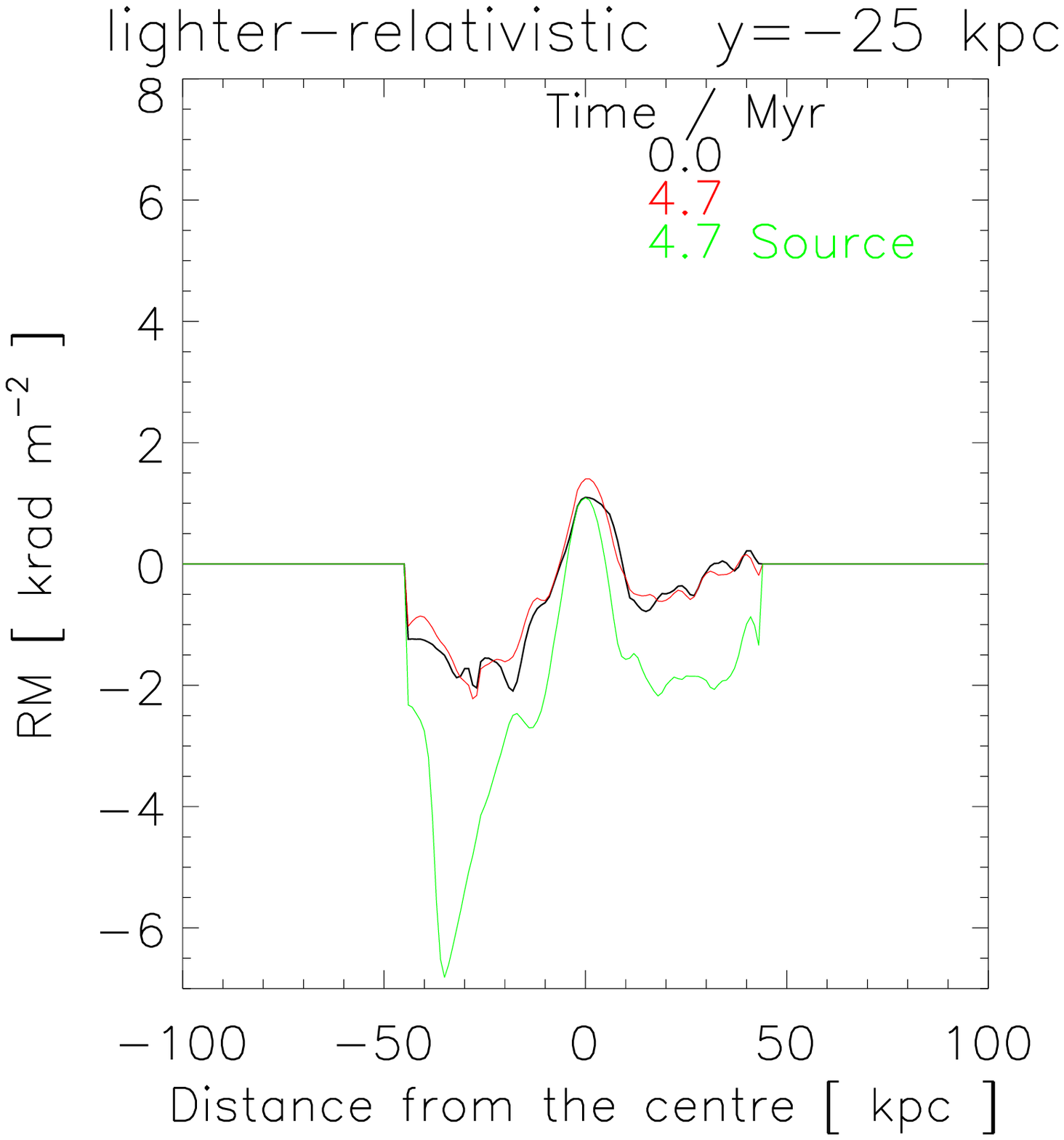}
  \includegraphics[width=.3\textwidth,bb =1.2in 0in 6.8in 6.51in,clip=]
{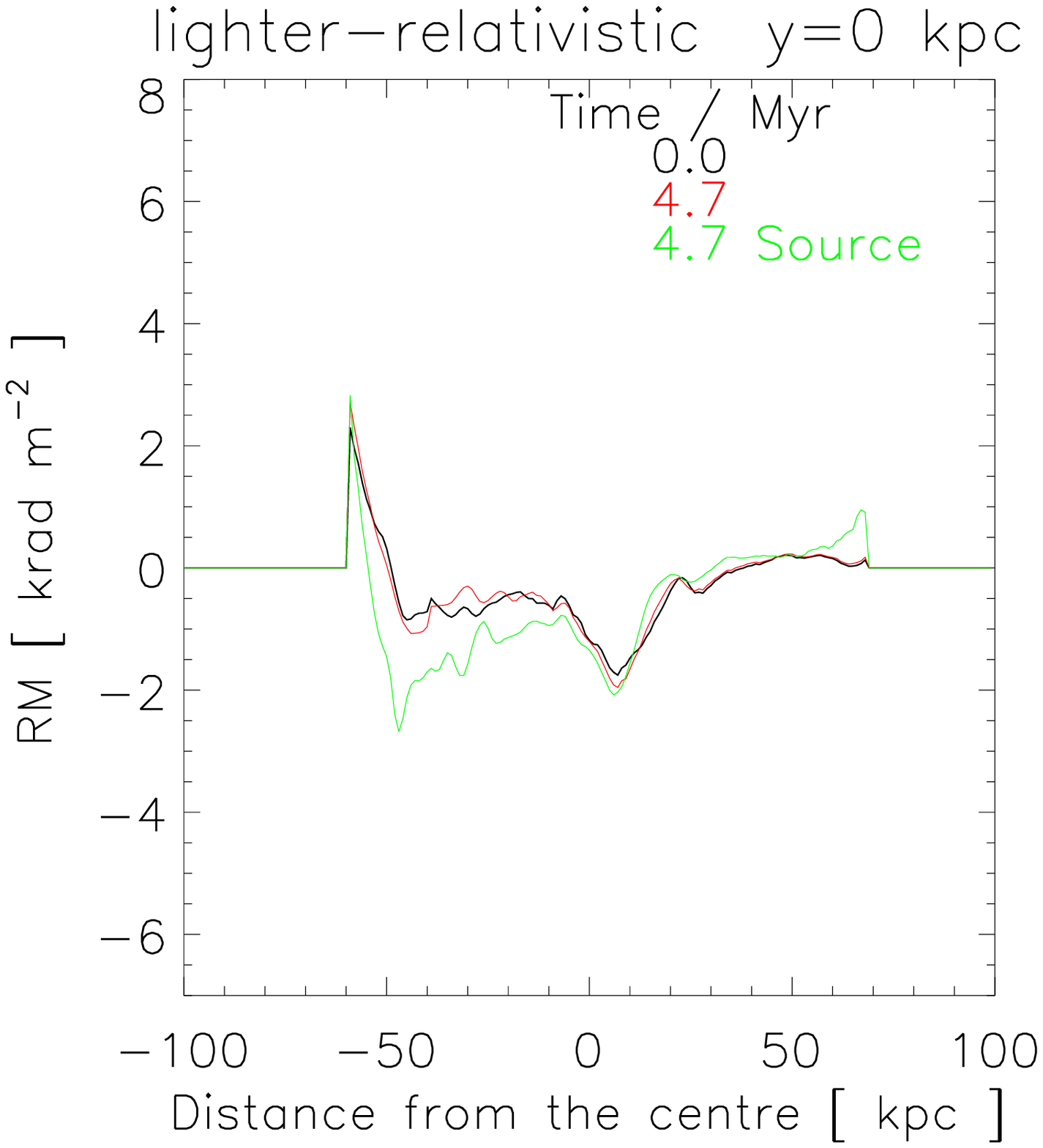}
  \includegraphics[width=.3\textwidth,bb =1.2in 0in 6.8in 6.51in,clip=]
{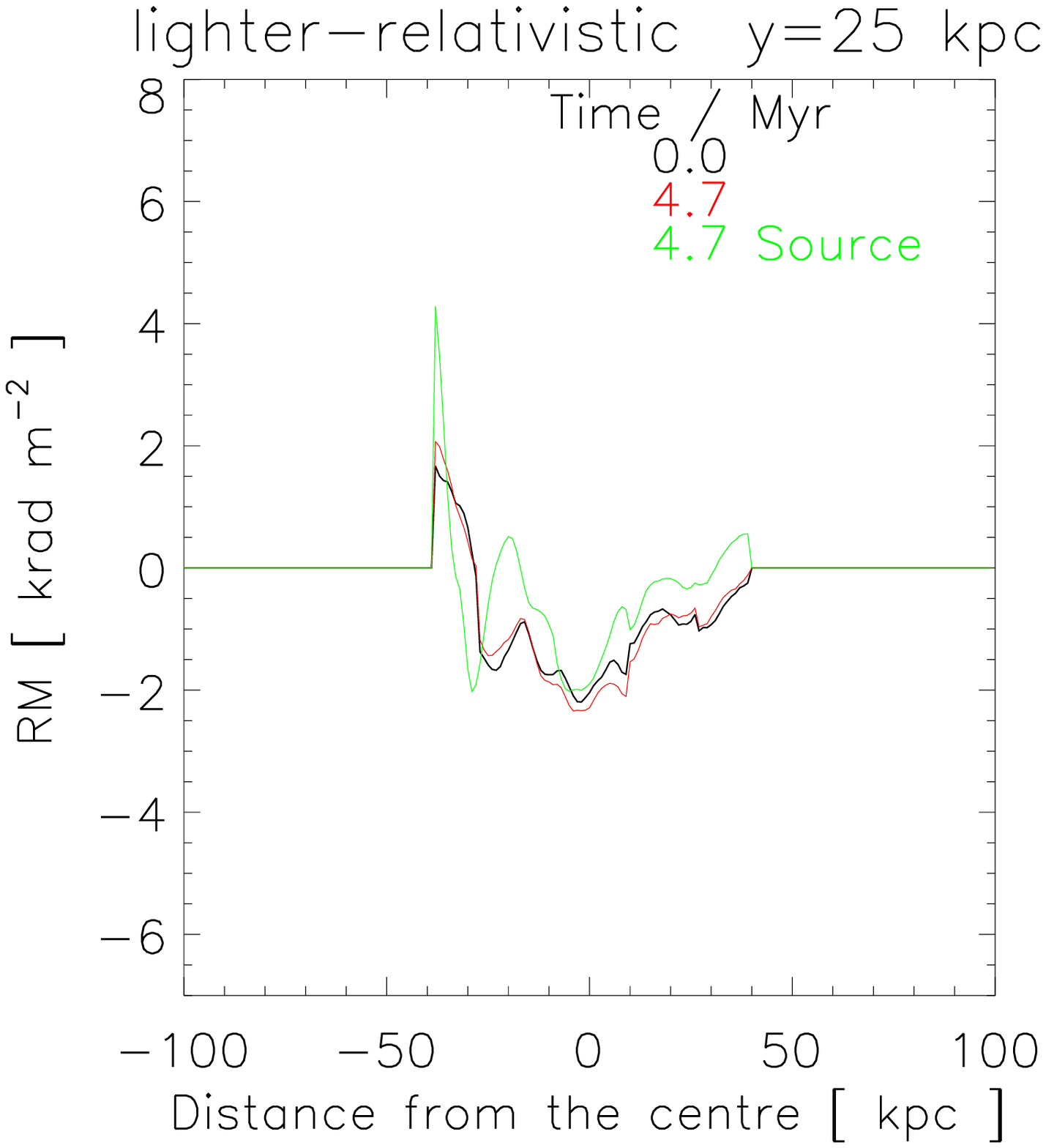}}
     \subfigure[RM histograms of the three maps in~(a). Colors are as in panel~(b).]
        {\label{RM-lighter-relativistic-45-hist}
  \includegraphics[width=.4\textwidth,bb =0.6in 0.25in 5.6in 5.505in,clip=]
{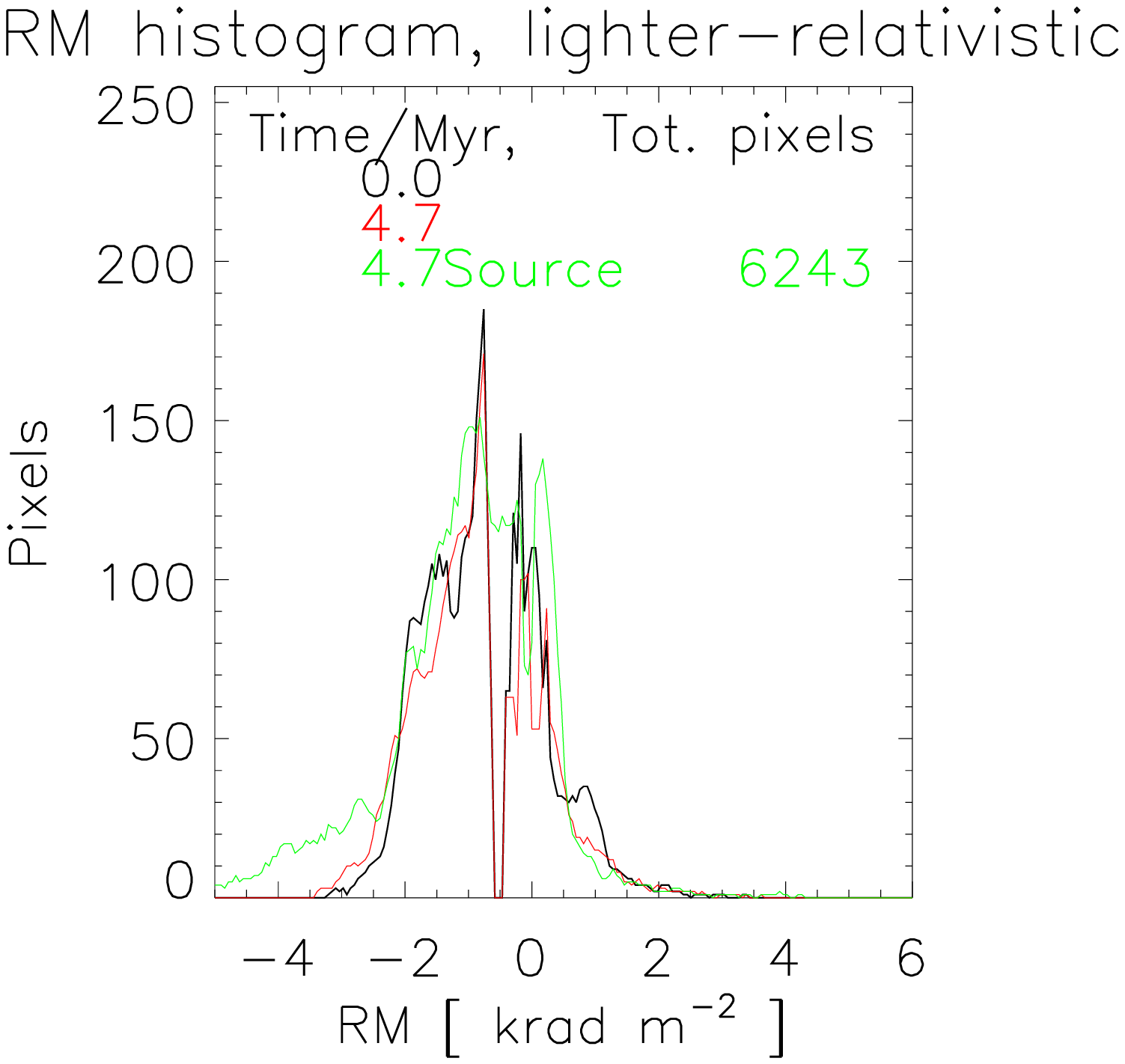}}
     \vspace*{0pt}
  \caption{Same as \fig{RM-lighter-fast90} but for the
lighter-faster 
jets and \hbox{$\theta_v=\,\,$45$^{\circ}$}. 
The approaching lobe is on the positive distance range.
}
\label{RM-lighter-relativistic45}
\end{figure*}

\begin{figure*}
  \centering
     \subfigure[No-source RM map at $t=\,$0\,Myr   (left), 
	             No-source RM map at $t=\,$4.5\,Myr (middle),
	                       RM map at $t=\,$4.4\,Myr (right).]
        {\label{RM-light-fast-90}
   \includegraphics[width=\textwidth]{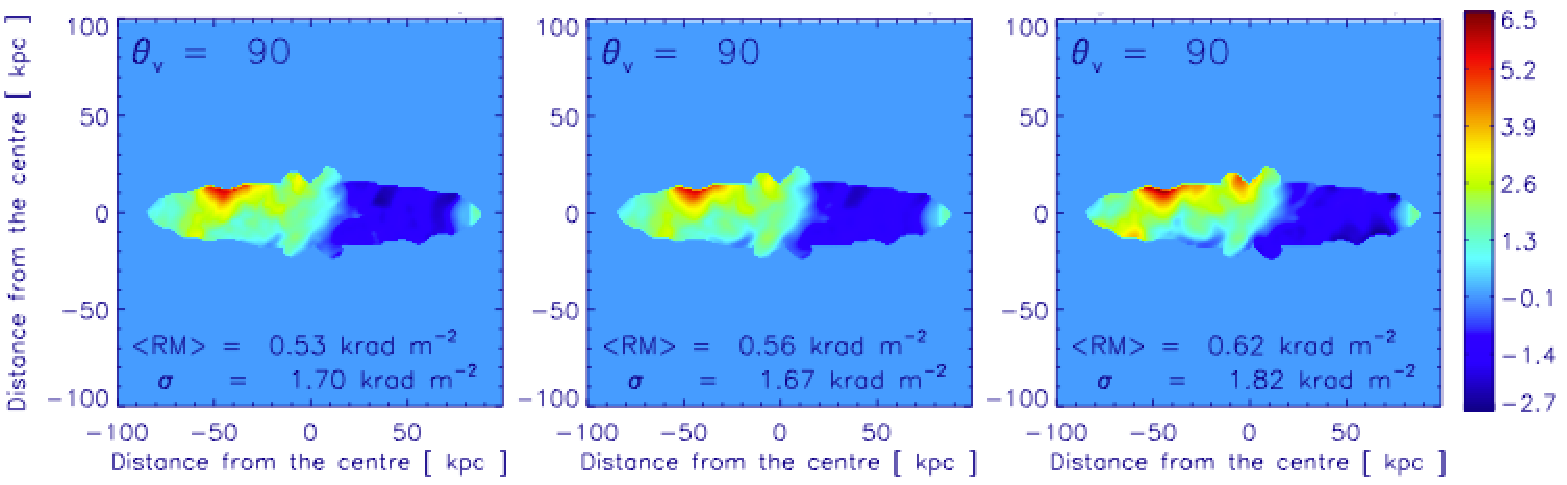}}
     \subfigure[RM at $y=-$12~(left), $y=$0~(middle) and $y=$12\,kpc~(right).
	  The black, red and green curves correspond to the left, the middle and
	  the right panels in (a), respectively.]
        {\label{RM-light-fast-90-cut}
  \includegraphics[width=.34\textwidth,bb =.5in 0in 6.8in 6.52in,clip=]
{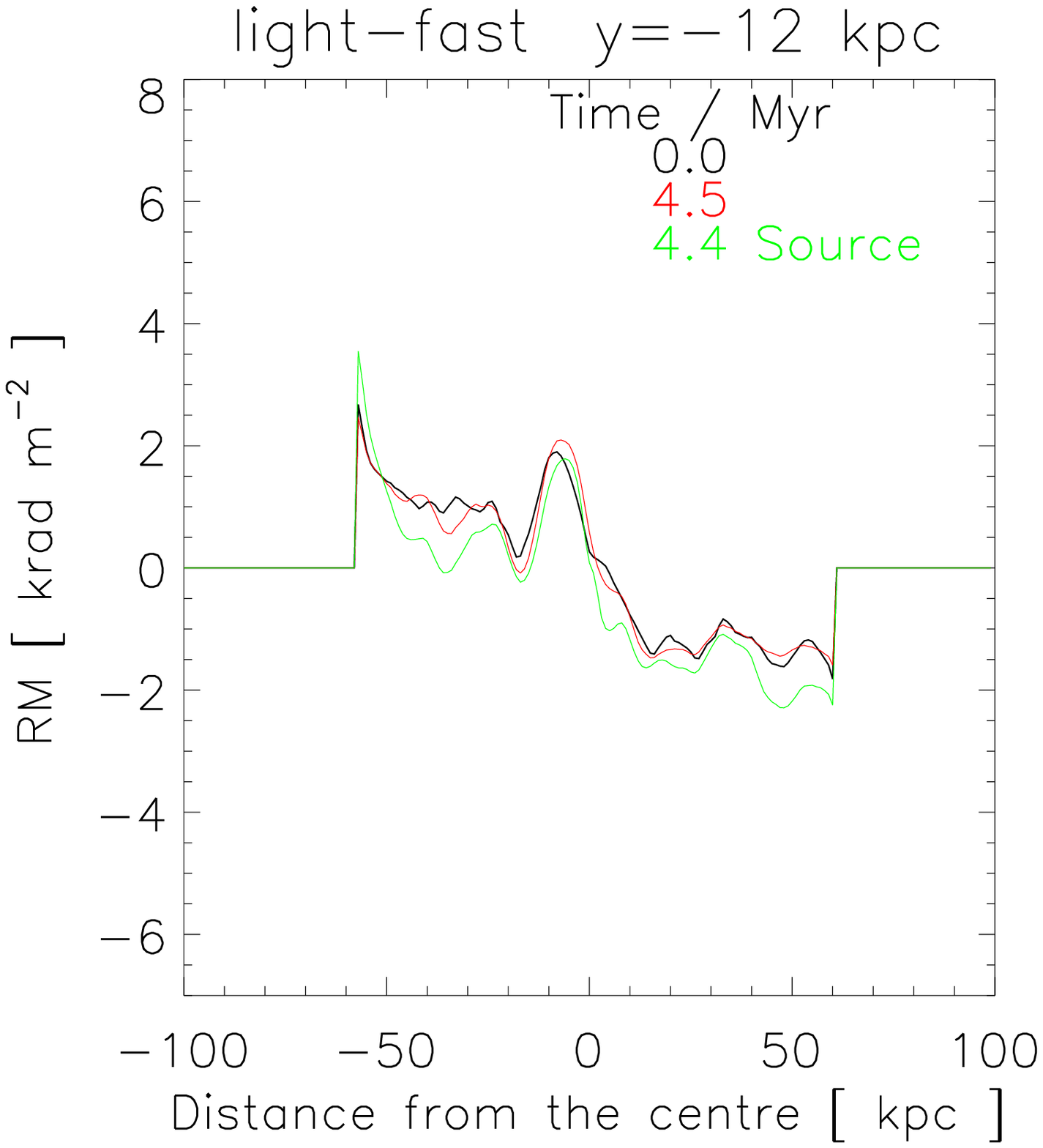}
  \includegraphics[width=.3\textwidth,bb =1.2in 0in 6.8in 6.51in,clip=]
{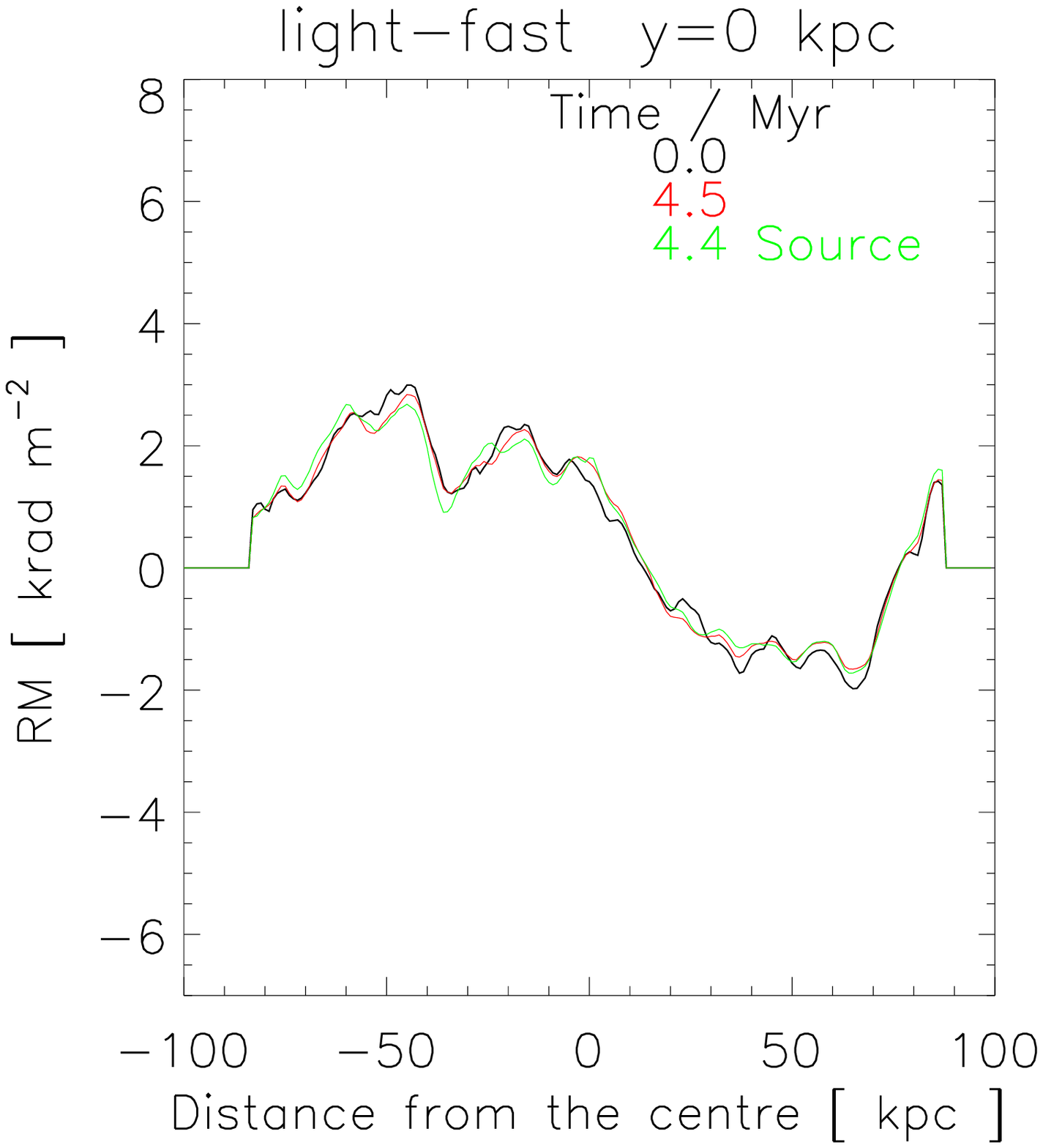}
  \includegraphics[width=.3\textwidth,bb =1.2in 0in 6.8in 6.51in,clip=]
{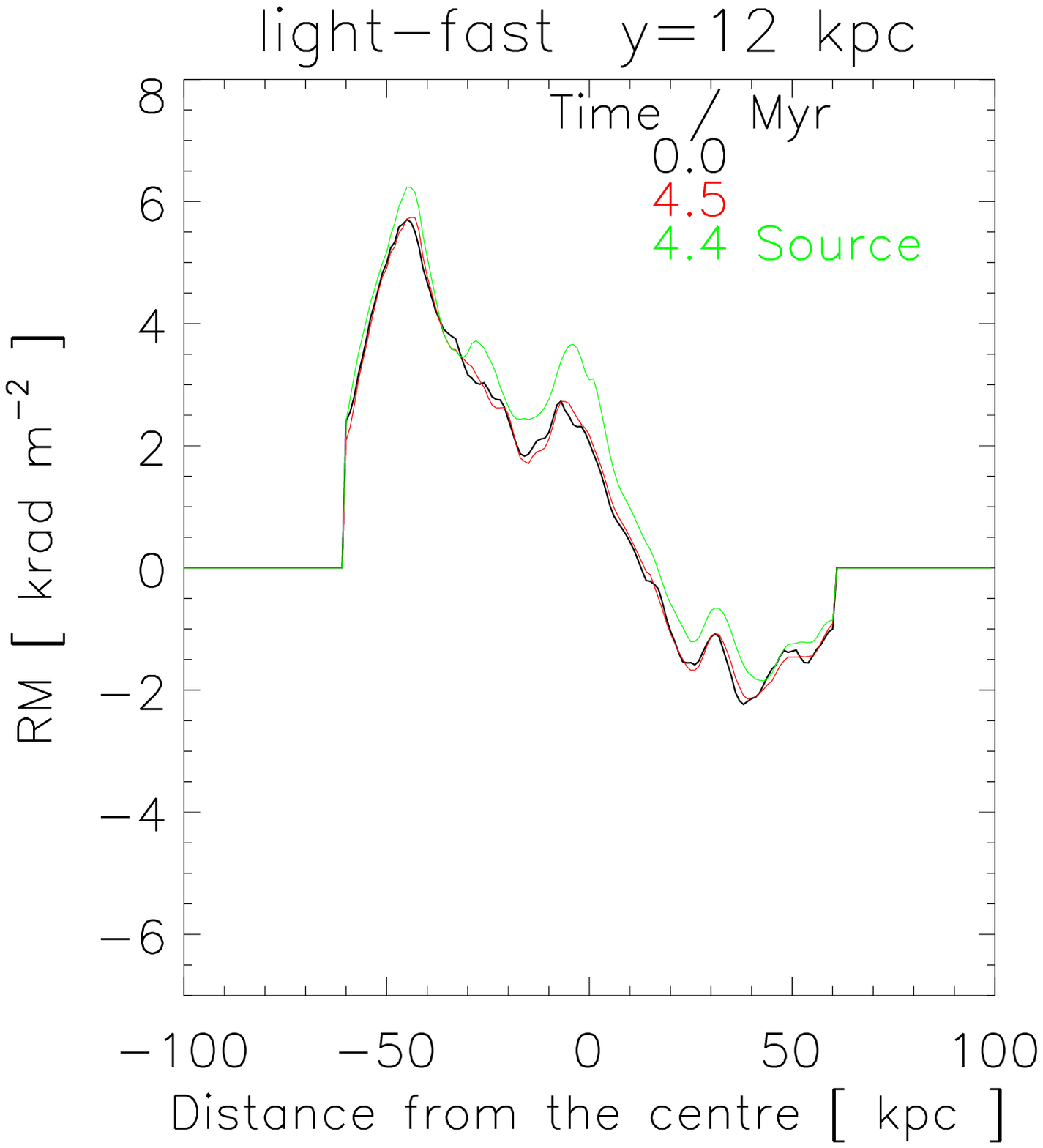}}
     \subfigure[RM histograms of the three maps in~(a). Colors are as in panel~(b).]
        {\label{RM-light-fast-90-hist}
  \includegraphics[width=.4\textwidth,bb =0.6in 0.25in 5.6in 5.505in,clip=]
{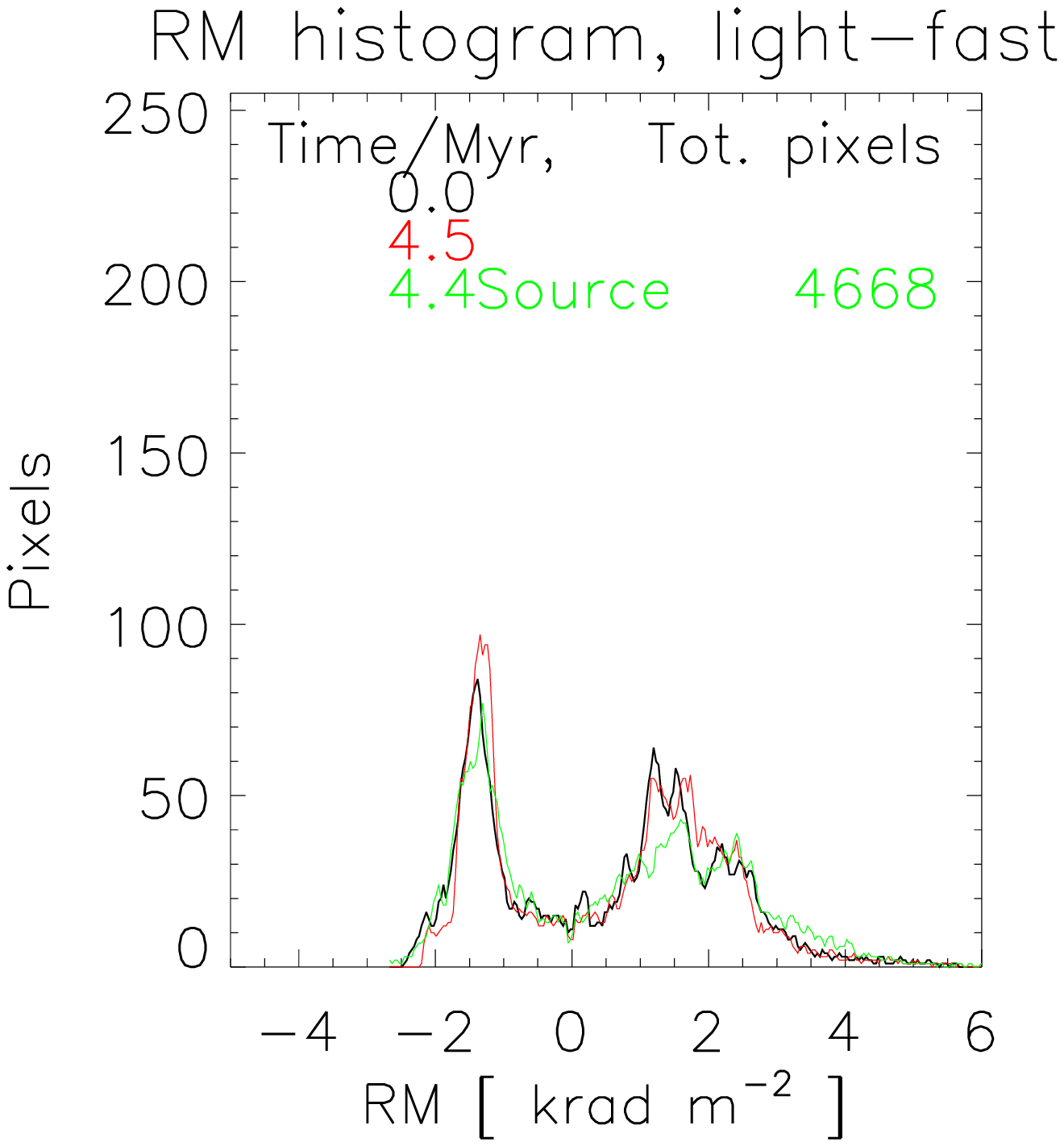}}
     \vspace*{0pt}
  \caption{Same as \fig{RM-lighter-fast90} but for the light-fast jets.
}
\label{RM-light-fast90}
\end{figure*}

\begin{figure*}
  \centering
     \subfigure[No-source RM map at $t=\,$0\,Myr   (left), 
	             No-source RM map at $t=\,$4.5\,Myr (middle),
	                       RM map at $t=\,$4.4\,Myr (right).]
        {\label{RM-light-fast-45}
   \includegraphics[width=\textwidth]{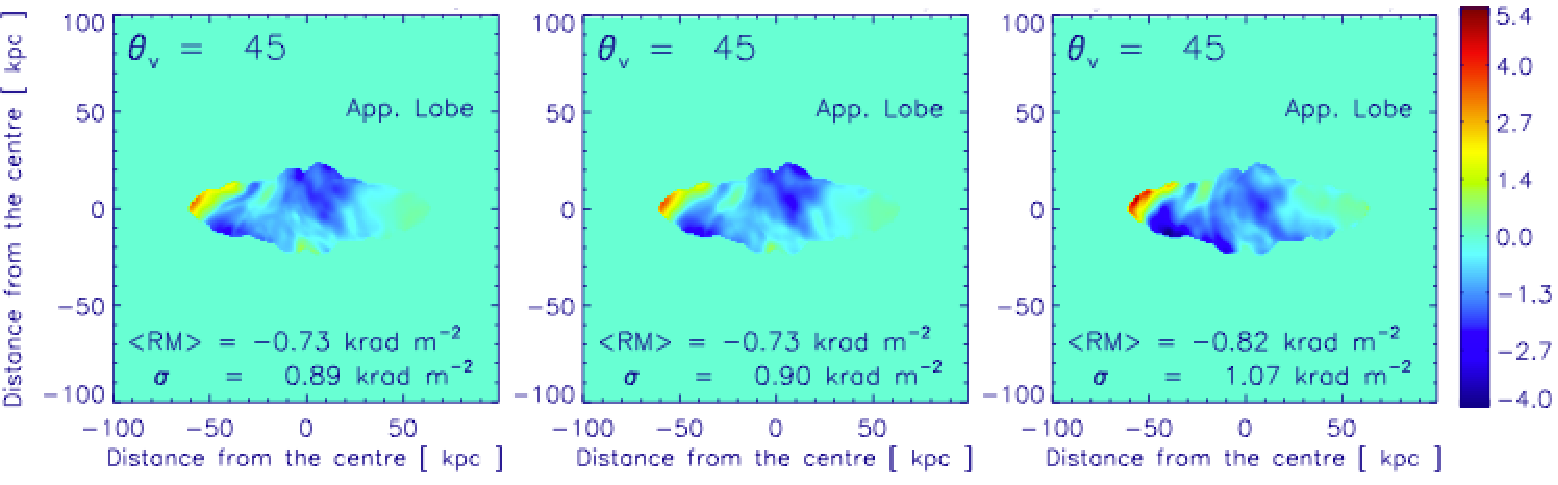}}
     \subfigure[RM at $y=-$12~(left), $y=$0~(middle) and $y=$12\,kpc~(right).
	  The black, red and green curves correspond to the left, the middle and
	  the right panels in (a), respectively.]
        {\label{RM-light-fast-45-cut}
  \includegraphics[width=.34\textwidth,bb =.5in 0in 6.8in 6.52in,clip=]
{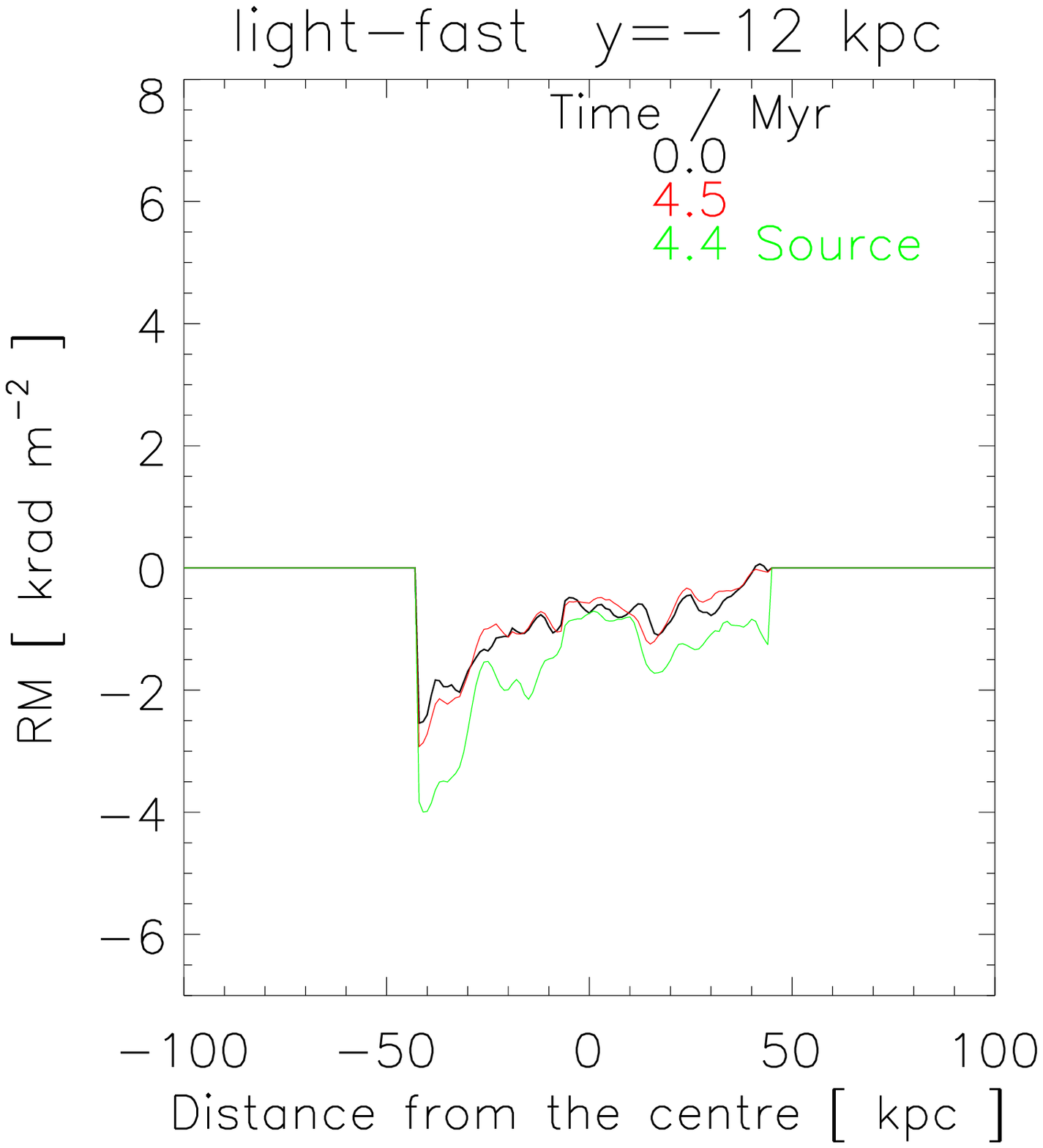}
  \includegraphics[width=.3\textwidth,bb =1.2in 0in 6.8in 6.51in,clip=]
{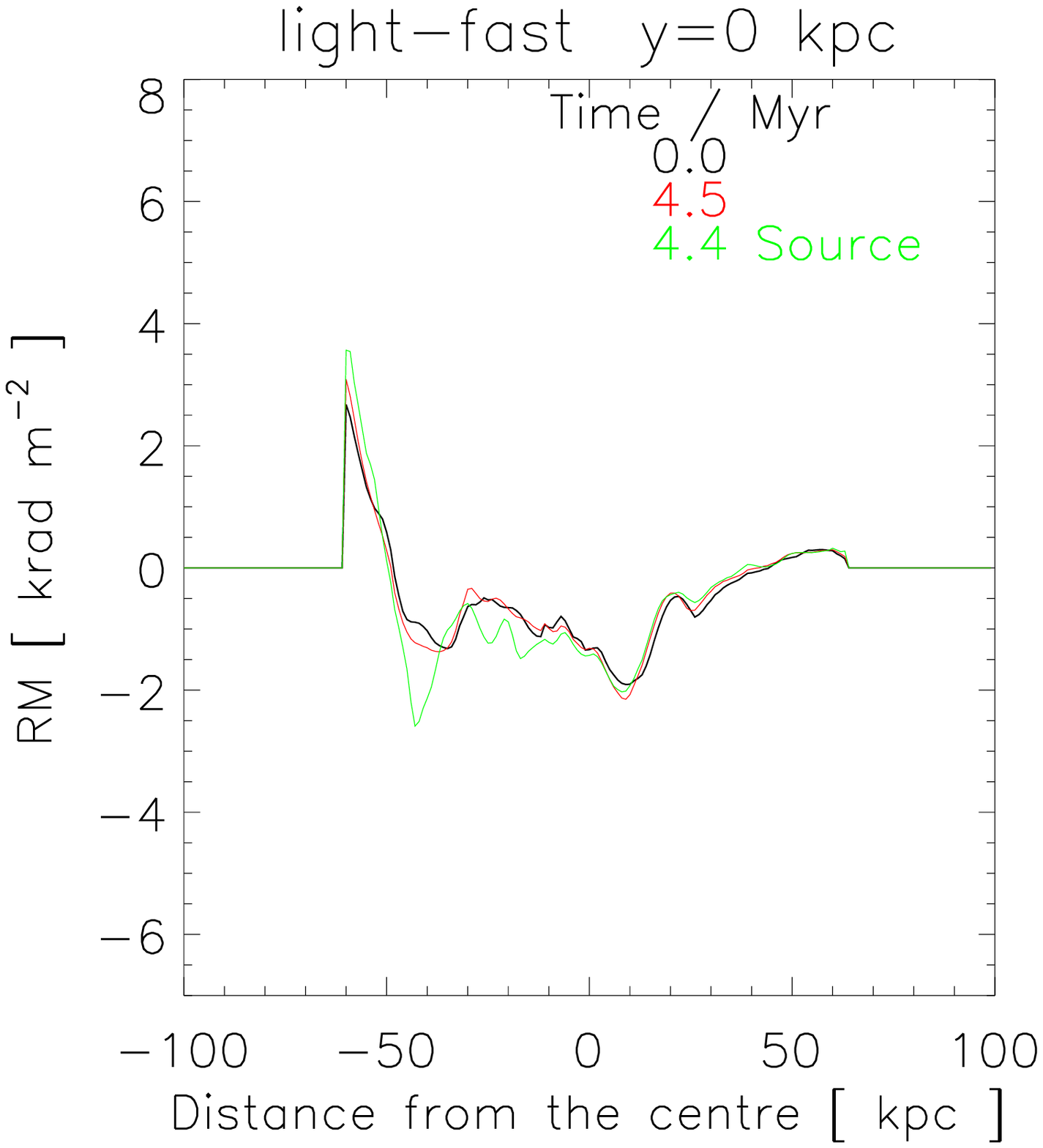}
  \includegraphics[width=.3\textwidth,bb =1.2in 0in 6.8in 6.51in,clip=]
{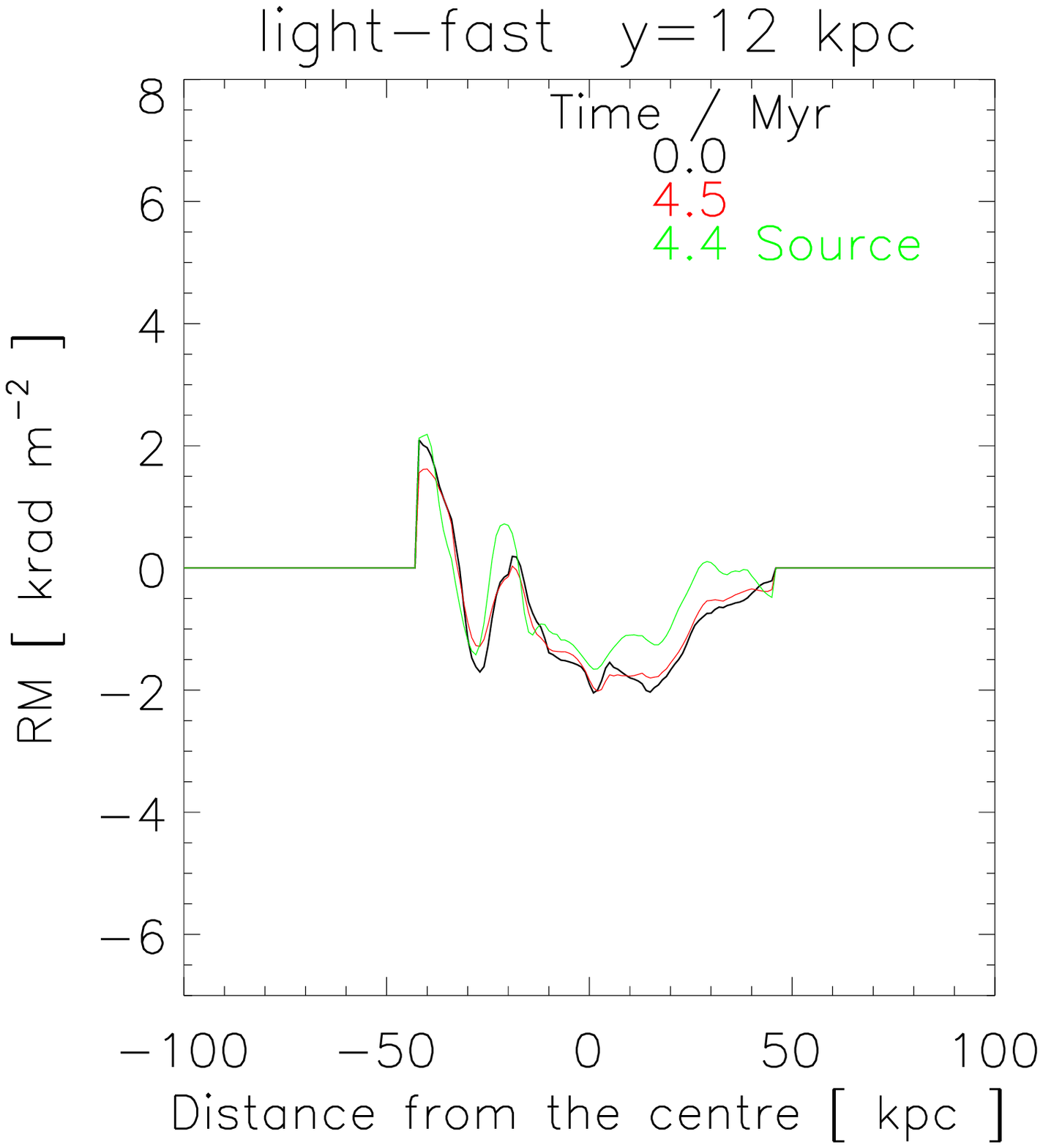}}
     \subfigure[RM histograms of the three maps in~(a). Colors are as in panel~(b).]
        {\label{RM-light-fast-45-hist}
  \includegraphics[width=.4\textwidth,bb =0.6in 0.25in 5.6in 5.505in,clip=]
{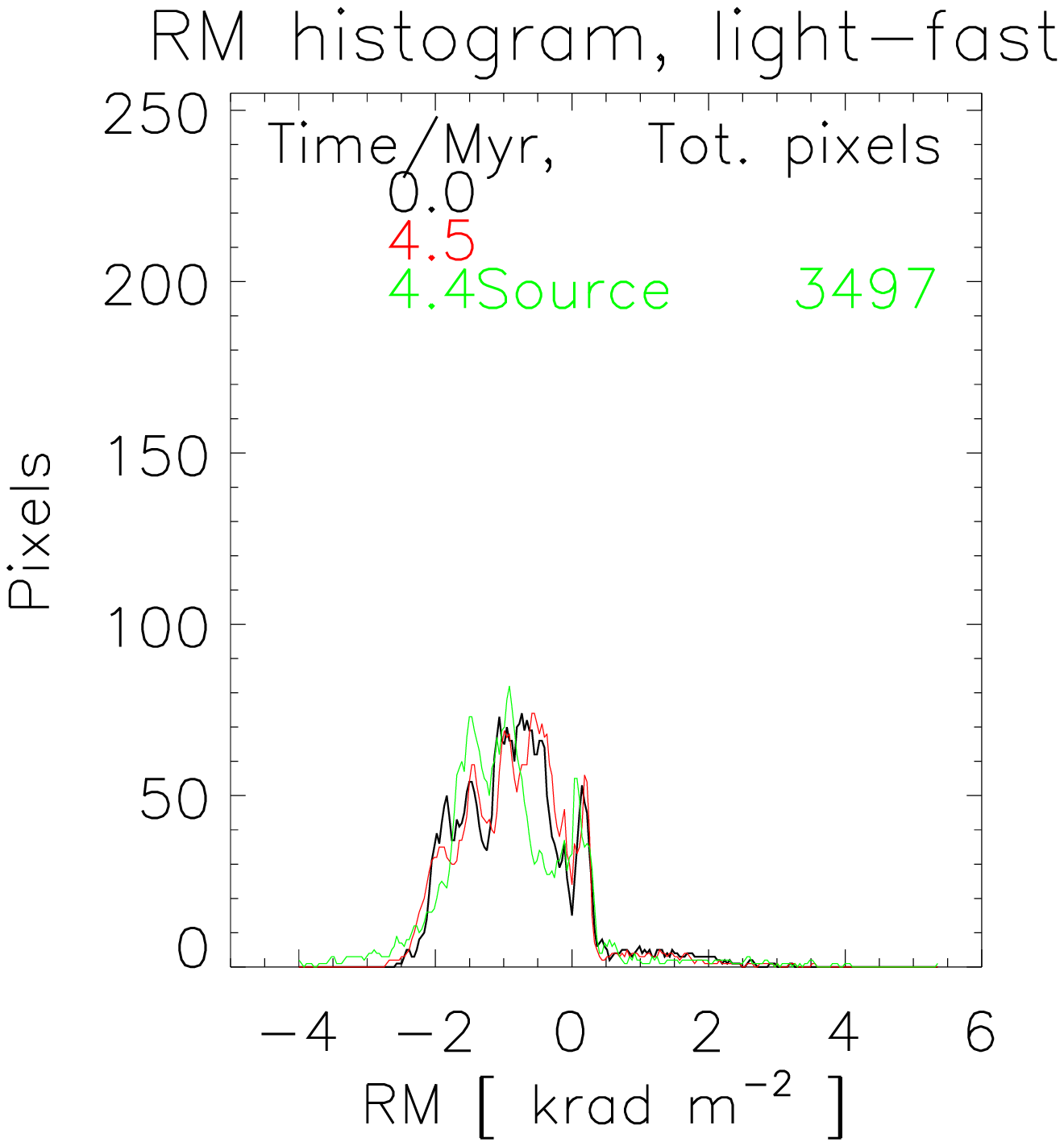}}
     \vspace*{0pt}
  \caption{Same as \fig{RM-lighter-fast90} but for the light-fast jets
and \hbox{$\theta_v=\,\,$45$^{\circ}$}. 
The approaching lobe is on the positive distance range.
}
\label{RM-light-fast45}
\end{figure*}

  Figures~\ref{RM-lighter-fast90}--\ref{RM-light-fast45} show our
  synthetic RM maps and their histograms. Each Figure corresponds 
  to one of the jets-simulations and a viewing angle (either
  45~or 90~degrees). Panels are arranged by rows.
  The top row shows (from left to right): the no-source RM map at
  $t_{\rmn{jet}}=\,$0\,Myr (using the cocoon coordinates of the
  corresponding source at $t_{\rmn{jet}}=t_{\rmn{e}}$); the no-source
  RM map at $t_{\rmn{jet}}=t_{\rmn{e}}$ (also using the cocoon coordinates
  of the corresponding source at $t_{\rmn{jet}}=t_{\rmn{e}}$); the
  RM map at $t_{\rmn{jet}}=t_{\rmn{e}}$.
  %
  The middle row, (b), has panels in which the black, the red and
  the green profiles correspond to horizontal lines through the
  left, the middle and the right maps in the top row, respectively.
  The middle panel in row~(b) shows 1-D RM profiles at $y=\,$0\,kpc.
  The left and right panels in~(b) show 1-D RM profiles at symmetric
  $y$-values with respect to $y=\,$0\,kpc, which go near the sources' edge.
  Finally, the bottom row,~(c), shows histograms of the three maps in 
  the top row --\,where (again) the black, the red and the green profiles
  in~(c) correspond to the left, the middle and the right maps in~(a),
  respectively.


Our synthetic RM maps consistently show that powerful jets enhance the RM distribution
locally. We see this process scales~with both the jet velocity
and density.
More spherical cocoons, such as those that are typically shown by very light
jets, generally yield higher RM values; they favour the
alignment of enhanced magnetic field components with the line of
sight.

The considered RM enhancement occurs because the expanding hypersonic sources
compress the ICM gas and magnetic fields in the shocked ambient
region
which is located just ahead of the cocoons' contact surface.
Magnetic field lines are compressed and stretched
perpendicularly to the source expansion direction
(\fig{figTopo}). 
We see that this RM increase is 
%
typically larger at the sources' edge,
where 
the alignment between the line of sight and the compressed
cluster magnetic field lines is more important. 
Moreover, shocks are stronger near the hotspots
than near the sides of the cocoons. Such structure is also evident
in the RM maps --\,e.g. see 
%
   the ``lighter-faster'' 
	source map, 
	Figure~\ref{RM-lighter-relativistic45}~(a), right panel.

By comparing the top left and top middle panels in 
Figures~\ref{RM-lighter-fast90}--\ref{RM-light-fast45},
as well as the black and the red profiles
in the rows~(b) and~(c), we see a modest RM 
variation in the ``no-source'' RM maps.
Relative to such small changes, the source produced RM
enhancements are larger, particularly 
%
	those of the
%
jets with 
velocities exceeding Mach~80.
The ``no-source'' RM maps produced at $\theta_v=\,\,$45$^{\circ}$ show 
the expected global asymmetry:
having a shorter total Faraday depth, the
approaching source lobes consistently show lower RM values and
variation than the receding lobes. 
%
   We see 
%
powerful radio jets enhance these
RM gradients and 
that 
the histograms in panel (c) of
Figures~\ref{RM-lighter-fast90}--\ref{RM-light-fast45} show 
that 
this effect happens at the highest RM values. When jets are on the plane of
the sky ($\theta_v=\,\,$90$^{\circ}$) the RM histograms tend to be
double peaked. 
Yet at $\theta_v=\,\,$45$^{\circ}$, the histograms are narrower and
single peaked; the Faraday depth is more uniform at
$\theta_v=\,\,$90$^{\circ}$. Further, the alignment of 
the 
compressed CMF components with the line of sight is more important for
observations at angles within 20--70~degrees due to the prolate~spheroid
geometry of cocoons. Therefore, we see higher RM values at 45~degrees
than at 90. The choice of 45~degrees is interesting because
it separates quasars from radio galaxies in the unification model
of \citet{barthel89}.

\subsubsection{RM evolution}
\label{RMevo}

\begin{figure*}
  \centering
  \includegraphics[width=.51\textwidth,bb=.7in .5in 5.7in 4.5in,clip=]{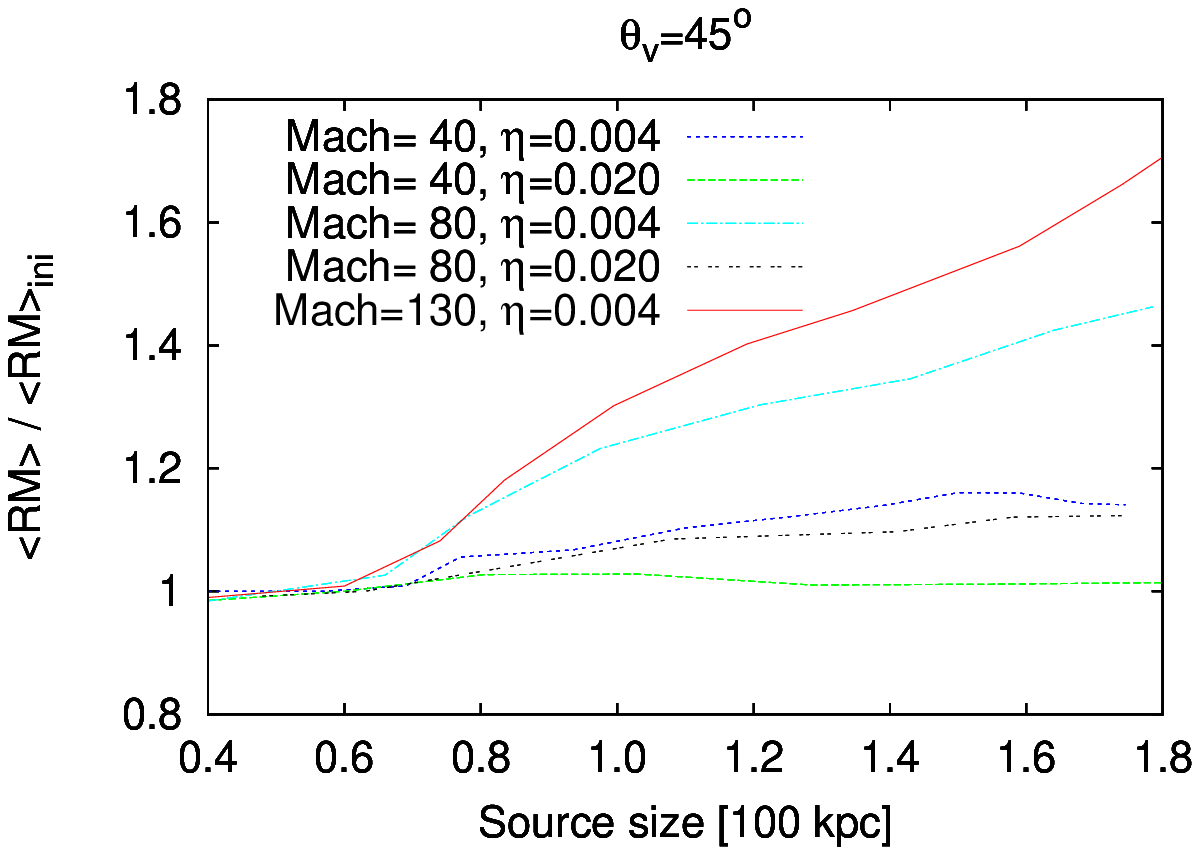}
  \includegraphics[width=.45\textwidth,bb=1.3in .5in 5.7in 4.5in,clip=]{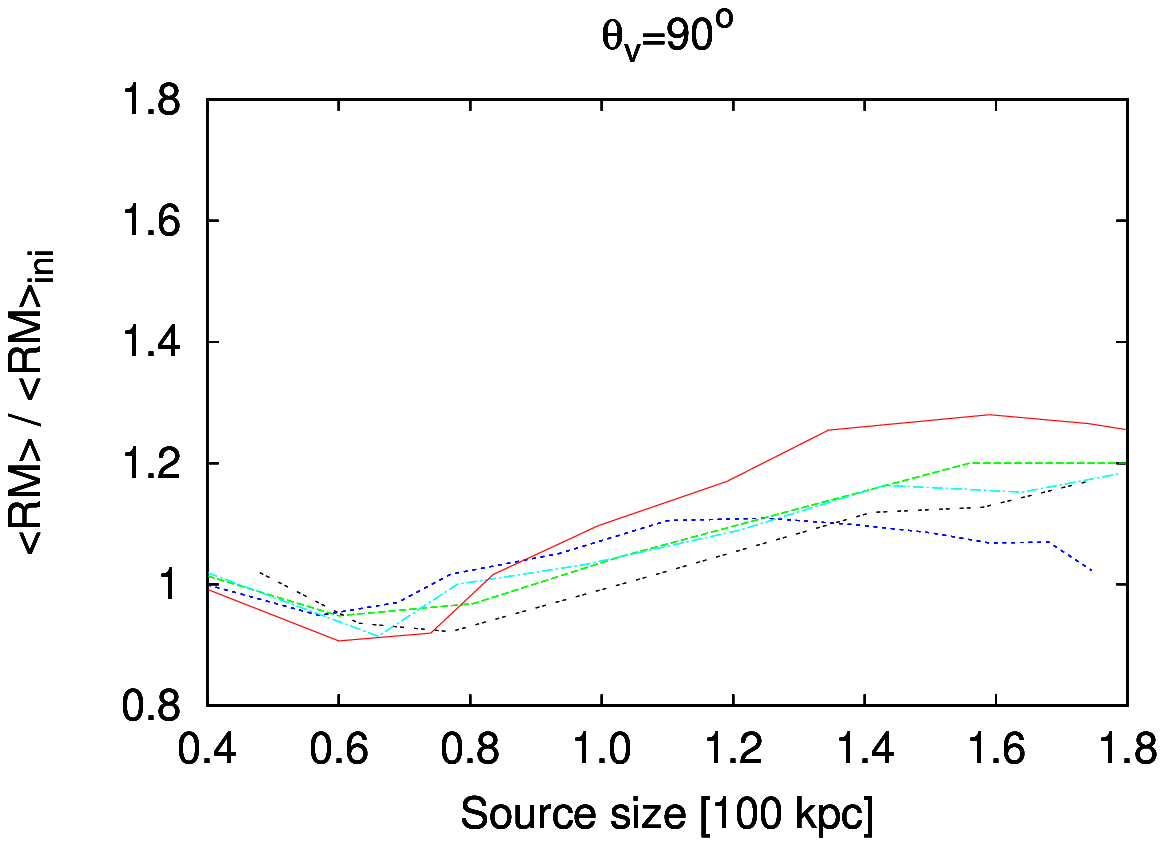}\\
\vskip.3cm                 
  \includegraphics[width=.51\textwidth,bb=.7in .5in 5.7in 4.5in,clip=]{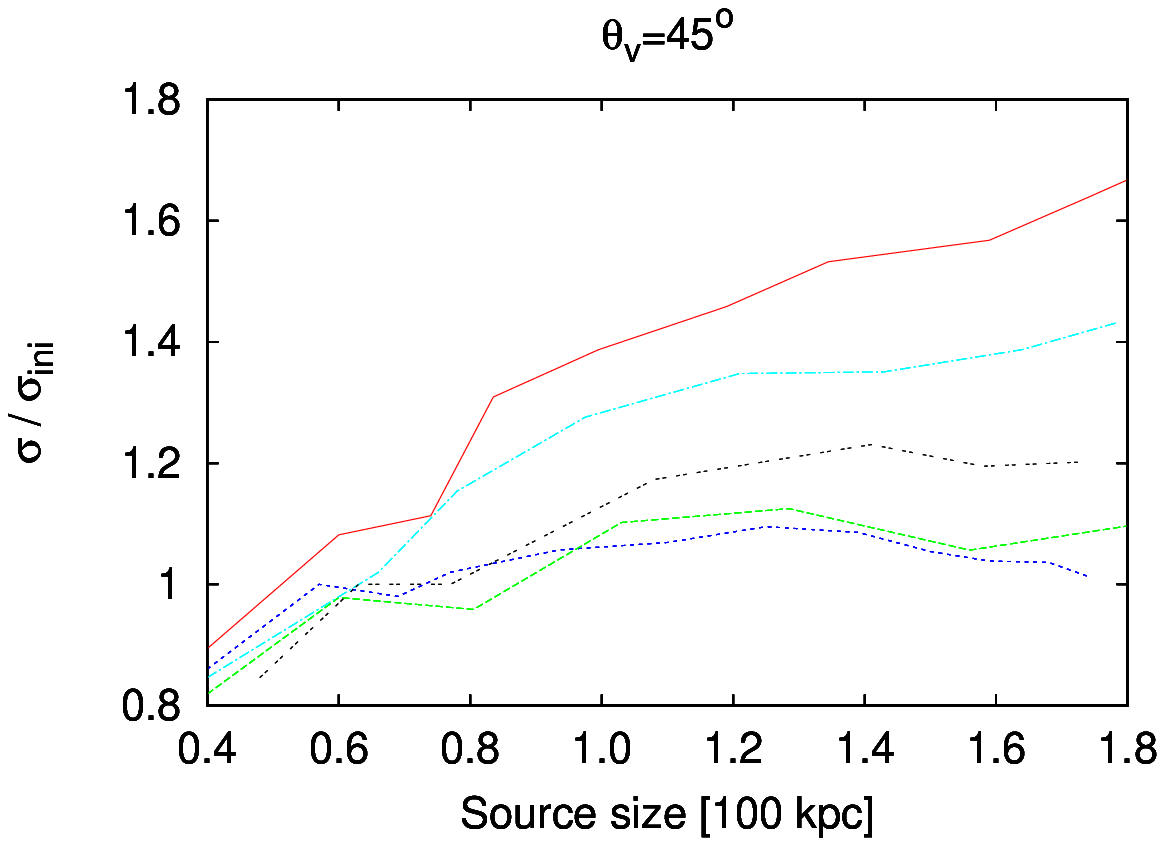}
  \includegraphics[width=.45\textwidth,bb=1.3in .5in 5.7in 4.5in,clip=]{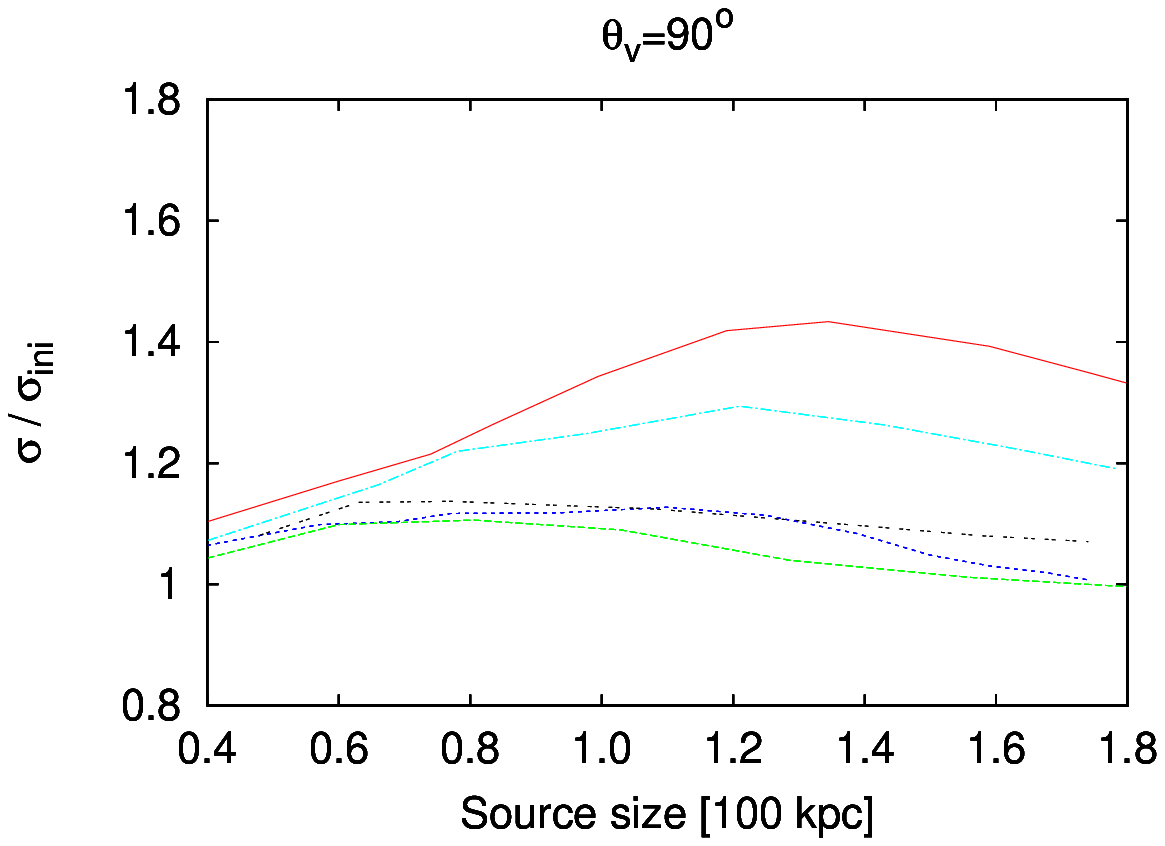}
     \vspace*{0pt}
  \caption{
Evolution of the RM statistics as radio sources expand. Panels at
the top row display the evolution of the mean RM, whereas panels in the
bottom show the evolution of the RM standard deviation. Panels in 
either the left or the right column correspond to synthetic observations
made at 45~or 90~degrees, respectively. The RM is computed using
expanding sources at corresponding timesteps. Profiles are normalised
to the mean RM, or the RM standard deviation, of corresponding 
no-source RM maps
at $t_{\rmn{jet}}=\,$0 (see Section~\ref{RMmapsB}).
}
\label{meanRMevo}
\end{figure*}

The time evolution of the RM statistics has been followed as the 
sources expand. 
In Figure~\ref{meanRMevo}
the top row displays the evolution of the mean RM.
The bottom row shows the evolution of the RM standard
deviation. The viewing angle is 45 and 90 degrees for the left and
right columns, respectively.
RM maps are produced using expanding sources at corresponding
timesteps. Profiles are normalised to their corresponding initial
values (i.e. to either the mean RM or the RM standard deviation
produced against Faraday screens of matching sources, but on the
gas and magnetic field distribution at $t_{\rmn{jet}}=\,$0, see
Section~\ref{RMmapsB}).

As the sources expand we see that the RM statistics scale~up with the
speed of the jets. Generally the ``lighter'' sources yield more
important changes on the statistics than the ``light'' sources, by
factors of about 2--7. We see the gradients in the RM statistics 
evolution profiles (\fig{meanRMevo})
follow similar trends to those of the ICM energetics profiles
(Figure~\ref{fig:energy}). Also, in agreement with the RM histograms
(panel \textit{c}, Figures~\ref{RM-lighter-fast90}--\ref{RM-light-fast45}),
the effect of the sources on the RM evolution appears to be more
significant for observations made at 45~degrees than at 90. This is, again
(see previous section), caused by the geometry of cocoons.

Gradients in the RM statistics evolution profiles (Figure~\ref{meanRMevo})
also show a relation with those in the pressure evolution profiles
%
	that we 
%
shown in Figure~\ref{pressEVO}. The latter show that the ``slow'' sources
reach pressure equilibrium with the ambient medium earlier
than faster sources ($v_{\rmn{j}}\ge\,$ 80\,Mach).  In agreement
we see that the ``slow'' sources cause modest changes on the RM,
relative to faster sources. We see that the ``lighter-faster'' 
source,
which has the most powerful jets amongst our models, enhances both
the mean RM and the RM standard deviation by approximately 70\% for
a viewing angle of 45~degrees (red curve, \fig{meanRMevo}, left
column). This effect is very important because, as we will discuss
in the next Section, the observed values of the RM dispersion are
used to infer the strength of CMFs.

\begin{table*}
\centering
    \begin{minipage}{120mm}
   \caption{Summary of the RM statistics evolution. These are
the maxima of the mean~RM and the RM standard deviation, normalized to
their corresponding initial values, reached 
in synthetic observations that capture the effects of powerful radio 
sources on CMFs, in terms of the jet power and $\theta_v$.} 
   \begin{tabular}{@{}lcccc@{}}
   \hline
   Simulation 
      &max($\left< RM \right> / \left< RM_{\rmn{ini}} \right>$)
         &max($\sigma/\sigma_{\rmn{ini}}$)
            &max($\left< RM \right> / \left< RM_{\rmn{ini}} \right>$)
               &max($\sigma/\sigma_{\rmn{ini}}$) \\
   
      &$\theta_v=\,\,$45$^{\circ}$
         &$\theta_v=\,\,$45$^{\circ}$
            &$\theta_v=\,\,$90$^{\circ}$
               &$\theta_v=\,\,$90$^{\circ}$ \\
   \hline
   lighter-slow	         &1.16    &1.09   &1.11    &1.32 \\ 
   light-slow	         &1.03    &1.13   &1.20    &1.11 \\ 
   lighter-fast	         &1.46    &1.43   &1.18    &1.29 \\ 
   light-fast	         &1.12    &1.23   &1.17    &1.12 \\ 
   lighter-faster        &1.73    &1.68   &1.28    &1.44 \\
   \hline
   \end{tabular}
   \end{minipage}
\label{table2}
\end{table*}

\subsection[]{Field structure evolution} 
\label{POWsection}

In order to investigate the effects that jets have on the structure of
CMFs, we calculate three-dimensional magnetic power spectra in Fourier 
space using the output from our simulations. Specifically, we use
the magnetic fields of the five jets-simulations 
which were
produced at $t=t_e$ (see Table~1).
The spectra are shown in \fig{POW-jets}. 
   For this computation we exclude
the volume that is occupied by the sources
(we use the spatial distribution of the jets' 
tracer, Section~\ref{cocoon}).

In general we find that the spectra preserve the initial condition of a power-law 
with a Kolmogorov index (Section~\ref{cmfs}). Yet the energy injection from
the jets produces power enhancements on scales that range from the resolution
limit, 1\,kpc, to the source size, $\sim\,$40\,kpc. 
An interesting feature that we see is
a flattening on the power spectra
at the scale defined by the cocoon size --\,\fig{POW-jets}.
 This effect 
 depends on
the power of the jets.
As radio sources expand, the energy of the jets is transferred
to the ICM gas and CMFs in the shocked ambient region. There, magnetic
field lines are compressed and stretched tangentially to the surface
area of the radio cocoons (Section~\ref{visit}). 
Hence, we find that fatter or more spherical cocoons, such as
the ones produced by the ``lighter'' jets, are more efficient in increasing
and redistributing the energy of the CMFs.

The scales at which the jets affect the CMFs are more evident in \fig{POW-ratio}.
There we plot the ratio of the power spectra in \fig{POW-jets} over
the corresponding spectra computed using the ``no-source'' 
simulation. i.e. 
%
   for each plot in \fig{POW-ratio}, we use the  
   corresponding profile in \fig{POW-jets} as the numerator.
   The spectrum in the denominator is calculated in the
   following way: (i) we combine the magnetic fields 
   of the ``no-source'' run at $t_{\rmn{jet}}=t_e$
   with the cocoon spatial distribution of the 
   corresponding jets-simulation 
   at $t_{\rmn{jet}}=t_e$.
   (ii) The magnetic fields inside the cocoon are
   removed. This yields a cube filled with magnetic fields, which have not
   been affected by the jets at all, and an empty cavity in the middle.
   The geometry of such cavity exactly matches the shape and orientation of the 
	cocoon's contact surface. (iii) We compute the 3D magnetic power spectrum 
	of these fields. 
%
%
We note the numerical diffusion of the simulations is cancelled in the
computation of the profiles shown in \fig{POW-ratio}.

\begin{figure}
\includegraphics[width=.5\textwidth]{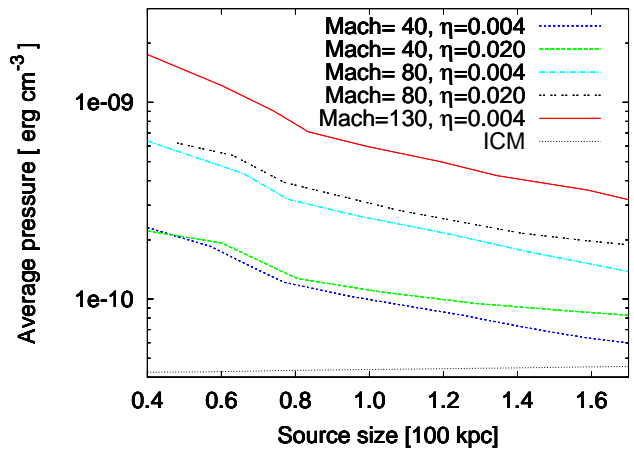}
     \vspace{0pt}
 \caption{Evolution of the average cocoon thermal pressure as a function 
of the source size measured from hotspot to hotspot. The ICM thermal pressure is also
shown (bottom dashed black line) for reference.
}
\label{pressEVO}
\end{figure}

\begin{figure}
  \centering
     \subfigure[Power spectra for jets with different properties.]
        {\label{POW-jets}
   \includegraphics[width=.40\textwidth]{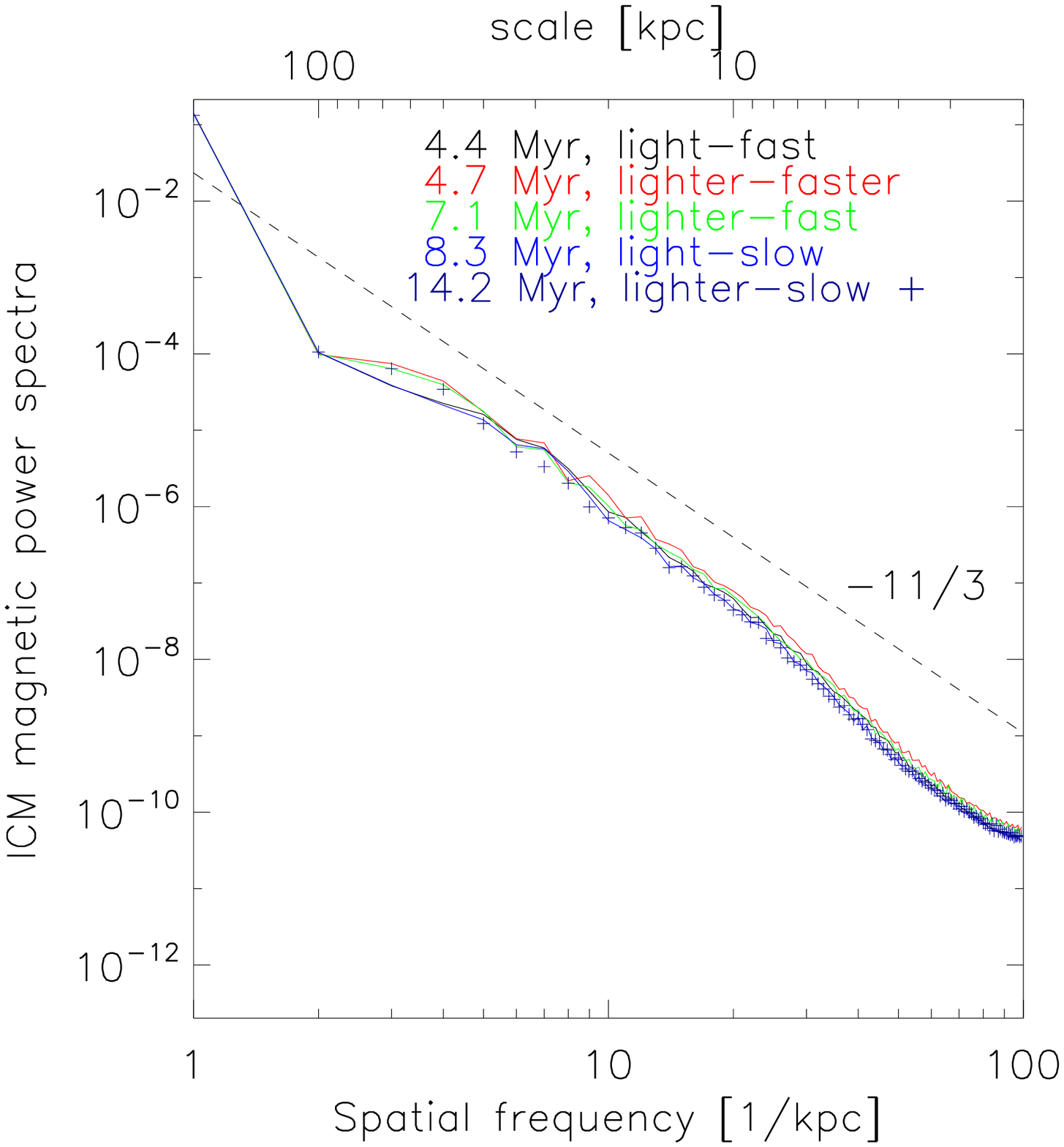}} \\
        \hskip.2cm
     \subfigure[Ratio of the power spectra in (a) over the power spectra of 
	  corresponding no-jets simulations (see Section~\ref{POWsection} for details).]
        {\label{POW-ratio}
   \includegraphics[width=.44\textwidth] {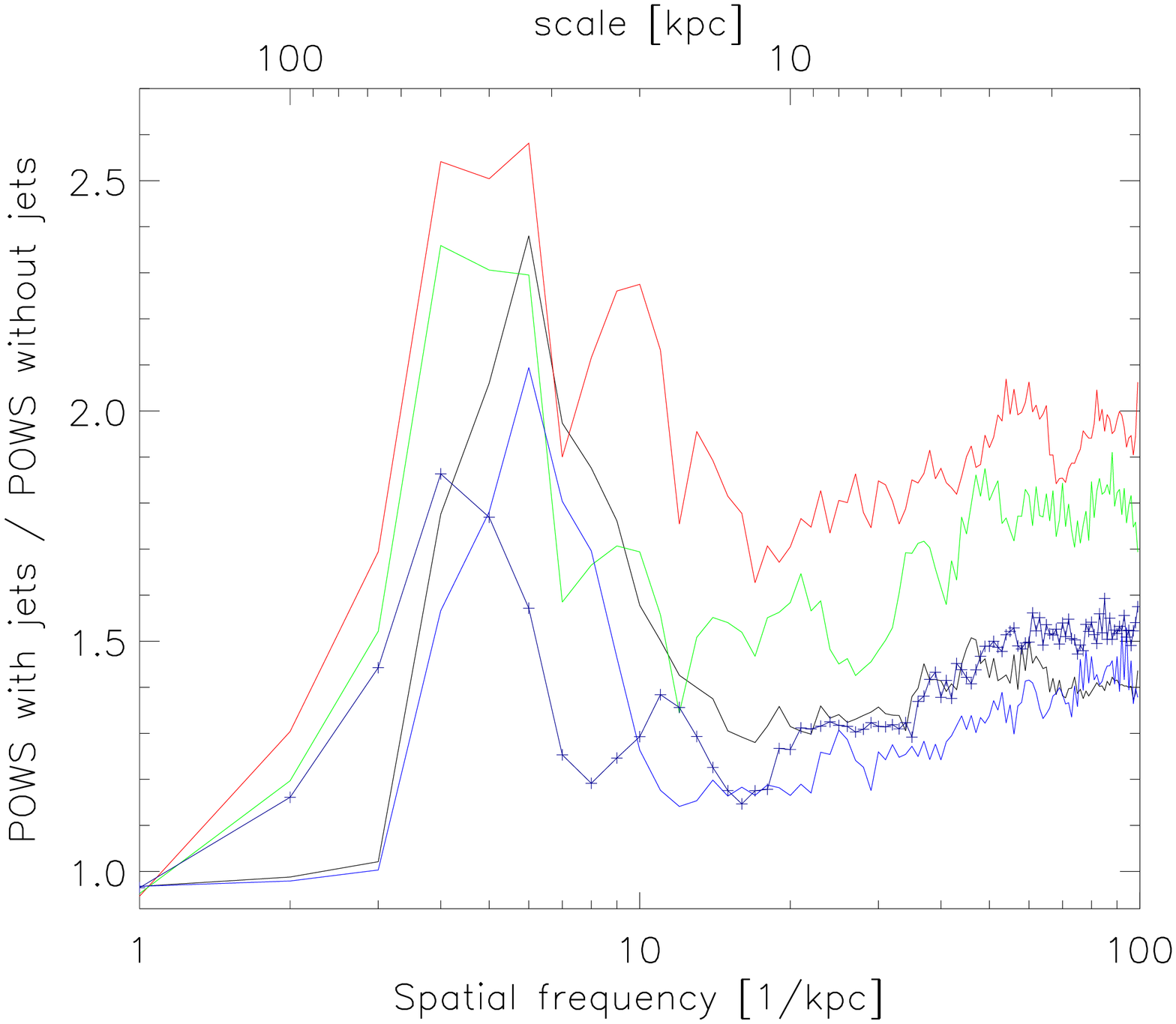}}
        \hskip.2cm
     \subfigure[RM structure function produced at $\theta_v=\,\,$45$^{\circ}$]
        {\label{struct}
	\includegraphics[width=.40\textwidth,bb= .4in .25in 6.6in 7in,clip=]{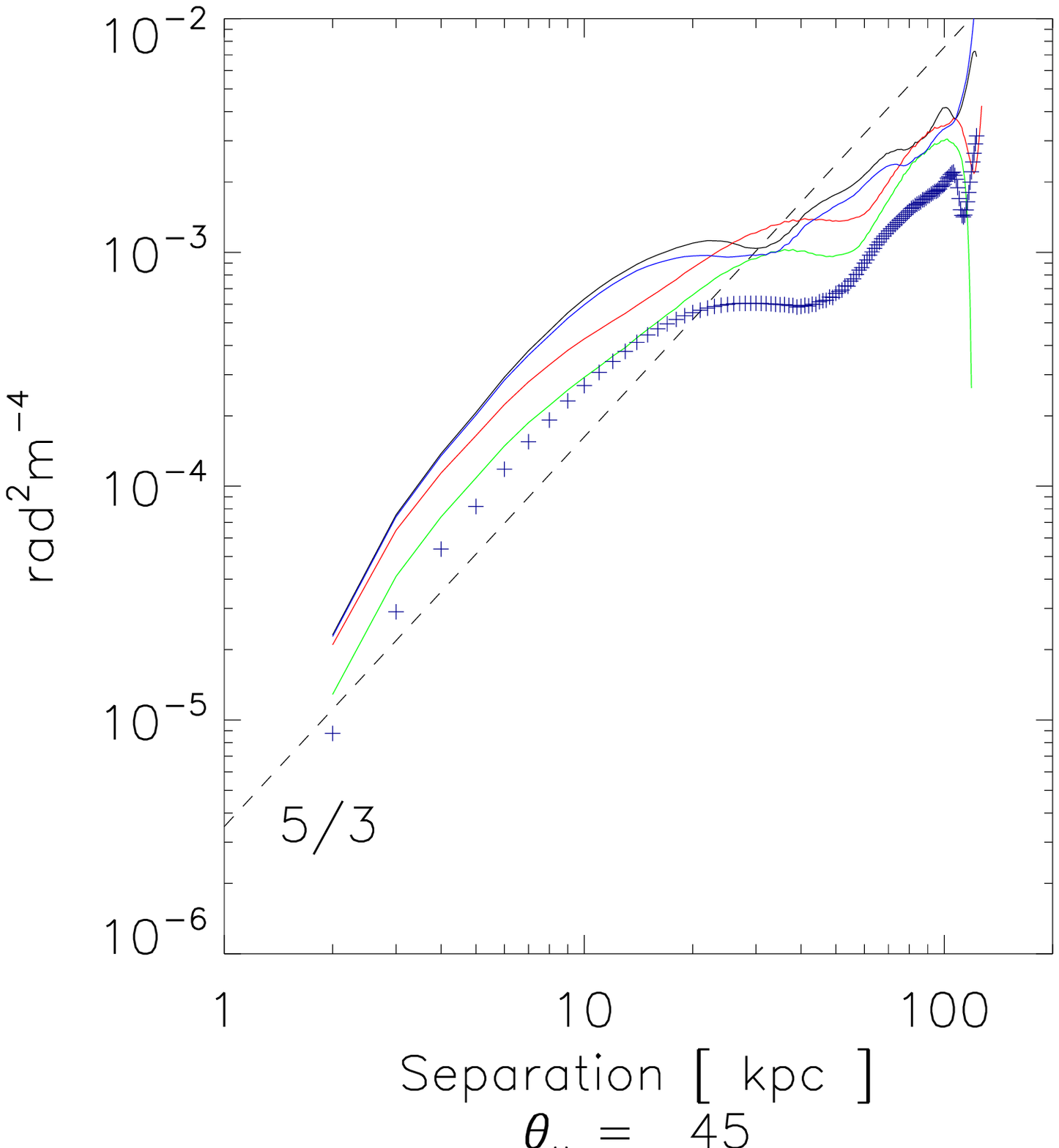}}
     \vspace*{0pt}
  \caption{Structure of CMFs. (a) and (b) are in the 
three-dimensional frequency space. (c) is in the two-dimensional 
scale space of the RM.
}
\label{POW}
\end{figure}

Gradients in these spectral ratio profiles persistently show
maxima close to~\hbox{40\,kpc}, as well as a roughly constant and progressive
increase from scales close to~10\,kpc 
down 
\hbox{to 1\,kpc} --\,see \fig{POW-ratio}. These features are related to
the source energy and momentum injection 
   scales and also to the thickness
   of the ambient medium swept-up region. This gas component expands as the sources
   evolve towards a thermal pressure equilibrium state with the ICM gas 
   (Figure~\ref{pressEVO}), and reaches scales of about~10\,kpc at 
   the end of the simulations.

We see that the expansion and size of our model cocoons are clearly
linked to the enhancement of the CMF power spectra at large-scales.
The cocoons which are formed by the lightest, fast jets are bigger
and have more isotropic gas and energy distributions (see
Section~\ref{energetics}).
\fig{figTopo} illustrates the deformation of the CMFs' topology 
that is caused 
by the ``lighter-fast'' source at $\theta_v=\,\,$90$^{\circ}$ and
$t_{\rmn{jet}}=\,$5.3\,Myr. The cocoon is marked by dots and it
clearly stretches and compresses the neighbouring magnetic field
lines, particularly ahead of the jets' working surface.

\citet{ensslin03} have demonstrated that 
there is a correlation between
the statistical properties of the RM distribution, in the real space, and
the structure of \mf\ in the three-dimensional frequency space. Based
on these facts we 
%
   compute the RM structure functions of those of our synthetic RM maps that
   correspond to 
%
$t_{\rmn{jet}}=t_e$ and \tff.  The structure function is given by
\begin{equation}
S(\Lambda) = \frac{ 1 }{ N(\Lambda) } \sum_{(x,y),(x',y')} 
  \, [ RM(x,y) - RM(x',y')  ]^2,
\end{equation}
\noindent where the sum is taken over the $N(\Lambda)$ non-zero RM pixels 
that are 
inside bins (annuli) centred at $(x,y)$ and have radii 
of $\Lambda=|(x,y) - (x',y')|$ \citep{simonetti84}.

\fig{struct} shows the RM structure functions in question. 
At the end of the simulations,
the structure functions generally 
still show the power law 
that we have used as the 
initial conditions of the simulations (Section~\ref{cmfs}). Yet
these curves 
considerably flatten at scales of 
order~tens of kpc. The structure functions rise
again from a certain scale on. We see that this scale is larger for that sources with
fatter cocoons. The flat regions therefore seem to be directly related
to the cocoon width.
This is consistent with 
our findings in 
the 3D magnetic power spectra at the frequency space --\,\fig{POW-jets}.

\section[]{Discussion}
\label{discussion}

A combination of 3D-MHD numerical simulations and synthetic RM
observations have been produced and presented. These 
simulations follow the evolution
and observational signatures of CMFs under the effects of powerful
jets from an AGN in the core of a non-cool core cluster. To understand
the basic physics of this interaction, the effects from other cluster
galaxies, synchrotron cooling, magnetic reconnections and observational
factors (such as beaming, depolarization and light travel effects)
have not been considered in the model.  
We have focussed on 
the role that the jet velocity, the jet density and the
observational viewing angle play
on the jets-CMFs interaction. 
Plasma energetics, RM statistics and magnetic power spectra consistently
indicate that CMFs are affected by AGN jets. The effect depends
on the jets' properties; the lighter and the faster the jets,
the more important the results they have on the ambient medium.
%

The shape of the CMF energetics profiles (Figure~\ref{fig:energy})
is consistent with the following arguments.
Radio sources expand due to the energy injection
from jets. The cocoons displace the ICM gas and magnetic fields from the
cavity, compressing them into the region located between the contact surface
and the bow shock (\figs{colormaps} and \ref{figTopo}). 
These CMF lines are also stretched in this process. The magnetic component tangential to
the surface of the radio cocoons is therefore enhanced. As the sources expand, the
cocoon pressure decays and the compression caused by bow shocks weakens
(Figure~\ref{pressEVO}). This is in good agreement with radio source evolution models
(e.g. see \citealp{kaiser97}). Once the sources with Mach~40~jets 
(green and dark-blue curves, Figure~\ref{pressEVO}) 
have expanded for about 100\,kpc, their
cocoon thermal pressure is comparable with that in the environment. Thus
only these relatively slow sources are not powerful enough to significantly
affect the magnetic energy of the ICM. 

Our RM integration neglects the contribution from ICM gas and magnetic fields
beyond the computational domain.
%
%
Jet-produced RM enhancements are local, important at the edge of the sources
and should not depend on this outer RM component. Observations
show that both the cluster ion density and the strength of the CMFs decreases
away from the cluster core (\citealp{kim91,clarke01};~\citealp{carilli02},
and references therein).  
Thus we expect to capture the major contribution to the
RM by our procedure. Since the region we use for the analysis is
spherical, any additional contribution 
to the RM is expected to be statistically spatially uniform.


\begin{figure}
  \centering
  \includegraphics[width=42mm,bb= .1in 0in 6.55in 6.515in,clip=] 
{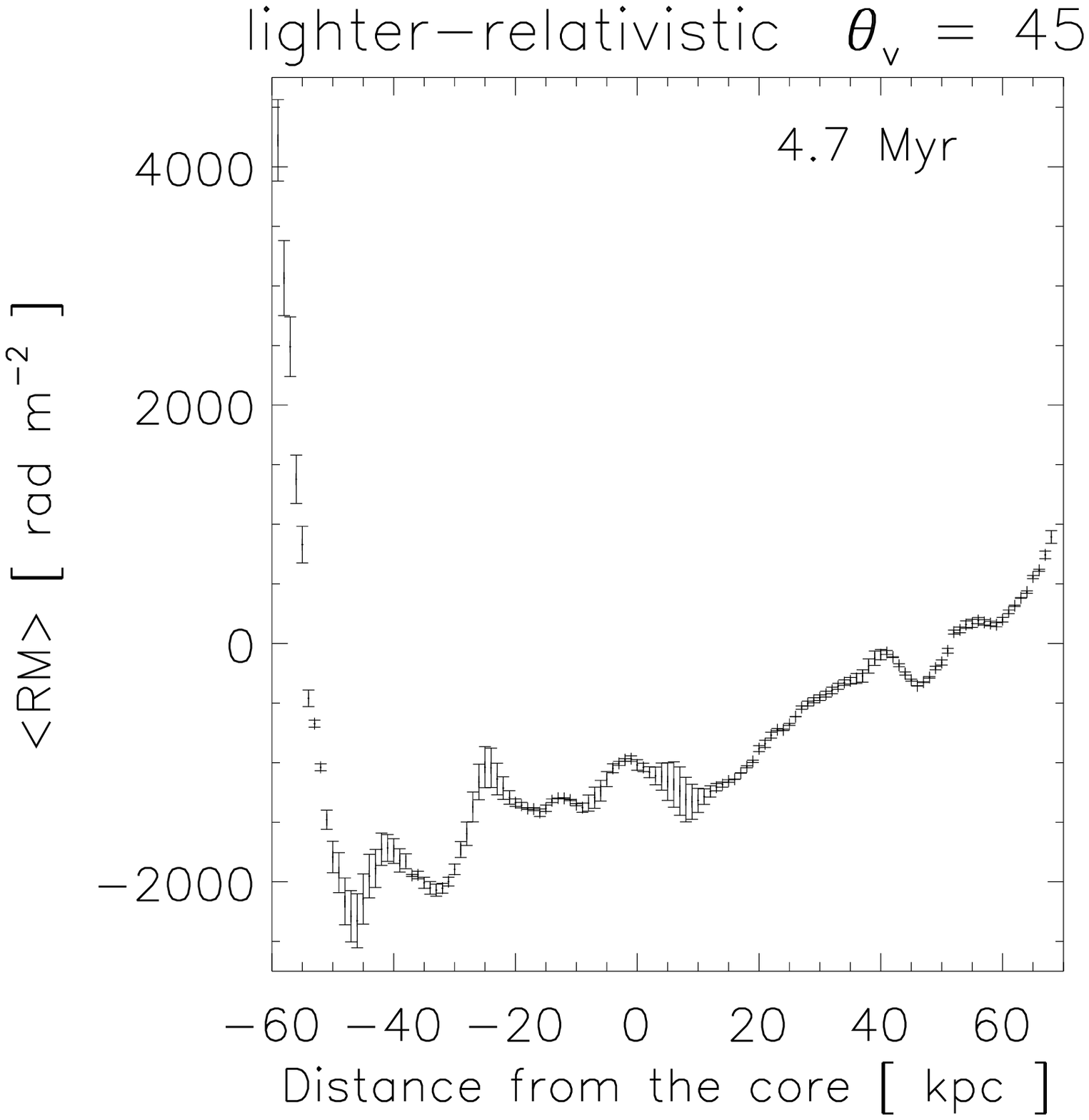}
  \includegraphics[width=40mm,bb= .4in 0in 6.55in 6.505in,clip=] 
{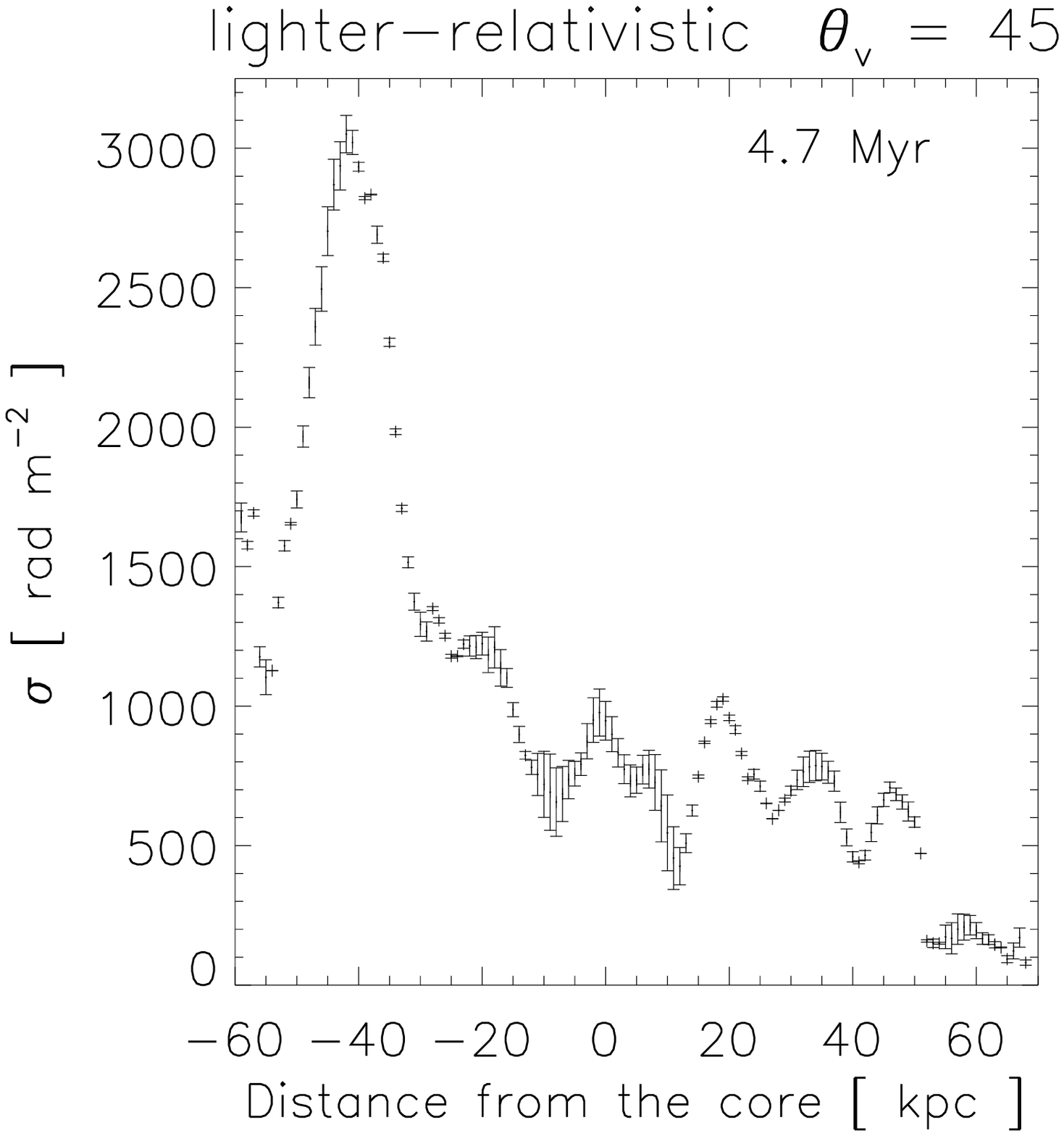}
  \caption{Profiles of RM statistical fluctuation projected onto the 
jet axis of our ``lighter-faster'' 
source. The viewing angle corresponding to these maps is 45$^{\circ}$.
The receding lobe is in the negative distance range where fluctuations
are the strongest. 20\,kpc represent 18.7\,arcsec approximately.
}
     \vspace*{20pt}
\label{fig:hydra}
\end{figure}

\subsection{Characteristic RM gradients} 
\label{RMgradients}

   \citet{guidetti11} have recently reported bands in the observed
	RM images of the FR~II radio sources 3C\,353 and Cygnus~A, as well
   as in the FR\,I sources 0206+35, M\,84 and 3C\,270. These bands are
   perpendicular to the major axis of the radio lobes and are
   characterized by very little small-scale RM structure. To explain the
   nature of such features, Guidetti et~al. present synthetic RM
   observations generated with 
	a simple model of gas with constant
   density and magnetic field distributions. This plasma is then affected by
   compression induced by an elliptical cavity that is artificially
   placed inside the gas. The model allows these authors
   to conclude that there are three magnetic field components associated
   with observed RM: an isotropic field which has large-scale fluctuations,
   plausibly associated with the undisturbed ICM; a well-ordered field
   draped around the leading edges of the radio lobes; a field with
   small-scale fluctuations which is located 
	in the shells of compressed gas surrounding the inner lobes.  
%
We do not use an ordered magnetic field component 
   in our model 
and do not observe RM bands in our synthetic images.
Thus we confirm the findings of Guidetti et~al., namely that without any 
ordered field component 
   in the ICM, 
the radio source expansion will not lead to banded RM structures.

   In addition, our model shows 
	that source-induced RM variations can be as large as
	70\%, are correlated with the radio jet properties, depend on time 
	and are stronger at the edge of the cocoons. This is a very important point
	because the observed RM signal is used to infer the strength of the CMFs
	\citep[e.g. see][]{feretti95,feretti99,taylor00,eilek02}.


\subsubsection{Asymmetrical RM gradients between radio lobes}
\label{hydra}

\begin{figure*}
  \centering
  \includegraphics[width=77mm,bb=.3in 0in 8.5in 7.5in,clip=]{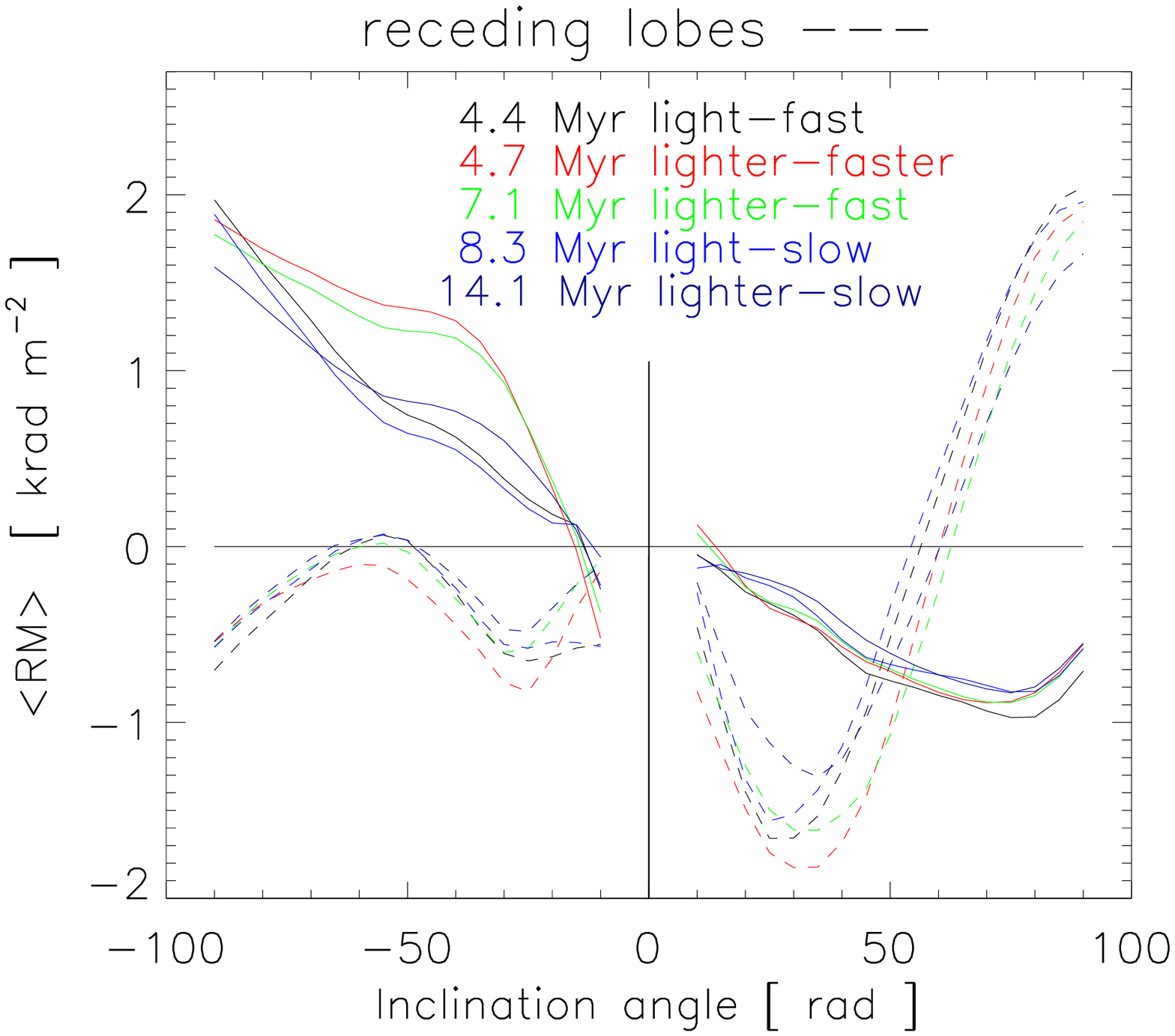}
        \hskip.2cm
  \includegraphics[width=75mm,bb=.3in 0in 8.3in 7.5in,clip=]{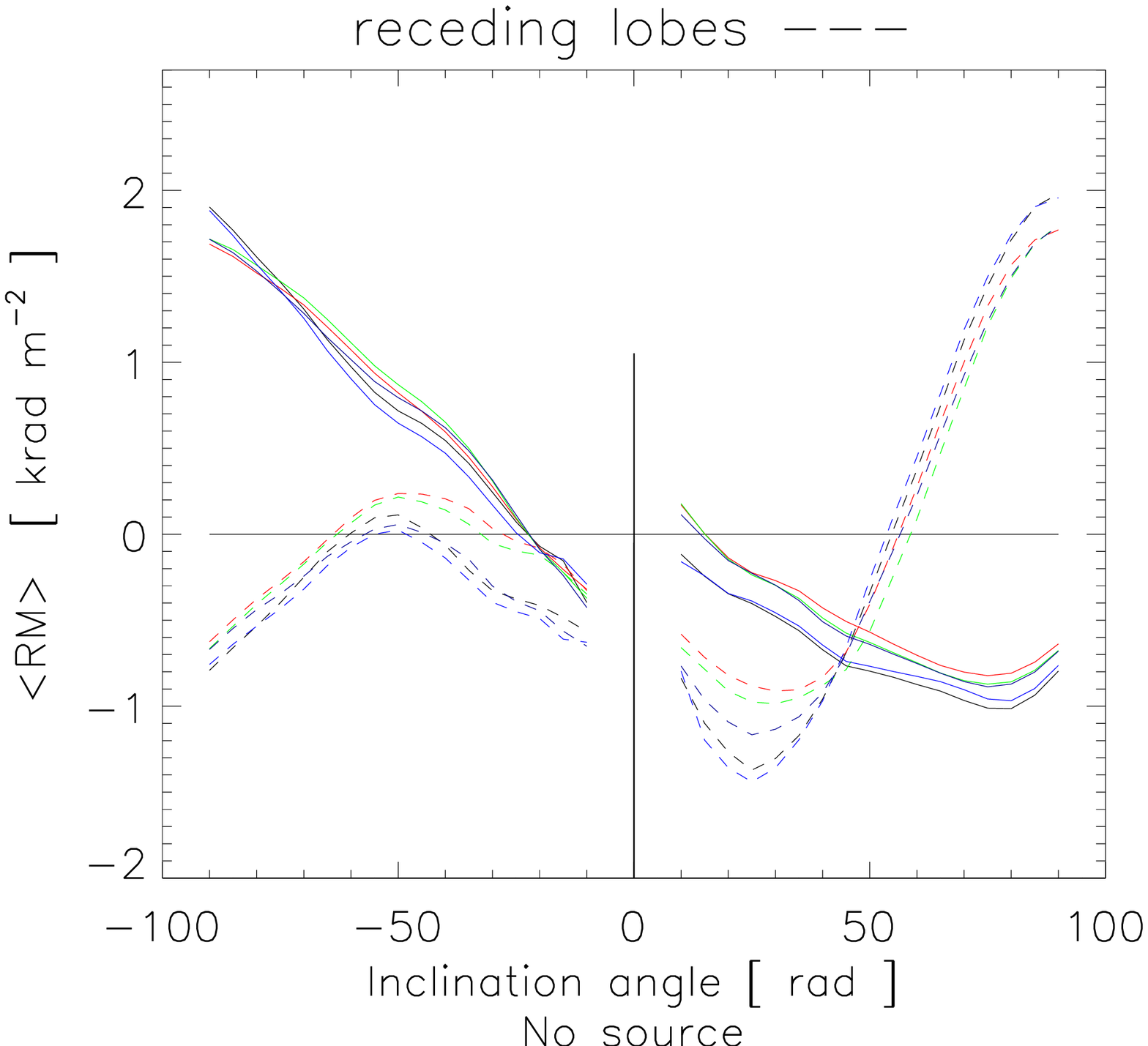} \\
\vskip.5cm
~~~
  \includegraphics[width=66mm,bb=0.5in 0in 8.5in 7.5in,clip=]{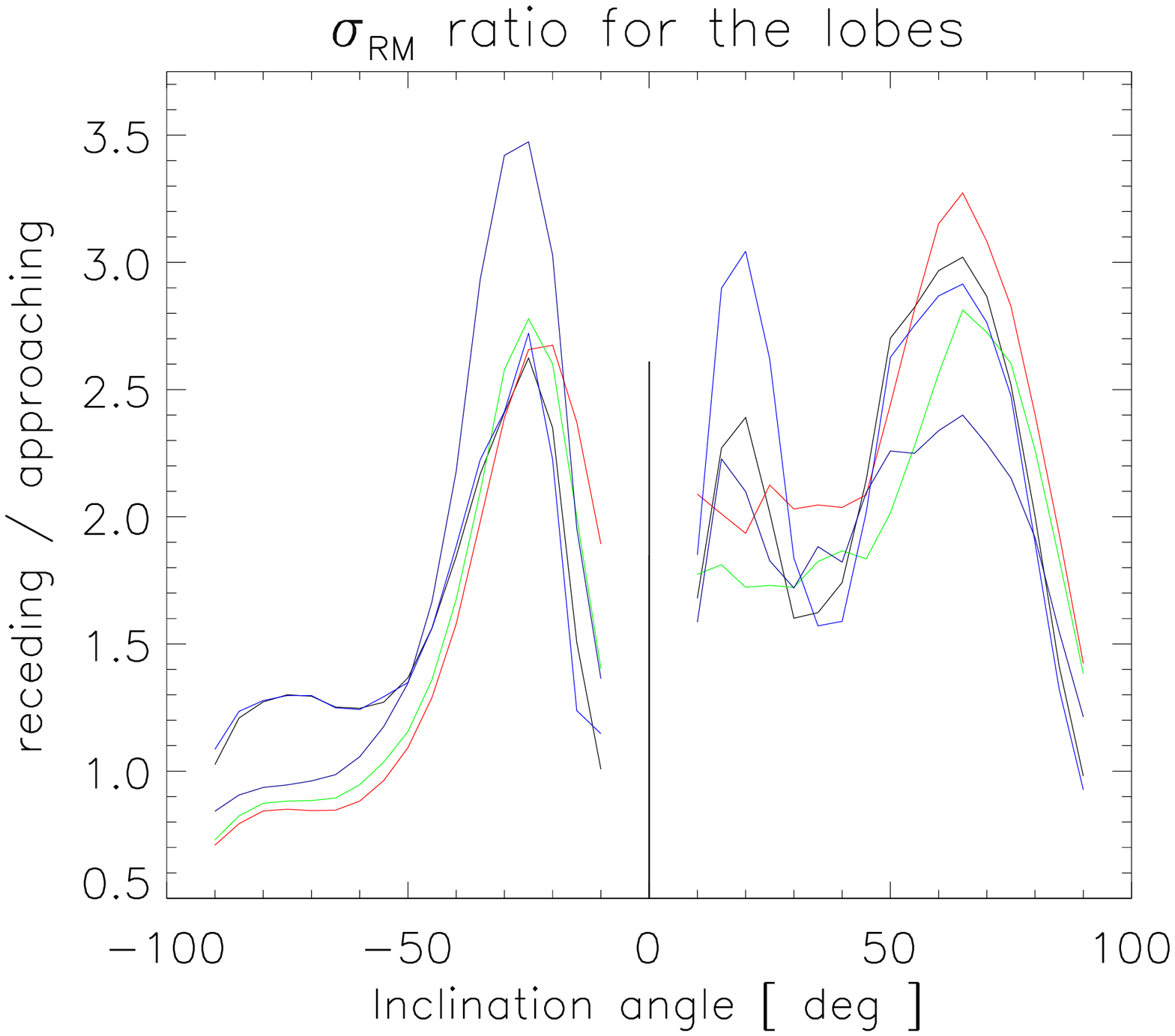}
~~~~~
        \hskip.2cm
  \includegraphics[width=65mm,bb=0.5in 0in 8.3in 7.5in,clip=]{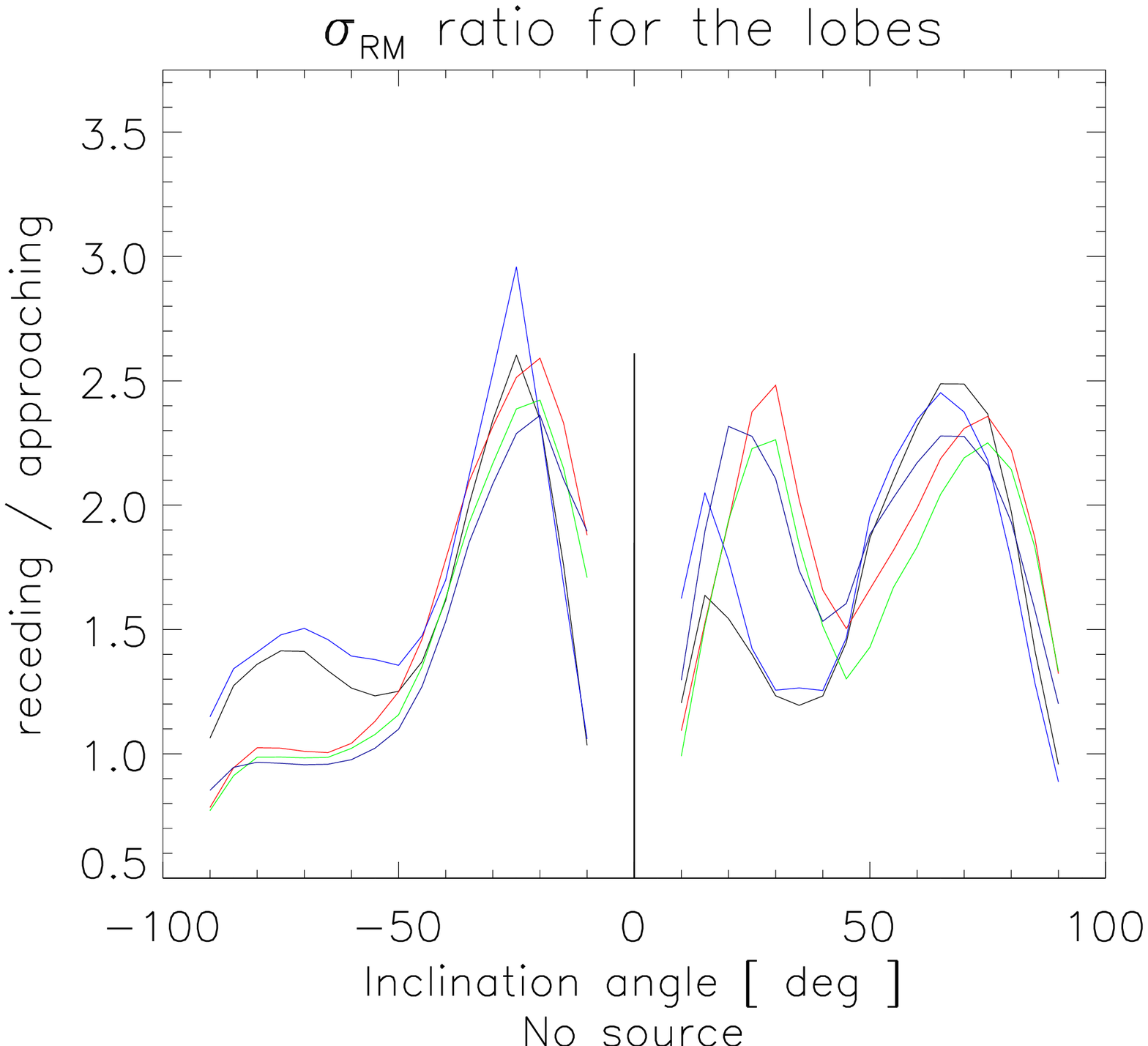}
  \caption{RM statistics comparison between the approaching and receding
radio lobes as a function of the angle between the jet axis and the
line of sight. Top: mean RM of the approaching (solid) and the
receding (dashed) lobes. Bottom: \hbox{$\sigma_{RM}$(receding lobe) $/
\sigma_{RM}$(approaching lobe)}. Positive and negative angles represent
RM integrations along and against the line of sight, respectively.  
Panels in the right column show data from no-source RM maps 
(Section~\ref{RMmapsB}).
}
     \vspace*{0pt}
\label{lobes}
\end{figure*}

Our synthetic RM maps persistently show pronounced gradients at regions
corresponding to the receding radio lobes (the left ones) 
relative to the approaching ones
(\figs{RM-lighter-fast90}--\ref{RM-light-fast45}). Such RM differences
have been observed in the FR~IIs
Cygnys~A \citep{dreher87,carilli94} and 3C\,353 \citep{guidetti11}
as well as in several FR~Is (e.g. 3C\,31 and Hydra\,A,
\citealp{laing08}; 3C\,449, \citealp{guidetti10}; 0206+53,
3C\,270 and M\,84, \citealp{guidetti11}).
Although we simulate RM enhancements caused by FR~II radio sources, FR~Is
--\,which are known to form ICM X-ray cavities as well\,-- may well
produce similar effects.
Hydra\,A is a well studied central cluster radio source. The Faraday
rotation in this case is thought to be dominated by the local ICM
contribution. 
This cluster presents an ICM magnetic
spectral index of \hbox{$n \, \approx$\,11$/$3} which has been 
inferred against the emission from the embedded luminous FR~I radio source
Hydra\,A. The inclination angle of this source has been inferred to be
\hbox{within 37--60\,degrees}
\citep{taylor93,ensslin06,laing08}.  

In terms of RM statistical
fluctuations that are asymmetrical between the source's lobes, we find
that our ``lighter-faster'' 
simulation (Table~1) agrees to a
certain degree with the
studies of \citet{taylor93} and \citet{laing08} on Hydra~A. This is
partly unexpected, as we have chosen a particular realisation for the
magnetic power spectrum of the CMFs. However, we focus here on the edge effects,
which is where we believe that radio sources have the strongest effect:
Two panels are shown in Figure~\ref{fig:hydra} where 
we present curves of $\left< RM \right>$ (left panel) and $\sigma_{\rmn{RM}}$
(right panel) as a function of position along a horizontal line between the lobes.
The mean and standard deviation of the RM are taken on vertical slices
of the map corresponding to the ``lighter-faster'' 
simulation
at $\theta_v=\,\,$45$^{\circ}$ and $t_{\rmn{jet}}=\,$4.7\,Myr.  
Vertical error bars are
computed by taking the difference between the red and the black curves
in Figure~\ref{RM-lighter-relativistic45}. 
Both statistical measures have large gradients near the left edge. This effect is produced
by the expanding radio source. Towards the right edge, the radio
source expansion has also strongly amplified the RM values. However,
the initial magnetic field in that region happened to be small, which
is why this region does not look spectacular.
This is quite comparable to the RM distribution in
observed radio sources,
   e.g. see \citet[][Figure~22, panels \textit{a} and \textit{b}]{laing08}
   who present the RM observations of Hydra~A made by \citet{taylor93}
   along with synthetic RM observation.
It is therefore well possible that such large RM-gradients towards the
edges of radio sources are strongly influenced by the lobe expansion.
Note that we consider here only one realisation for the 
%
random 
%
Faraday screen. A detailed match of the considered RM gradient with any
observation is therefore not intended. 

Our model jets inject energy at scales that range from the resolution
limit to the source size, but preferentially near the latter. 
As we have seen, this results in a flattening in both 
the magnetic power spectra and the RM structure
functions at large-scales.  
\citet{laing08},
\citet{guidetti10} and \citet{guidetti11} have reported such flat gradients in the observed
structure functions of the FR~I sources 3C~31, 3C\,270, 3C~449 and Hydra\,A.
Our simulations show that the physics of radio source expansion may
explain these structure function gradients naturally. Thus conclusions
about the nature of the ICM turbulence may not be drawn from the
gradients in question.

\subsubsection{The Laing-Garrington effect}
\label{screens}

We have calculated the mean and the standard
deviation of the RM for the approaching
(\meanrm$^{\rmn{ap}}$ and $\sigma_{\rmn{RM}}^{\rmn{ap}}$)
and the receding
(\meanrm$^{\rmn{re}}$ and $\sigma_{\rmn{RM}}^{\rmn{re}}$)
source lobes, as a function of the viewing angle.
   In the top left panel of \fig{lobes} we show \meanrm$^{\rmn{ap}}$
   in solid lines and \meanrm$^{\rmn{re}}$ in dashed lines. In all of these
   plots, positive (negative) angles correspond to integrations along
   (against) the line of sight, and again jets are
   on the plane of the sky at \hbox{$\pm$90$^{\circ}$}.  In the
   bottom left panel of \fig{lobes} we show
   $\sigma_{\rmn{RM}}^{\rmn{re}}/$$\sigma_{\rmn{RM}}^{\rmn{ap}}$.
   In addition, to demonstrate the effect of the jets on the Faraday screens,
	which will in turn affect the RM computations, we produce
	analogous (\meanrm$^{\rmn{re}}$, \meanrm$^{\rmn{ap}}$ and
	$\sigma_{\rmn{RM}}^{\rmn{re}}/$$\sigma_{\rmn{RM}}^{\rmn{ap}}$)
	plots but use the data from the no-source RM maps
	(Section~\ref{RMmapsB}). We show these no-source RM plots
   in the right column of \fig{lobes}, side by side to
	their \hbox{(with-)} source counterparts.  
	
Together, all of these profiles show that
the general dependence of the lobe RM statistics on the viewing
angle is shaped by the intrinsic nature of the Faraday screens, as it is
generally assumed in the interpretation of the Laing-Garrington
effect. The result of the radio source expansion is to amplify several RM features.

Only for positive 
inclinations our realisation results in stronger
average RMs in the receding lobes (\fig{lobes}, top panel). We see strong amplification at
inclinations of $20^\circ<\theta_\mathrm{v}<70^\circ$.
$\sigma_\mathrm{RM}$ is linked to the depolarisation \citep{burn66,krause07}. Already
intrinsically, the depolarisation is always higher for the receding
lobe
($\sigma_{\rmn{RM}}^{\rmn{re}}/$$\sigma_{\rmn{RM}}^{\rmn{ap}}>$1).
   Panels in the bottom row (\fig{lobes}) show that
this trend is mostly amplified by the radio sources, apart from the fatter (lighter)
ones for which the $\sigma$-ratio
drops below unity near $\theta_\mathrm{v}=\pm$90$^\circ$.
Thus the Laing-Garrington effect is only moderately affected by the radio
source expansion, in such a way that the associated trends tend to be amplified.

We have shown that powerful radio sources affect the structure of
CMFs which is characterized by the magnetic power spectrum index~$n$.
This process is important because
the ICM energy transformation,
either from or to magnetic form, as well as the observations
\citep{murgia04}, depend on $n$.
Hence we agree with
\citet{laing08} in that the structure of CMFs inferred by using the
emission of cluster radio sources should take a piecewise defined
form over the range of the spatial frequencies.  Our simulations
show, moreover, that such functional forms change with time (see
Figure~\ref{POW}a,~b; note that the profiles in panel \textit{b}
would be completely flat otherwise). Faraday rotation measurements from clusters 
with broad differences in age will be key in the determination
of evolving magnetic structure functions of the CMFs.

\section{Summary and conclusions}
\label{conclu}

We have presented a combination of 3D-MHD numerical simulations and
synthetic RM observations to model the evolution and observational
signatures of cluster magnetic fields under the effects of powerful
AGN jets at the core of a non-cool core galaxy cluster.  We prescribe
the cluster magnetic field as a Gaussian random field with 
a power law energy spectrum tuned to 
%
	resemble the expectations for 
%
turbulence.
We focus on how both the jets' power, in terms of their density 
and velocity, and the observational angle affect the observed RM signal.
Such signal is used by observational studies to infer the strength of
cluster magnetic fields. Our results are very consistent and show 
the following.

Powerful jets increase the ICM energy in proportion to their velocity
and in inverse proportion to their density.
Light jets are more efficient in this process because they inflate
fatter cocoons than relatively heaver jets. 
%
	Only jets with Mach numbers of~80 and~130, which are over-pressured 
	with respect 
	to the ambient medium, significantly increase the ICM magnetic energy.
%
We see that the time evolution of the CMF energetics is correlated with that
of the cocoon pressure.

Jets compress and displace the magnetised ICM. This results in a
flattening of the 3D magnetic power spectra and the associated
field RM structure functions. This happens at scales from the
resolution limit to the source size, but preferentially near the
latter. We see the effect is correlated with the jet velocity and
pronounced for fat cocoons. 
The RM structure functions of 
 3C~31, 3C\,270, 3C~449 and Hydra\,A show flat gradients at large scales, and the
cause of such shapes has been previously modeled 
\citep{laing08,guidetti10,guidetti11}
in terms of ICM turbulence.
Our model suggests, however, that the physics of radio source
expansion may as well explain this feature.

The general distribution of the Faraday depth is determined by the
intrinsic nature of the ICM.
Powerful jets enhance the RM locally by factors up to~$\sim\,$1.73.
We find this process is proportional to the jet velocity and in inversely
proportional to the beam density. 
Expanding hypersonic sources compress the ICM gas and magnetic fields
in the shocked ambient region. The strength of the fields
that are perpendicular to the local lobe expansion direction increases, 
specially near the hotspots.  
The alignment of compressed CMF components and the line of
sight is favoured (i) at the cocoon edge (ii) by sources with a fat cocoon
(iii) at $\theta_v$ within 20--70~degrees, due to the cocoon geometry.
Jet-produced RM enhancements are thus more important under these
conditions and should be more evident in radio galaxies than in
quasars. 

Both $\left< RM \right>$ and $\sigma_{RM}$
continuously change and tend to increase as the radio sources develop. 
``Lighter'' sources with jet velocities~$\ge\,$40\,Mach yield $\left<RM\right>$ and
$\sigma_{RM}$ enhancement of factors within~1.2 to~1.7, in proportion
to the jet velocity. 
These RM enhancements may therefore lead to
overestimations of the CMF strength by up to about 70\%.

\section*{Acknowledgements}

It is a pleasure to thank Robert Laing and Malcolm Longair for
useful discussions about this paper, as well as the anonymous referee
for comments that helped to improve this work. The software used in these
investigations was in part developed by the DOE-supported ASC /
Alliance Center for Astrophysical Thermonuclear Flashes at the
University of Chicago. MHE acknowledges: financial support from
CONACyT (The Mexican National Council of Science and Technology,
196898/217314); useful discussions with David~Titterington and
Jongsoo~Kim; Dongwook~Lee for the 3D-USM-MHD solver of Flash3.1;
Volker Gaibler for the routines to map magnetic field geometry.


\bsp

\label{lastpage}
 
\end{document}